\documentclass[conference,compsoc]{IEEEtran}

\usepackage{amssymb}     
\usepackage{amsmath}
\usepackage{pifont}    

\usepackage{graphicx}    
\usepackage{booktabs}
\usepackage{multirow}
\usepackage{array}
\usepackage{longtable}
\usepackage{tabularx}

\usepackage{hyperref}
\usepackage[noabbrev,nameinlink]{cleveref}
\hypersetup{
    colorlinks = true,
    linkcolor = blue,
    anchorcolor = blue,
    citecolor = blue,
    filecolor = blue,
    urlcolor = blue
}
\usepackage[frozencache=false,cachedir=minted-cache]{minted}
\usepackage[minted,most]{tcolorbox}
\usepackage{xcolor}
\usepackage{minted}
\setminted[solidity]{
    autogobble=true,
    framesep=2mm,
    fontsize=\scriptsize,
    breaklines=true,
    bgcolor=gray!4
} 
\usepackage{xspace}
\usepackage{listings}
\usepackage{enumerate}
\usepackage{enumitem}
\usepackage{caption}
\usepackage{graphicx}
\usepackage{subcaption} 
\usepackage[framemethod=TikZ]{mdframed}

\newcommand{\smartbugs}{\textit{Smart-bugs}\xspace}
\newcommand{\primevulc}{\textit{PrimeVul}\xspace}
\newcommand{\reposvulcpp}{\textit{ReposVul\_cpp}\xspace}
\newcommand{\reposvulpy}{\textit{ReposVul\_py}\xspace}

\newcommand{\downgrade}{ {\textit{downgrade}}\xspace}
\newcommand{\upgrade}{ {\textit{upgrade}}\xspace}

\definecolor{deepseek-r}{RGB}{91, 181, 172}    
\definecolor{deepseek-nr}{RGB}{139, 218, 210}   
\definecolor{qwen-r}{RGB}{216, 179, 101}        
\definecolor{qwen-nr}{RGB}{235, 203, 136}       
\definecolor{llama-r}{RGB}{222, 82, 108}        
\definecolor{llama-nr}{RGB}{243, 126, 148}      
\definecolor{lightblue}{rgb}{0.678, 0.847, 0.902}

\usepackage{colortbl}
\usepackage{xintexpr, xinttools}
\usepackage{hhline}

\newcommand{\ccell}[1]{ 
    \cellcolor{blue!\xinttheiexpr 30*#1\relax} {#1}
}

\newcommand{\myframed}[2]{%
  \begin{mdframed}[linecolor=#1!90,backgroundcolor=#1!4,roundcorner=2pt,linewidth=1.5pt]
  #2
  \end{mdframed}
}

\newcounter{findingcounter}
\newcounter{suggestioncounter}

\newcommand{\findbox}[1]{%
  \stepcounter{findingcounter} 
  \myframed{deepseek-nr}{\textbf{Finding \#\thefindingcounter: }#1} 
}

\begin{document}


\title{\Large \bf A Systematic Study of Code Obfuscation Against LLM-based Vulnerability Detection}

\author{
    \IEEEauthorblockN{
        Xiao Li$^{1}$, Yue Li$^{1}$, Hao Wu$^{1}$,
        Yue Zhang$^{2}$, Yechao Zhang$^{3}$,
        Fengyuan Xu$^{1}$\textsuperscript{{\footnotesize *}},
        Sheng Zhong$^{1}$
        }
    \IEEEauthorblockA{
        $^{1}$National Key Lab for Novel Software Technology, Nanjing University \\
        $^{2}$Shandong University 
        $^{3}$Nanyang Technological University
    }
}
\maketitle 
\renewcommand{\thefootnote}{\fnsymbol{footnote}} 
\footnotetext[1]{Corresponding author: Fengyuan Xu (fengyuan.xu@nju.edu.cn)}

\renewcommand{\thefootnote}{\arabic{footnote}}

\maketitle

\begin{abstract}

As large language models (LLMs) are increasingly adopted for code vulnerability detection, their reliability and robustness across diverse vulnerability types have become a pressing concern. In traditional adversarial settings, code obfuscation has long been used as a general strategy to bypass auditing tools, preserving exploitability without tampering with the tools themselves. Numerous efforts have explored obfuscation methods and tools, yet their capabilities differ in terms of supported techniques, granularity, and programming languages, making it difficult to systematically assess their impact on LLM-based vulnerability detection.  To address this gap, we provide a structured systematization of obfuscation techniques and evaluate them under a unified framework. Specifically, we categorize existing obfuscation methods into three major classes (layout, data flow, and control flow) covering 11 subcategories and 19 concrete techniques. We implement these techniques across four programming languages (Solidity, C, C++, and Python) using a consistent LLM-driven approach, and evaluate their effects on 15 LLMs spanning four model families (DeepSeek, OpenAI, Qwen, and LLaMA), as well as on two coding agents (GitHub Copilot and Codex).  Our findings reveal both positive and negative impacts of code obfuscation on LLM-based vulnerability detection, highlighting conditions under which obfuscation leads to performance improvements or degradations. We further analyze these outcomes with respect to vulnerability characteristics, code properties, and model attributes. Finally, we outline several open problems and propose future directions to enhance the robustness of LLMs for real-world vulnerability detection.

\end{abstract}
\section{Introduction}


Large language models (LLMs) have evolved from natural language understanding to sophisticated code reasoning, enabling automated vulnerability detection and repair across multiple programming languages. Models such as GPT-4~\cite{openai2023gpt4}, DeepSeek~\cite{liu2024deepseek}, and LLaMA~\cite{touvron2023llama} demonstrate strong semantic reasoning by jointly analyzing natural language descriptions and program logic. Moreover, rapidly emerging coding agents claim to possess advanced code review capabilities that can identify critical vulnerabilities before deployment~\cite{github-copilot, codex}. Yet as these models become increasingly deployed in software auditing pipelines, their reliability and robustness against adversarial perturbations remain open concerns.

A fundamental and long-standing challenge in this context is code obfuscation, a process that alters a program’s structure while preserving its functionality~\cite{collberg1997taxonomy}. In traditional settings, obfuscation has been used defensively to protect intellectual property and offensively to evade malware detection. In the LLM era, this practice exposes a new vulnerability: since LLM-based detectors depend heavily on statistical regularities in identifiers, layout, and data representations rather than explicit symbolic execution, obfuscation that disrupts such regularities can substantially change model predictions without altering program semantics. Even simple transformations such as renaming variables~\cite{yang2022natural, tian2023code, zhang2023black, du2023extensive, zhao2025adversarial, huang2025iterative, nikiema2025codebarrier, yefet2020adversarial, li2024aacegen, jha2023codeattack, zhang2022towards, zhang2020generating} or using misleading comments~\cite{swindle2024evaluation} have been shown to confuse models that rely on surface-level cues instead of deeper structural reasoning.

Despite this emerging evidence, existing studies remain fragmented. Prior work often explores only a limited subset of obfuscation methods, primarily those affecting superficial code layout~\cite{zhang2020generating,yefet2020adversarial,yang2022natural,zhang2022towards,tian2023code,zhang2023black,du2023extensive,jha2023codeattack,swindle2024evaluation,li2024aacegen,zhao2025adversarial,huang2025iterative,nikiema2025codebarrier}. Richer transformations involving data-flow encoding~\cite{nikiema2025codebarrier, swindle2024evaluation, tian2023code, zhao2025adversarial, yan2024llm} or control-flow restructuring have not been systematically examined. Similarly, evaluations usually target one or two models and rely on coarse binary judgments of whether a vulnerability exists, neglecting the more detailed aspects of identifying its type or location. Dataset diversity is also limited, with most experiments confined to one language and dominated by a few vulnerability classes. Moreover, the lack of investigation into coding agents leaves their robustness against obfuscation largely unknown to the security community. These constraints prevent a comprehensive understanding of how obfuscation interacts with model architecture, reasoning ability, and data complexity.
To address these gaps, our study is guided by five research questions focusing on obfuscation methods, model diversity, detection granularity, dataset diversity, and the performance of coding agents.


To systematically answer these questions, we build a unified experimental framework that integrates 19 representative obfuscation techniques grouped into 11 subcategories and three major classes: layout, data flow, and control flow. These transformations are applied to four programming languages (Solidity, C, C++, and Python) drawn from publicly available vulnerability datasets. We evaluate 15 LLMs across four major model families (DeepSeek, OpenAI, Qwen, and LLaMA), covering reasoning and non-reasoning variants and spanning parameter scales from 7B to 671B. Our evaluation employs a fine-grained four-level scoring scheme to measure not only whether vulnerabilities are detected but also whether their type and location are correctly identified.

Our analysis reveals several key findings. \textit{First}, surprisingly, obfuscation exerts a dual influence: while expected performance degradation (downgrade) is common, certain transformations unexpectedly improve detection accuracy (upgrade) by removing misleading surface cues. \textit{Second}, control-flow virtualization and mixed-programming-language transformations have the strongest degrading effect. \textit{Third}, model capacity defines a clear robustness boundary: models smaller than 8B parameters show pronounced instability, whereas those larger than 8B maintain higher resilience, though additional scaling yields diminishing returns. \textit{Fourth}, reasoning-augmented models perform better on unobfuscated code but are paradoxically more sensitive to obfuscation, revealing a trade-off between reasoning power and generalization stability. \textit{Fifth}, dataset properties strongly mediate these effects, with vulnerability types involving pointer safety, reentrancy, and access control showing the largest fluctuations under obfuscation. \textit{Sixth}, although coding agents exhibit higher detection success rates than general-purpose LLMs, they still experience both downgrade and upgrade effects under code obfuscation, particularly when facing inline assembly and virtualization techniques. \textit{Finally}, hot-plugging a new model into the agent framework can reduce its effectiveness in transferring vulnerability-detection knowledge, leading to the partial comprehension issue noted in DeepSeek-V3 $\times$ GitHub Copilot. \looseness=-1

These observations point to a broader implication for both attackers and defenders. For adversaries, effective evasion lies not in arbitrary complexity but in semantics-preserving transformations that exploit the gap between statistical and symbolic reasoning within LLMs. For defenders, robustness cannot be achieved merely through scale or reasoning enhancement; it requires cross-layer understanding of code semantics, diversity-aware training, and consistency validation across model families. Accordingly, strengthening the underlying models themselves along with developing hot-plugging mechanisms that preserve precision when integrating them into coding agents will be essential for advancing the next generation of robust, agent-based vulnerability detectors. Our findings thus chart a path toward building more resilient LLM-based vulnerability detectors by revealing how obfuscation reshapes the boundary between code semantics and model perception. {Our key contributions are as follows:}

\begin{itemize}[leftmargin=*]
   
\item \textbf{Comprehensive Taxonomy.} We categorize existing obfuscation techniques into three major classes—layout, data flow, and control flow—covering 11 subcategories and 19 concrete methods, enabling structured comparison across languages and use cases.

\item \textbf{Unified Evaluation Framework.} We implement\footnote{Source code, dataset and results are available at \url{https://github.com/oxygen-hunter/SoK-Code-Obfuscation-in-LLM-VD-arxiv}} these transformations across four programming languages (Solidity, C, C++, and Python) and evaluate their impact on 15 LLMs spanning four model families (DeepSeek, OpenAI, Qwen, and LLaMA), as well as two coding agents (GitHub Copilot and Codex), covering a wide range of parameter scales and reasoning configurations.

\item \textbf{Systematic Findings.} We reveal a nuanced landscape: while obfuscation can predictably degrade performance, certain transformations paradoxically improve detection by removing misleading cues. We uncover model- and dataset-specific vulnerabilities, delineate an 8B-parameter robustness boundary, and highlight a trade-off between reasoning ability and generalization under perturbations. We find that coding agents, while generally stronger than standalone LLMs, still experience robustness drops when the underlying model is weak on certain datasets or when hot-plugging introduces precision loss.
\end{itemize}


\section{Background}\label{sec:background}

\subsection{LLM for Vulnerability Detection}

\begin{table*}[t]
\scriptsize
\renewcommand{\arraystretch}{1.15}
\setlength{\tabcolsep}{2pt}
\centering
\caption{Overview of existing code obfuscation methods. 
\textbf{Refs/Tools}: representative obfuscation works and tools. 
\textbf{Attacks on LLM}: obfuscation-related attacks against LLMs. 
\textbf{Potency}: extent to which transformation $T$ changes program complexity. 
\textbf{Resilience}: resistance against reverse engineering. 
\textbf{Cost}: overhead or required resources.~\cite{collberg1997taxonomy}}
\label{tab:taxonomy}
\begin{tabularx}{\textwidth}{ccXcXXccc}
\toprule
\textbf{Category} & \textbf{Subcategory} & \textbf{Technique} & \textbf{Abbr} & \textbf{Refs / Tools} & \textbf{Attacks} & \textbf{Potency} & \textbf{Resilience} & \textbf{Cost} \\
\midrule
\multirow{8}{*}{\textbf{Layout}} 
 & Identifier obfuscation & meaningful name $\rightarrow$ meaningless name & L1 & \cite{zhang2023bian, 0sir1ss2022Anubis, DashingSoft2017Pyarmor, Liftoff2014Pyminifier, H4wkst3r2022InvisibilityCloak, PELock2021JObfuscator, JavascriptObfuscator2016, Pkfr2018YakproPo, Naneau2014PhpObfuscator, Ph72017ObfuscatorClass, Hnfull2020IntensioObfuscator, PyObfx2018, Secureyourself72019PowerShellObfuscation, Kkar2014VBSObfuscatorPython, Dentrax2019Z00bfuscator, Whoward32020CCodeObfuscator, EvilBytecode2025EbyteGoMorpher, JKerbin2024SCOTool, Uxebu2015Confusion, PELock2016AutoItObfuscator, Artemixer2024Gofuscator, Yardenlaif2024Balagan, Wufhex2024PyDelta, Domchen2017UglifyTS, Avilum2022Jsafer} & \cite{swindle2024evaluation} & med & one-way & free \\ 
\cline{3-9}
 &  & identifier name $\rightarrow$ searched AE name  & -- & N/A & \cite{yang2022natural, tian2023code, zhang2023black, du2023extensive, zhao2025adversarial, huang2025iterative, nikiema2025codebarrier, yefet2020adversarial, li2024aacegen, jha2023codeattack, zhang2022towards, zhang2020generating} & dep & one-way & free \\
\cline{2-9}
 & Comments obfuscation & Delete comments / docstrings & L2 & \cite{zhang2023bian, 0sir1ss2022Anubis, H4wkst3r2022InvisibilityCloak, PELock2021JObfuscator, Pkfr2018YakproPo, Hnfull2020IntensioObfuscator, Secureyourself72019PowerShellObfuscation, Whoward32020CCodeObfuscator, JKerbin2024SCOTool, LiuYuancheng2025PyCodeObfuscator, Yardenlaif2024Balagan, Avilum2022Jsafer} & N/A & high & one-way & free \\
\cline{3-9}
 &  & Use misleading comments / docstrings & -- & N/A & \cite{swindle2024evaluation} & dep & one-way & free \\
\cline{2-9}
 & Code formatting & Remove spaces / indent / braces & L3 & \cite{zhang2023bian, Liftoff2014Pyminifier, Pkfr2018YakproPo, Naneau2014PhpObfuscator, Ph72017ObfuscatorClass, Whoward32020CCodeObfuscator, JKerbin2024SCOTool, Avilum2022Jsafer} & \cite{swindle2024evaluation} & low & one-way & free \\
\cline{2-9}
 & Prog.~lang. (single) & for $\rightarrow$ while & L4 & \cite{Pkfr2018YakproPo} & N/A & low & trivial & free \\
\cline{3-9} 
 &  & for $\rightarrow$ while (searched AE) & -- & N/A & \cite{tian2023code, zhao2025adversarial} & low & trivial & free \\
\cline{3-9}
 &  & if-else $\rightarrow$ switch & L5 & \cite{Pkfr2018YakproPo} & N/A & low & trivial & free \\
\cline{3-9}
 &  & if-else $\rightarrow$ switch (searched AE) & -- & N/A & \cite{tian2023code} & low & trivial & free \\
\cline{3-9}
 &  & loop $\rightarrow$ recursion & L6 & \cite{Pkfr2018YakproPo} & \cite{swindle2024evaluation} & low & one-way & free \\
\cline{3-9}
 &  & loop $\rightarrow$ recursion (searched AE) & -- & N/A & N/A & low & one-way & free \\
\cline{2-9}
 & Prog.~lang. (mix) & Inline assembly & L7 & \cite{romano2022wobfuscator} & N/A & high & one-way & free \\
\cline{3-9}
 &  & External call another lang. & L8 & \cite{DashingSoft2017Pyarmor} & N/A & high & one-way & free \\
\midrule
\multirow{8}{*}{\textbf{Data flow}} 
 & Data encoding & Arithmetic const. substitution & D1 & \cite{zhang2023bian, collberg2012distributed, JavascriptObfuscator2016, PyObfx2018, EvilBytecode2025EbyteGoMorpher, JKerbin2024SCOTool, PELock2016AutoItObfuscator, Artemixer2024Gofuscator, Avilum2022Jsafer} & \cite{nikiema2025codebarrier, swindle2024evaluation, tian2023code, zhao2025adversarial, yan2024llm} & dep & dep & dep \\ 
\cline{3-9}
 &  & Boolean const. substitution & D2 & \cite{zhang2023bian, PyObfx2018, EvilBytecode2025EbyteGoMorpher, PELock2016AutoItObfuscator, Artemixer2024Gofuscator} & \cite{yan2024llm} & dep & dep & dep \\ 
\cline{3-9}
 &  & String const. substitution & D3 & \cite{Scrt2020Avcleaner, H4wkst3r2022InvisibilityCloak, PELock2021JObfuscator, JavascriptObfuscator2016, Pkfr2018YakproPo, QQuick2015Opy, PyObfx2018, EvilBytecode2025EbyteGoMorpher, LiuYuancheng2025PyCodeObfuscator, PELock2016AutoItObfuscator, Artemixer2024Gofuscator, Wufhex2024PyDelta, Avilum2022Jsafer} & \cite{nikiema2025codebarrier, swindle2024evaluation, yan2024llm} & dep & dep & dep \\
\cline{2-9}
 & Data structure & Aggregation (scalar $\rightarrow$ vector) & D4 & \cite{zhang2023bian, PELock2021JObfuscator, Uxebu2015Confusion, Avilum2022Jsafer} & N/A & low & strong & free \\
\cline{3-9}
 &  & Splitting (vector $\rightarrow$ scalar) & D5 & \cite{zhang2023bian, JKerbin2024SCOTool} & N/A & low & strong & free \\
\cline{3-9}
 &  & Reorder scalars/fields & D6 & \cite{zhang2023bian, JavascriptObfuscator2016} & N/A & low & one-way & free \\
\cline{3-9}
 &  & Change scope & D7 & \cite{zhang2023bian} & N/A & low & one-way & free \\
\cline{2-9}
 & Data storage/access & Static access $\rightarrow$ dynamic access & D8 & \cite{zhang2023bian, collberg1997taxonomy, PELock2016AutoItObfuscator} & N/A & dep & dep & dep \\
\midrule
\multirow{3}{*}{\textbf{Control flow}} 
 & Extend control flow & Insert opaque predicates & C1 & \cite{zhang2023bian, 0sir1ss2022Anubis, collberg2012distributed, JKerbin2024SCOTool, PELock2016AutoItObfuscator} & N/A & dep & dep & dep \\ 
\cline{3-9}
 &  & Insert dead code (randomly) & -- & \cite{JavascriptObfuscator2016, Hnfull2020IntensioObfuscator, LiuYuancheng2025PyCodeObfuscator} & \cite{nikiema2025codebarrier, swindle2024evaluation} & dep & dep & dep \\
\cline{3-9}
 &  & Insert dead code (searched AE) & -- & N/A & \cite{zhao2025adversarial, yefet2020adversarial, zhang2022towards} & dep & dep & dep \\
\cline{3-9}
 &  & Insert dead code (hign-attention flashboom) & -- & N/A & \cite{li2025flashboom} & dep & dep & dep \\
\cline{2-9}
 & Restructure control flow & Flatten control flow & C2 & \cite{zhang2023bian, collberg2012distributed, PELock2021JObfuscator, JavascriptObfuscator2016, PELock2016AutoItObfuscator, Kaftejiman2021Ejja, Avilum2022Jsafer} & N/A & dep & dep & dep \\ 
\cline{2-9}
 & Replace control flow & Virtualization & C3 & \cite{0sir1ss2022Anubis, collberg1997taxonomy, I2rys2021SBPNO, Wodxgod2020SimpleObfuscator, Ph72017ObfuscatorClass, Chrisrands2019Emojify, LiuYuancheng2025PyCodeObfuscator, Darsyn2016Obfuscate, Wufhex2024PyDelta} & N/A & dep & dep & dep \\
\bottomrule
\end{tabularx}
\vspace{-3mm}
\end{table*}

Large Language Models have rapidly advanced from natural language processing to code reasoning, enabling tasks such as vulnerability detection and automated repair~\cite{brown2020language,openai2023gpt4,chen2021codex,roziere2023code}. 
Early efforts, including CodeBERT~\cite{feng2020codebert} and GraphCodeBERT~\cite{guo2021graphcodebert}, leveraged code-specific pretraining to capture program structure and semantics, but their small parameter sizes and limited context windows restricted effectiveness.The emergence of large-scale models such as Codex~\cite{chen2021codex}, GPT-4~\cite{openai2023gpt4}, and LLaMA-2~\cite{touvron2023llama} has substantially boosted vulnerability detection by jointly reasoning over natural language specifications and source code, enabling accurate identification across diverse languages~\cite{li2024empirical,fu2023llm4vuln}. Specialized frameworks (e.g., LLM4Vuln~\cite{fu2023llm4vuln}, VulnHunter~\cite{ding2024vulnhunter}) further enhance detection accuracy by combining LLMs with prompt engineering, few-shot learning, and program analysis. 
Recent work also highlights hybrid approaches that integrate LLMs with static and dynamic analysis~\cite{pearce2022asleep,wang2024evaluation}, leveraging symbolic reasoning and statistical learning to uncover subtle vulnerabilities. 
Recent advances have introduced coding agents (e.g. ~\cite{github-copilot} and ~\cite{codex}) which claim to identify common vulnerabilities and suggest fixes directly within the IDE. Compared to general-purpose LLMs, these agents are typically optimized for code-related tasks through specialized fine-tuning and prompt engineering, and are often supported by auxiliary toolchains that enhance their reasoning over code. 
\looseness=-1

\subsection{Obfuscation Impacts on LLM  Detection }

Code obfuscation refers to the process of transforming a program into a form that is more complex to understand or analyze while preserving its original functionality~\cite{collberg1997taxonomy,Collberg1998OpaqueConstructs}. In essence, an obfuscating transformation $T$ maps a program $P$ to an equivalent program $P' = T(P)$, where $P$ and $P'$ produce the same externally observable behavior, even though $P'$ may appear radically different in structure, readability, or efficiency. The key idea is to maintain semantic equivalence while introducing syntactic or structural diversity.
Because of this property, obfuscation has long served as a double-edged tool in software and security domains. On one side, defenders employ obfuscation to protect intellectual property, deter reverse engineering, and safeguard critical logic. On the other side, adversaries exploit it to disguise malware, evade detection systems, and generate adversarial variants that confuse automated program analysis.

With the emergence of LLM-based vulnerability detection, this dual-use nature takes on a new dimension. Unlike classical static or dynamic analyzers that rely on rigid syntactic rules, LLMs depend heavily on statistical regularities, token distributions, and structural cues learned from massive corpora. Obfuscation deliberately disrupts these cues (by altering code layout, restructuring data representations, or modifying control flow) while preserving functional equivalence. As a result, semantically identical yet syntactically divergent code can systematically mislead LLMs, leading to missed vulnerabilities, inconsistent predictions, or elevated false positives.
To systematically examine this phenomenon, as shown in \autoref{tab:taxonomy}, we organize existing obfuscation methods into a taxonomy based on three dimensions:  (i) layout flow, (ii) data flow, and (iii) control flow, and analyze how each class interferes with LLM-based detection:

\begin{itemize}
    \item \textbf{Layout Obfuscation.} Layout obfuscation removes or distorts surface-level features such as identifiers, comments, formatting, or syntactic constructs.
 These transformations do not change data or control flow, but alter human-readable cues.
 Since LLMs learn token distributions, identifier semantics, and formatting regularities from large corpora, layout obfuscation directly disrupts their priors. For instance, name obfuscation eliminates semantic hints like \texttt{isAdmin}, comment deletion removes context for intent inference, and mixed-language code forces the model to reason across unfamiliar token spaces. As reported in existing work~\cite{yang2022natural, nikiema2025codebarrier}, even trivial transformations such as renaming can cause significant performance degradation.

\item \textbf{Data Flow Obfuscation.} Data flow obfuscation alters how information is encoded, structured, or accessed, while leaving the overall control structure intact.
 Techniques include replacing constants with complex expressions, fragmenting strings, aggregating or splitting variables, and switching from static to dynamic data access.
 These changes obscure the semantic equivalence between code fragments and weaken the statistical patterns LLMs rely on. For example, arithmetic substitution or boolean extension creates rarely seen token sequences, confusing attention mechanisms; string fragmentation disrupts pattern recognition for security-sensitive strings; and scope modifications alter context visibility. Studies such as~\cite{nikiema2025codebarrier, swindle2024evaluation} show that data encoding can reduce detection accuracy by masking vulnerable values behind atypical representations.

\item \textbf{Control Flow Obfuscation.}
Control flow obfuscation restructures program execution order while preserving functionality.
 Methods include inserting opaque predicates, flattening control-flow graphs, and virtualizing code execution through custom interpreters.
 These transformations disrupt logical dependencies that LLMs capture via attention over structured patterns. Junk code and opaque branches dilute meaningful signal, flattening erases hierarchical cues essential for reasoning, and virtualization replaces recognizable constructs with opaque bytecode-like patterns. For LLMs trained on standard source code, such changes collapse their ability to track execution paths, leading to higher false negatives in vulnerability detection.
 
\end{itemize}

 \section{Research Gaps and Our Research Questions}\label{sec:code_obfuscation}

\subsection{Gaps in Obfuscation-related LLM Studies}

Recent research has begun to explore the interaction between code obfuscation and LLMs in security-critical tasks such as vulnerability detection and malware analysis. These studies demonstrate that LLMs, while powerful in semantic reasoning, remain susceptible to perturbations at the syntactic or structural level. For example, simple transformations such as identifier renaming or comment deletion have been shown to reduce detection accuracy in models like CodeBERT and GPT-based detectors, highlighting their reliance on surface-level token distributions rather than deep semantic understanding ~\cite{yang2022natural, jha2023codeattack, zhang2022towards}. Similarly, work on adversarial programming (e.g., obfuscated reentrancy or buffer-overflow code snippets) shows that even trivial layout changes can mislead LLMs into overlooking known vulnerabilities~\cite{li2025flashboom}. 
 
\newcommand{\tickYes}{\ding{51}} 
\newcommand{\tickNo}{\color{lightgray}{\ding{55}}}  

\newcommand\featuretext[1]{%
  \llap{\vrule width.35pt height2pt depth2.5pt\kern1pt}%
  \rlap{\rotatebox{25}{\textbf{#1}}}%
}

\begin{table}[t]
\scriptsize
\setlength{\tabcolsep}{4.2pt}
\centering
\caption{Comparison of obfuscation-related attacks on LLMs. }
\vspace{-5mm}
\begin{tabular}{rl llll llll lll}
 
\textbf{Work} &
\featuretext{Layout} &
\featuretext{Data-flow} &
\featuretext{Control-flow} &
\featuretext{Single Model} &
\featuretext{Cross-Series} &
\featuretext{Size Analysis} &
\featuretext{Reasoning vs Non} &
\featuretext{Binary YES/NO} &
\featuretext{Vuln Type} &
\featuretext{Vuln Location} &
\featuretext{Lang} \\
\midrule
MHM~\cite{zhang2020generating} (2020) & \tickYes & \tickNo & \tickNo & \tickNo & \tickYes & \tickNo & \tickNo & \tickNo & \tickNo & \tickNo & \tickNo \\
DAMP~\cite{yefet2020adversarial} (2020) & \tickYes & \tickNo & \tickYes & \tickNo & \tickYes & \tickNo & \tickNo & \tickNo & \tickNo & \tickNo & \tickNo \\
ALERT~\cite{yang2022natural} (2022)  & \tickYes & \tickNo & \tickNo & \tickNo & \tickNo & \tickNo & \tickNo & \tickYes & \tickNo & \tickNo & \tickNo \\
CARROT~\cite{zhang2022towards} (2022) & \tickYes & \tickNo & \tickYes & \tickNo & \tickYes & \tickNo & \tickNo & \tickNo & \tickNo & \tickNo & \tickNo \\
CODA~\cite{tian2023code} (2023) & \tickYes & \tickYes & \tickNo & \tickNo & \tickYes & \tickNo & \tickNo & \tickYes & \tickNo & \tickNo & \tickNo \\
RNNS~\cite{zhang2023black} (2023) & \tickYes & \tickNo & \tickNo & \tickNo & \tickYes & \tickNo & \tickNo & \tickYes & \tickNo & \tickNo & \tickNo \\
BeamAttack~\cite{du2023extensive} (2023) & \tickYes & \tickNo & \tickNo & \tickNo & \tickYes & \tickNo & \tickNo & \tickYes & \tickNo & \tickNo & \tickNo \\
CodeAttack~\cite{jha2023codeattack} (2023) & \tickYes & \tickNo & \tickNo & \tickNo & \tickYes & \tickNo & \tickNo & \tickNo & \tickNo & \tickNo & \tickNo \\
Swindle et al.~\cite{swindle2024evaluation}
 (2024) & \tickYes & \tickYes & \tickYes & \tickNo & \tickYes & \tickNo & \tickNo & \tickNo & \tickNo & \tickNo & \tickNo \\
CODEBREAKER~\cite{yan2024llm} (2024) & \tickNo & \tickYes & \tickNo & \tickNo & \tickNo & \tickNo & \tickNo & \tickNo & \tickYes & \tickNo & \tickNo \\
AaceGEN~\cite{li2024aacegen} (2024) & \tickYes & \tickNo & \tickNo & \tickNo & \tickYes & \tickNo & \tickNo & \tickYes & \tickNo & \tickNo & \tickNo \\
Zhao et al.~\cite{zhao2025adversarial} (2025) & \tickYes & \tickYes & \tickYes & \tickNo & \tickYes & \tickNo & \tickNo & \tickYes & \tickNo & \tickNo & \tickNo \\
ITGen~\cite{huang2025iterative} (2025) & \tickYes & \tickNo & \tickNo & \tickNo & \tickYes & \tickNo & \tickNo & \tickYes & \tickNo & \tickNo & \tickNo \\
Nikiema et al.~\cite{nikiema2025codebarrier} (2025) & \tickYes & \tickYes & \tickYes & \tickNo & \tickYes & \tickNo & \tickNo & \tickNo & \tickNo & \tickNo & \tickNo \\
Flashboom~\cite{li2025flashboom} (2025) & \tickNo & \tickNo & \tickYes & \tickNo & \tickYes & \tickNo & \tickNo & \tickYes & \tickYes & \tickNo & \tickYes \\

\bottomrule
\end{tabular}
\vspace{-2mm}
\end{table}

However, most existing works focus on a very limited set of obfuscation methods, often confined to cosmetic or layout-level transformations. While traditional program obfuscation has long included techniques such as control-flow flattening, data-flow encoding, or virtualization, few LLM-targeted studies have systematically incorporated these categories. This results in evaluations that are not representative of the broader landscape of obfuscation techniques. 
For example, ALERT~\cite{yang2022natural} is restricted to identifier renaming. CODA~\cite{tian2023code} extends this scope slightly by introducing data-encoding substitutions such as boolean or arithmetic transformations. More recently, CODEBREAKER~\cite{yan2024llm} manipulates the dataflow of vulnerable lines to evade detection, yet still leaves control-flow and virtualization-based obfuscation unexplored. Such a narrow focus overlooks richer obfuscation strategies and limits our understanding of how LLMs perform under diverse and realistic adversarial transformations.

In terms of model evaluation, prior work typically benchmarks only one or two mainstream models (e.g., GPT-3.5 or CodeBERT), with little consideration of architectural diversity. Comparative analyses across different LLM families (e.g., OpenAI, LLaMA, Qwen, DeepSeek), different parameter scales, or reasoning versus non-reasoning variants are largely absent. Without these dimensions, it is unclear whether observed weaknesses are model-specific quirks or generalizable vulnerabilities of LLMs. 
For example, CODA mainly targets CodeBERT and GraphCodeBERT, while Zhao et al.~\cite{zhao2025adversarial} only consider CodeBERT and Qwen-7B-Code~\cite{hui2024qwen2}. Similarly, CODEBREAKER uniquely evaluates GPT-3.5 and GPT-4. Such fragmented coverage underscores the lack of systematic evaluation across model families, scales, and reasoning capabilities. 

From the detection perspective, most evaluations reduce the task to a binary question: whether the model can recognize the presence of a vulnerability. This yes/no framing overlooks finer-grained aspects, such as whether the model can correctly identify the type of vulnerability (e.g., CWE classification) or its location in the code. As a result, it remains ambiguous whether LLMs actually understand the underlying security flaw or are merely providing shallow pattern matches.  For example, RNNS~\cite{zhang2023black} adopts a binary yes/no evaluation, whereas CODEBREAKER collects more detailed outputs from models, such as the vulnerability type. Such binary-oriented evaluations are overly simplistic, capturing only the existence of a flaw while failing to reflect deeper diagnostic capabilities that are critical for practical vulnerability detection.

Regarding data quality, the datasets used in these studies are often limited in both diversity and representativeness. Many works focus on one or two programming languages, such as Python or C, and the distribution of vulnerability types is typically imbalanced, dominated by a handful of CWE categories. Furthermore, few studies investigate the role of code complexity (e.g., function nesting depth, lines of code) in shaping detection robustness, despite its importance in both obfuscation and model performance.  For example, ITGen~\cite{huang2025iterative} is conducted on C/C++ programs, while BeamAttack~\cite{du2023extensive} relies solely on Java code from the OWASP Benchmark. Such language-constrained settings limit the generalizability of findings across programming ecosystems.

Finally, existing studies have not examined what happens when code obfuscation meets strategies for improving LLM-based vulnerability detection. As shown in \autoref{tab:compare-agent-finetune-rag-prompt}, coding agents simultaneously inherit all four advantages of fine-tuning, retrieval augmentation, and prompt engineering, making them the most powerful option available today. These advantages include: \textbf{1) an optimized underlying model:} for example, Codex leverages GPT-5-Codex, trained specifically for agentic coding tasks; \textbf{2) access to auxiliary toolchains:} for example, GitHub Copilot retrieves large-scale code from GitHub repositories to improve contextual accuracy;\textbf{ 3) low cost:} coding agents are typically priced at only tens of dollars per month; \textbf{4) ease of use:} users can issue natural-language commands---including vulnerability detection---without crafting specialized prompts or possessing deep technical knowledge. Given these strengths, coding agents warrant dedicated study. Unfortunately, no prior work has examined how code obfuscation affects vulnerability detection performed by coding agents, leaving their robustness against obfuscation attacks largely unexplored.

\begin{table}[t]
\scriptsize
\setlength{\tabcolsep}{3.5pt}
\centering
\caption{Comparison of strategies for improving LLM-based vulnerability detection. CA = Coding Agent, FT = Fine-Tuning, RA = Retrieval Augmentation, PE = Prompt Engineering }

\scriptsize
\setlength\tabcolsep{5pt}
\begin{tabularx}{0.48\textwidth}{r XXXX}
\toprule
\textbf{Work}
& \multicolumn{1}{c}{\textbf{CA}}
& \multicolumn{1}{c}{\textbf{FT}}
& \multicolumn{1}{c}{\textbf{RA}}
& \multicolumn{1}{c}{\textbf{PE}} \\
\midrule

Example & \parbox{2.5cm}{
GitHub-Copilot \\Codex}& \parbox{2.5cm}{
\cite{wen2023less} \cite{yang2023does} \\ \cite{liu2024pre}  \cite{wang2024combining} } &  \parbox{2.5cm}{
\cite{du2024vul} \cite{liu2023software} \\ \cite{wen2024vuleval}  \cite{zhou2024large} }& \parbox{2.5cm} {
\cite{fu2023chatgpt} \cite{yin2024pros} \\ \cite{zhang2024prompt} \cite{zhou2024large}  }\\

\midrule
Model opt.      & \multicolumn{1}{c}{\tickYes} & \multicolumn{1}{c}{\tickYes} & \multicolumn{1}{c}{\tickNo} & \multicolumn{1}{c}{\tickNo} \\
Auxiliary toolchain & \multicolumn{1}{c}{\tickYes} & \multicolumn{1}{c}{\tickNo} & \multicolumn{1}{c}{\tickYes} & \multicolumn{1}{c}{\tickNo} \\
Cheap           & \multicolumn{1}{c}{\tickYes} & \multicolumn{1}{c}{\tickNo} & \multicolumn{1}{c}{\tickYes} & \multicolumn{1}{c}{\tickYes} \\
Ease of use  & \multicolumn{1}{c}{\tickYes} & \multicolumn{1}{c}{\tickNo} & \multicolumn{1}{c}{\tickNo} & \multicolumn{1}{c}{\tickYes} \\

\bottomrule
\end{tabularx}

\label{tab:compare-agent-finetune-rag-prompt}
\vspace{-3mm}
\end{table}

\subsection{Research Questions}
\label{subsec:rq}

As discussed in the literature review, prior studies on obfuscation-related attacks against LLMs suffer from several key limitations: the set of obfuscation techniques examined is narrow, the diversity of victim models is limited, detection tasks are often oversimplified, the datasets used lack language coverage, balance, and complexity analysis, and the robustness of coding agents has not yet been investigated. To address these gaps, we formulate the following research questions to guide our study:

\begin{itemize}[left=0.1cm]
    \item \textbf{RQ1 (Obfuscation Methods).} To what extent do different obfuscation strategies, including layout, data-flow, control-flow, and multi-language transformations, affect the robustness of LLM-based vulnerability detection?

    \item \textbf{RQ2 (Model Diversity).} How do variations in LLM families (e.g., OpenAI, LLaMA, Qwen, DeepSeek), parameter scales, and reasoning capabilities influence their resilience against obfuscation?

    \item \textbf{RQ3 (Detection Granularity).} Beyond binary existence checks, can LLMs accurately identify both the type and the location of vulnerabilities in obfuscated code?

    \item \textbf{RQ4 (Dataset Diversity).} How do dataset characteristics, such as programming language coverage, vulnerability distribution balance, and code complexity, shape the impact of obfuscation on LLM-based detection?

    \item \textbf{RQ5 (Coding Agent).} How does code obfuscation affect the performance of coding agents when detecting vulnerabilities?
\end{itemize}

These questions allow us to move beyond prior fragmented evaluations and build a more systematic and comprehensive understanding of how obfuscation interacts with LLMs in the context of vulnerability detection.
\section{Understanding Obfuscation-Induced Shifts in LLM Vulnerability Detection}\label{sec:experiment}



\subsection{Experiment Setup}\label{sec:expriment-setup}

\noindent\textbf{Datasets.} Our dataset consists of a vulnerability benchmark designed to evaluate the impact of code obfuscation on LLM-based vulnerability detection.  As summarized in ~\autoref{tab:datasets}, the benchmark covers four programming languages: Solidity smart contracts (\smartbugs\cite{durieux2020empiricalsmartbugs}), C projects (\primevulc\cite{ding2024primevul}), C++ projects (\reposvulcpp\cite{wang2024repository}), and Python projects (\reposvulpy\cite{wang2024repository}). Following prior work~\cite{li2025flashboom} of vulnerability detection, to avoid exceeding the inference capacity of the LLMs, we filter out code files with more than 500 lines of code. For the C, C++, and Python datasets, we further limit the number of vulnerable samples in each CWE category to a maximum of five files per language, in order to mitigate data imbalance and the overrepresentation of particular vulnerabilities. \looseness=-1

\begin{table}[ht]
\caption{All datasets used in our experiment. Avg LOC stands for the average line of code. Avg Func stands for the average number of functions.}
\vspace{-1mm}
\label{tab:datasets}
\centering
\scriptsize
\setlength\tabcolsep{6pt}
\begin{tabular}{@{}lrrrrr@{}}
\toprule
\textbf{Dataset} & \textbf{Lang} & \textbf{Vuln Type} & \textbf{Count} & \textbf{Avg LOC} & \textbf{Avg Func} \\
\midrule
\smartbugs   & Solidity & 9    & 128 & 35  & 4  \\
\primevulc   & C        & 119 & 143 & 256 & 9  \\
\reposvulcpp & C++      & 35 & 84  & 172 & 10 \\
\reposvulpy  & Python   & 89 & 210 & 223 & 5  \\
\bottomrule
\end{tabular}
\vspace{-1mm}
\end{table}

\noindent\textbf{Obfuscation Technique.} We apply all the code obfuscation techniques described in \autoref{tab:taxonomy} to our datasets. To avoid overly fine-grained categorization that would lead to experiments of limited significance, we restrict the obfuscation granularity to the subcategory level rather than the technique level. An exception is Programming language obfuscation – mix-language, where the implementation requires both inline assembly and another programming language; therefore, \texttt{L7} and \texttt{L8} are treated as two separate combos. In total, we construct ten obfuscation combos: \texttt{L1}, \texttt{L2}, \texttt{L3}, \texttt{L4+L5+L6}, \texttt{L7}, \texttt{L8}, \texttt{D1+D2+D3}, \texttt{D4+D5+D6+D7}, \texttt{C1}, \texttt{C2}, and \texttt{C3}. We employ \texttt{gpt-4o} for code obfuscation to ensure consistency and thoroughness, with the detailed prompt template provided in Appendix \ref{sec:prompt-for-obfuscation}. 

\vspace{2mm}
\noindent\textbf{LLM-based Vulnerability Detection.} 
We now describe our approach to selecting and configuring the LLMs used in our experiments. As summarized in Table~\autoref{tab:models}, we curated a set of 15 models across 4 families, guided by three principles: (1) inclusion of both reasoning-oriented and non-reasoning counterparts, (2) coverage of a wide range of parameter scales, and (3) diversity in architectural design. To ensure stability and reproducibility, the temperature was fixed at $10^{-5}$, effectively eliminating randomness. Additionally, we select GitHub Copilot (powered by GPT-5) and Codex (powered by GPT-5-Codex), two widely used and accessible coding agents in the experimental region, as the subjects of our evaluation.

\begin{table}[ht]
    \centering
    \scriptsize 
    \renewcommand{\arraystretch}{1.15}
    \caption{Models Evaluated. Abbr. denotes the abbreviated names of the models used in this paper. DS-R1-Dist refers to DeepSeek-R1-Distill, and Inst stands for Instruction.}
    \resizebox{\linewidth}{!}{%
    \footnotesize
    \begin{tabular}{ll ll r}
        \toprule
        \multicolumn{2}{c}{\textbf{Non-reasoning LLM}} & \multicolumn{2}{c}{\textbf{Reasoning LLM}} & \textbf{Param} \\
        Model & Abbr. & Model & Abbr. &  \\
        \midrule
        \multicolumn{5}{l}{\textbf{Qwen Series}} \\
        Qwen2.5-7B-Inst   & \textbf{qn-7b}   & DS-R1-Dist-Qwen-7B   & \textbf{r1-qn-7b}   & 7B  \\
        Qwen2.5-14B-Inst  & \textbf{qn-14b}  & DS-R1-Dist-Qwen-14B  & \textbf{r1-qn-14b}  & 14B \\
        Qwen2.5-32B-Inst  & \textbf{qn-32b}  & DS-R1-Dist-Qwen-32B  & \textbf{r1-qn-32b}  & 32B \\
        \midrule
        \multicolumn{5}{l}{\textbf{Llama Series}} \\
        Llama-3.1-8B-Inst  & \textbf{lm-8b}   & DS-R1-Dist-Llama-8B   & \textbf{r1-lm-8b}   & 8B  \\
        Llama-3.3-70B-Inst & \textbf{lm-70b}  & DS-R1-Dist-Llama-70B  & \textbf{r1-lm-70b}  & 70B \\
        \midrule
        \multicolumn{5}{l}{\textbf{DeepSeek Series}} \\
        DeepSeek-V3       & \textbf{ds-v3}   & DeepSeek-R1             & \textbf{ds-r1}     & 671B\\
        \midrule
        \multicolumn{5}{l}{\textbf{OpenAI Series}} \\
        GPT-3.5-turbo   & \textbf{gpt-3.5}   & - & - & - \\
        GPT-4o          & \textbf{gpt-4o}   & - & - & - \\
        -           & \textbf{-} & o3-mini                 & \textbf{o3-mini}   & -   \\
        \bottomrule
    \end{tabular}
    }
    \label{tab:models}
    \vspace{-3mm}
\end{table}

\begin{table*}[h]
\centering
\caption{Detection successful rate across datasets. LLM Score 3,4 = Positive, Score 1,2 = Negative.}
\vspace{-1mm}
{
\scriptsize
\setlength\tabcolsep{3pt}
\renewcommand{\arraystretch}{0.6}
\begin{tabular}{l lccccccccccccccc}
\toprule
dataset & series & \multicolumn{6}{c}{qwen} & \multicolumn{4}{c}{llama} & \multicolumn{2}{c}{deepseek} & \multicolumn{3}{c}{openai} \\
\cmidrule(lr){3-8} \cmidrule(lr){9-12} \cmidrule(lr){13-14} \cmidrule(lr){15-17}
 & model & qn-7b & qn-14b & qn-32b & ds-qn-7b & ds-qn-14b & ds-qn-32b & lm-8b & lm-70b & ds-lm-8b & ds-lm-70b & ds-v3 & ds-r1 & gpt-3.5 & gpt-4o & o3-mini \\
\midrule
\multirow{13}{*}{\rotatebox{90}{\textbf{\smartbugs}}}
    & original & \ccell{0.48} & \ccell{0.62} & \ccell{0.45} & \ccell{0.29} & \ccell{0.52} & \ccell{0.58} & \ccell{0.46} & \ccell{0.55} & \ccell{0.46} & \ccell{0.59} & \ccell{0.83} & \ccell{0.78} & \ccell{0.57} & \ccell{0.68} & \ccell{0.89} \\
\cmidrule(lr){2-17}
    & L1 & \ccell{0.45} & \ccell{0.48} & \ccell{0.40} & \ccell{0.22} & \ccell{0.55} & \ccell{0.56} & \ccell{0.46} & \ccell{0.47} & \ccell{0.34} & \ccell{0.52} & \ccell{0.77} & \ccell{0.75} & \ccell{0.49} & \ccell{0.66} & \ccell{0.84} \\
    & L2 & \ccell{0.45} & \ccell{0.60} & \ccell{0.48} & \ccell{0.32} & \ccell{0.56} & \ccell{0.65} & \ccell{0.51} & \ccell{0.53} & \ccell{0.42} & \ccell{0.57} & \ccell{0.88} & \ccell{0.79} & \ccell{0.58} & \ccell{0.70} & \ccell{0.82} \\
    & L3 & \ccell{0.44} & \ccell{0.64} & \ccell{0.48} & \ccell{0.27} & \ccell{0.56} & \ccell{0.65} & \ccell{0.49} & \ccell{0.55} & \ccell{0.41} & \ccell{0.57} & \ccell{0.77} & \ccell{0.79} & \ccell{0.55} & \ccell{0.65} & \ccell{0.88} \\
    & L4+L5+L6 & \ccell{0.42} & \ccell{0.59} & \ccell{0.45} & \ccell{0.23} & \ccell{0.48} & \ccell{0.58} & \ccell{0.45} & \ccell{0.45} & \ccell{0.40} & \ccell{0.57} & \ccell{0.74} & \ccell{0.77} & \ccell{0.55} & \ccell{0.69} & \ccell{0.85} \\
    & L7 & \ccell{0.42} & \ccell{0.65} & \ccell{0.55} & \ccell{0.25} & \ccell{0.56} & \ccell{0.54} & \ccell{0.46} & \ccell{0.49} & \ccell{0.45} & \ccell{0.52} & \ccell{0.81} & \ccell{0.74} & \ccell{0.56} & \ccell{0.63} & \ccell{0.73} \\
    & L8 & \ccell{0.55} & \ccell{0.56} & \ccell{0.45} & \ccell{0.22} & \ccell{0.50} & \ccell{0.55} & \ccell{0.43} & \ccell{0.49} & \ccell{0.41} & \ccell{0.52} & \ccell{0.87} & \ccell{0.80} & \ccell{0.50} & \ccell{0.63} & \ccell{0.86} \\
    & D1+D2+D3 & \ccell{0.51} & \ccell{0.48} & \ccell{0.47} & \ccell{0.25} & \ccell{0.49} & \ccell{0.55} & \ccell{0.45} & \ccell{0.51} & \ccell{0.41} & \ccell{0.59} & \ccell{0.80} & \ccell{0.77} & \ccell{0.51} & \ccell{0.63} & \ccell{0.88} \\
    & D4+D5+D6+D7 & \ccell{0.52} & \ccell{0.60} & \ccell{0.50} & \ccell{0.24} & \ccell{0.58} & \ccell{0.62} & \ccell{0.47} & \ccell{0.55} & \ccell{0.41} & \ccell{0.52} & \ccell{0.84} & \ccell{0.77} & \ccell{0.56} & \ccell{0.66} & \ccell{0.87} \\
    & D8 & \ccell{0.43} & \ccell{0.55} & \ccell{0.45} & \ccell{0.30} & \ccell{0.55} & \ccell{0.60} & \ccell{0.50} & \ccell{0.52} & \ccell{0.41} & \ccell{0.62} & \ccell{0.78} & \ccell{0.74} & \ccell{0.53} & \ccell{0.62} & \ccell{0.88} \\
    & C1 & \ccell{0.49} & \ccell{0.52} & \ccell{0.45} & \ccell{0.24} & \ccell{0.52} & \ccell{0.62} & \ccell{0.42} & \ccell{0.51} & \ccell{0.39} & \ccell{0.56} & \ccell{0.80} & \ccell{0.76} & \ccell{0.49} & \ccell{0.67} & \ccell{0.88} \\
    & C2 & \ccell{0.46} & \ccell{0.59} & \ccell{0.48} & \ccell{0.23} & \ccell{0.52} & \ccell{0.58} & \ccell{0.44} & \ccell{0.49} & \ccell{0.41} & \ccell{0.56} & \ccell{0.74} & \ccell{0.81} & \ccell{0.48} & \ccell{0.55} & \ccell{0.83} \\
    & C3 & \ccell{0.52} & \ccell{0.62} & \ccell{0.57} & \ccell{0.14} & \ccell{0.41} & \ccell{0.44} & \ccell{0.41} & \ccell{0.49} & \ccell{0.28} & \ccell{0.45} & \ccell{0.80} & \ccell{0.74} & \ccell{0.48} & \ccell{0.75} & \ccell{0.76} \\

\midrule
\multirow{13}{*}{\rotatebox{90}{\textbf{\reposvulcpp}}}
    & original & \ccell{0.13} & \ccell{0.17} & \ccell{0.12} & \ccell{0.14} & \ccell{0.14} & \ccell{0.19} & \ccell{0.24} & \ccell{0.26} & \ccell{0.15} & \ccell{0.19} & \ccell{0.29} & \ccell{0.24} & \ccell{0.08} & \ccell{0.26} & \ccell{0.26} \\
\cmidrule(lr){2-17}
    & L1 & \ccell{0.15} & \ccell{0.21} & \ccell{0.19} & \ccell{0.06} & \ccell{0.15} & \ccell{0.17} & \ccell{0.13} & \ccell{0.19} & \ccell{0.11} & \ccell{0.12} & \ccell{0.17} & \ccell{0.18} & \ccell{0.10} & \ccell{0.19} & \ccell{0.31} \\
    & L2 & \ccell{0.13} & \ccell{0.19} & \ccell{0.17} & \ccell{0.10} & \ccell{0.14} & \ccell{0.23} & \ccell{0.29} & \ccell{0.24} & \ccell{0.11} & \ccell{0.17} & \ccell{0.32} & \ccell{0.20} & \ccell{0.13} & \ccell{0.25} & \ccell{0.30} \\
    & L3 & \ccell{0.13} & \ccell{0.24} & \ccell{0.18} & \ccell{0.11} & \ccell{0.15} & \ccell{0.18} & \ccell{0.23} & \ccell{0.25} & \ccell{0.13} & \ccell{0.23} & \ccell{0.35} & \ccell{0.23} & \ccell{0.12} & \ccell{0.21} & \ccell{0.25} \\
    & L4+L5+L6 & \ccell{0.10} & \ccell{0.25} & \ccell{0.18} & \ccell{0.13} & \ccell{0.11} & \ccell{0.17} & \ccell{0.24} & \ccell{0.20} & \ccell{0.12} & \ccell{0.13} & \ccell{0.26} & \ccell{0.17} & \ccell{0.18} & \ccell{0.25} & \ccell{0.23} \\
    & L7 & \ccell{0.15} & \ccell{0.15} & \ccell{0.20} & \ccell{0.10} & \ccell{0.15} & \ccell{0.21} & \ccell{0.14} & \ccell{0.25} & \ccell{0.13} & \ccell{0.12} & \ccell{0.29} & \ccell{0.10} & \ccell{0.11} & \ccell{0.17} & \ccell{0.17} \\
    & L8 & \ccell{0.11} & \ccell{0.18} & \ccell{0.14} & \ccell{0.04} & \ccell{0.13} & \ccell{0.17} & \ccell{0.15} & \ccell{0.17} & \ccell{0.15} & \ccell{0.10} & \ccell{0.15} & \ccell{0.07} & \ccell{0.05} & \ccell{0.11} & \ccell{0.15} \\
    & D1+D2+D3 & \ccell{0.10} & \ccell{0.15} & \ccell{0.17} & \ccell{0.07} & \ccell{0.15} & \ccell{0.13} & \ccell{0.12} & \ccell{0.23} & \ccell{0.08} & \ccell{0.15} & \ccell{0.20} & \ccell{0.17} & \ccell{0.07} & \ccell{0.17} & \ccell{0.29} \\
    & D4+D5+D6+D7 & \ccell{0.12} & \ccell{0.19} & \ccell{0.19} & \ccell{0.11} & \ccell{0.13} & \ccell{0.21} & \ccell{0.20} & \ccell{0.24} & \ccell{0.12} & \ccell{0.18} & \ccell{0.25} & \ccell{0.18} & \ccell{0.15} & \ccell{0.26} & \ccell{0.23} \\
    & D8 & \ccell{0.12} & \ccell{0.23} & \ccell{0.17} & \ccell{0.11} & \ccell{0.11} & \ccell{0.23} & \ccell{0.21} & \ccell{0.24} & \ccell{0.23} & \ccell{0.13} & \ccell{0.29} & \ccell{0.20} & \ccell{0.13} & \ccell{0.24} & \ccell{0.30} \\
    & C1 & \ccell{0.20} & \ccell{0.13} & \ccell{0.20} & \ccell{0.07} & \ccell{0.13} & \ccell{0.19} & \ccell{0.13} & \ccell{0.19} & \ccell{0.08} & \ccell{0.12} & \ccell{0.23} & \ccell{0.17} & \ccell{0.10} & \ccell{0.23} & \ccell{0.24} \\
    & C2 & \ccell{0.17} & \ccell{0.11} & \ccell{0.15} & \ccell{0.11} & \ccell{0.20} & \ccell{0.18} & \ccell{0.18} & \ccell{0.23} & \ccell{0.15} & \ccell{0.11} & \ccell{0.30} & \ccell{0.24} & \ccell{0.11} & \ccell{0.23} & \ccell{0.27} \\
    & C3 & \ccell{0.05} & \ccell{0.13} & \ccell{0.13} & \ccell{0.02} & \ccell{0.07} & \ccell{0.07} & \ccell{0.11} & \ccell{0.06} & \ccell{0.04} & \ccell{0.04} & \ccell{0.06} & \ccell{0.05} & \ccell{0.06} & \ccell{0.02} & \ccell{0.05} \\

\midrule
\multirow{13}{*}{\rotatebox{90}{\textbf{\reposvulpy}}}
    & original & \ccell{0.11} & \ccell{0.14} & \ccell{0.16} & \ccell{0.13} & \ccell{0.13} & \ccell{0.14} & \ccell{0.17} & \ccell{0.23} & \ccell{0.17} & \ccell{0.17} & \ccell{0.24} & \ccell{0.20} & \ccell{0.16} & \ccell{0.27} & \ccell{0.14} \\
\cmidrule(lr){2-17}
    & L1 & \ccell{0.18} & \ccell{0.21} & \ccell{0.19} & \ccell{0.08} & \ccell{0.13} & \ccell{0.13} & \ccell{0.16} & \ccell{0.20} & \ccell{0.12} & \ccell{0.16} & \ccell{0.28} & \ccell{0.23} & \ccell{0.12} & \ccell{0.20} & \ccell{0.20} \\
    & L2 & \ccell{0.08} & \ccell{0.17} & \ccell{0.17} & \ccell{0.11} & \ccell{0.14} & \ccell{0.13} & \ccell{0.15} & \ccell{0.20} & \ccell{0.15} & \ccell{0.20} & \ccell{0.29} & \ccell{0.23} & \ccell{0.15} & \ccell{0.25} & \ccell{0.15} \\
    & L3 & \ccell{0.15} & \ccell{0.19} & \ccell{0.18} & \ccell{0.13} & \ccell{0.14} & \ccell{0.12} & \ccell{0.20} & \ccell{0.21} & \ccell{0.18} & \ccell{0.19} & \ccell{0.31} & \ccell{0.25} & \ccell{0.24} & \ccell{0.27} & \ccell{0.19} \\
    & L4+L5+L6 & \ccell{0.08} & \ccell{0.16} & \ccell{0.17} & \ccell{0.12} & \ccell{0.16} & \ccell{0.12} & \ccell{0.17} & \ccell{0.18} & \ccell{0.12} & \ccell{0.17} & \ccell{0.29} & \ccell{0.20} & \ccell{0.12} & \ccell{0.23} & \ccell{0.18} \\
    & L7 & \ccell{0.06} & \ccell{0.15} & \ccell{0.13} & \ccell{0.11} & \ccell{0.12} & \ccell{0.13} & \ccell{0.12} & \ccell{0.15} & \ccell{0.12} & \ccell{0.13} & \ccell{0.24} & \ccell{0.17} & \ccell{0.17} & \ccell{0.13} & \ccell{0.12} \\
    & L8 & \ccell{0.10} & \ccell{0.16} & \ccell{0.14} & \ccell{0.11} & \ccell{0.12} & \ccell{0.12} & \ccell{0.14} & \ccell{0.17} & \ccell{0.14} & \ccell{0.15} & \ccell{0.19} & \ccell{0.15} & \ccell{0.11} & \ccell{0.11} & \ccell{0.12} \\
    & D1+D2+D3 & \ccell{0.11} & \ccell{0.14} & \ccell{0.16} & \ccell{0.12} & \ccell{0.13} & \ccell{0.12} & \ccell{0.15} & \ccell{0.21} & \ccell{0.15} & \ccell{0.17} & \ccell{0.25} & \ccell{0.23} & \ccell{0.14} & \ccell{0.20} & \ccell{0.18} \\
    & D4+D5+D6+D7 & \ccell{0.09} & \ccell{0.18} & \ccell{0.17} & \ccell{0.10} & \ccell{0.16} & \ccell{0.17} & \ccell{0.17} & \ccell{0.21} & \ccell{0.12} & \ccell{0.16} & \ccell{0.29} & \ccell{0.25} & \ccell{0.18} & \ccell{0.28} & \ccell{0.17} \\
    & D8 & \ccell{0.11} & \ccell{0.18} & \ccell{0.19} & \ccell{0.10} & \ccell{0.13} & \ccell{0.13} & \ccell{0.15} & \ccell{0.23} & \ccell{0.15} & \ccell{0.16} & \ccell{0.28} & \ccell{0.25} & \ccell{0.17} & \ccell{0.25} & \ccell{0.18} \\
    & C1 & \ccell{0.07} & \ccell{0.12} & \ccell{0.16} & \ccell{0.09} & \ccell{0.14} & \ccell{0.16} & \ccell{0.10} & \ccell{0.19} & \ccell{0.10} & \ccell{0.17} & \ccell{0.21} & \ccell{0.23} & \ccell{0.12} & \ccell{0.23} & \ccell{0.15} \\
    & C2 & \ccell{0.07} & \ccell{0.17} & \ccell{0.18} & \ccell{0.12} & \ccell{0.14} & \ccell{0.15} & \ccell{0.15} & \ccell{0.21} & \ccell{0.16} & \ccell{0.16} & \ccell{0.26} & \ccell{0.22} & \ccell{0.16} & \ccell{0.20} & \ccell{0.19} \\
    & C3 & \ccell{0.06} & \ccell{0.20} & \ccell{0.14} & \ccell{0.08} & \ccell{0.15} & \ccell{0.11} & \ccell{0.14} & \ccell{0.15} & \ccell{0.15} & \ccell{0.12} & \ccell{0.19} & \ccell{0.16} & \ccell{0.14} & \ccell{0.18} & \ccell{0.16} \\

\midrule
\multirow{13}{*}{\rotatebox{90}{\textbf{\primevulc}}}
    & original & \ccell{0.08} & \ccell{0.08} & \ccell{0.08} & \ccell{0.08} & \ccell{0.07} & \ccell{0.18} & \ccell{0.13} & \ccell{0.16} & \ccell{0.26} & \ccell{0.10} & \ccell{0.17} & \ccell{0.29} & \ccell{0.08} & \ccell{0.16} & \ccell{0.27} \\
\cmidrule(lr){2-17}
    & L1 & \ccell{0.07} & \ccell{0.17} & \ccell{0.15} & \ccell{0.02} & \ccell{0.08} & \ccell{0.09} & \ccell{0.11} & \ccell{0.08} & \ccell{0.09} & \ccell{0.12} & \ccell{0.15} & \ccell{0.24} & \ccell{0.08} & \ccell{0.07} & \ccell{0.27} \\
    & L2 & \ccell{0.05} & \ccell{0.23} & \ccell{0.15} & \ccell{0.09} & \ccell{0.11} & \ccell{0.10} & \ccell{0.18} & \ccell{0.15} & \ccell{0.14} & \ccell{0.10} & \ccell{0.22} & \ccell{0.35} & \ccell{0.13} & \ccell{0.11} & \ccell{0.31} \\
    & L3 & \ccell{0.05} & \ccell{0.28} & \ccell{0.15} & \ccell{0.06} & \ccell{0.10} & \ccell{0.10} & \ccell{0.17} & \ccell{0.15} & \ccell{0.20} & \ccell{0.16} & \ccell{0.20} & \ccell{0.27} & \ccell{0.09} & \ccell{0.10} & \ccell{0.33} \\
    & L4+L5+L6 & \ccell{0.08} & \ccell{0.18} & \ccell{0.18} & \ccell{0.06} & \ccell{0.10} & \ccell{0.09} & \ccell{0.14} & \ccell{0.13} & \ccell{0.10} & \ccell{0.09} & \ccell{0.18} & \ccell{0.20} & \ccell{0.08} & \ccell{0.08} & \ccell{0.36} \\
    & L7 & \ccell{0.06} & \ccell{0.18} & \ccell{0.15} & \ccell{0.10} & \ccell{0.12} & \ccell{0.13} & \ccell{0.14} & \ccell{0.11} & \ccell{0.19} & \ccell{0.13} & \ccell{0.18} & \ccell{0.28} & \ccell{0.11} & \ccell{0.12} & \ccell{0.28} \\
    & L8 & \ccell{0.03} & \ccell{0.17} & \ccell{0.16} & \ccell{0.04} & \ccell{0.10} & \ccell{0.07} & \ccell{0.13} & \ccell{0.12} & \ccell{0.13} & \ccell{0.10} & \ccell{0.13} & \ccell{0.16} & \ccell{0.06} & \ccell{0.06} & \ccell{0.20} \\
    & D1+D2+D3 & \ccell{0.05} & \ccell{0.12} & \ccell{0.05} & \ccell{0.06} & \ccell{0.14} & \ccell{0.10} & \ccell{0.09} & \ccell{0.14} & \ccell{0.10} & \ccell{0.20} & \ccell{0.15} & \ccell{0.23} & \ccell{0.05} & \ccell{0.08} & \ccell{0.34} \\
    & D4+D5+D6+D7 & \ccell{0.06} & \ccell{0.25} & \ccell{0.15} & \ccell{0.05} & \ccell{0.13} & \ccell{0.12} & \ccell{0.15} & \ccell{0.13} & \ccell{0.13} & \ccell{0.17} & \ccell{0.22} & \ccell{0.28} & \ccell{0.08} & \ccell{0.09} & \ccell{0.34} \\
    & D8 & \ccell{0.06} & \ccell{0.21} & \ccell{0.16} & \ccell{0.08} & \ccell{0.11} & \ccell{0.10} & \ccell{0.15} & \ccell{0.12} & \ccell{0.13} & \ccell{0.10} & \ccell{0.19} & \ccell{0.29} & \ccell{0.08} & \ccell{0.11} & \ccell{0.36} \\
    & C1 & \ccell{0.06} & \ccell{0.19} & \ccell{0.12} & \ccell{0.04} & \ccell{0.12} & \ccell{0.10} & \ccell{0.09} & \ccell{0.08} & \ccell{0.07} & \ccell{0.18} & \ccell{0.17} & \ccell{0.26} & \ccell{0.05} & \ccell{0.11} & \ccell{0.35} \\
    & C2 & \ccell{0.06} & \ccell{0.19} & \ccell{0.10} & \ccell{0.06} & \ccell{0.10} & \ccell{0.10} & \ccell{0.13} & \ccell{0.14} & \ccell{0.09} & \ccell{0.27} & \ccell{0.22} & \ccell{0.30} & \ccell{0.13} & \ccell{0.15} & \ccell{0.35} \\
    & C3 & \ccell{0.00} & \ccell{0.09} & \ccell{0.04} & \ccell{0.00} & \ccell{0.00} & \ccell{0.01} & \ccell{0.06} & \ccell{0.02} & \ccell{0.03} & \ccell{0.05} & \ccell{0.01} & \ccell{0.05} & \ccell{0.03} & \ccell{0.03} & \ccell{0.05} \\

\bottomrule
\end{tabular}
}
\label{tab:detection-result-all-datasets-type-evaluation}
\vspace{-4mm}
\end{table*}

\vspace{2mm}
\noindent\textbf{Evaluation Metrics.} Our metric for evaluating the impact of code obfuscation on LLM-based vulnerability detection is the change in detection grade. For a given code snippet containing a target vulnerability, a \downgrade occurs when the LLM detects the target vulnerability in the unobfuscated code but fails to detect it in the obfuscated version. Conversely, an \upgrade occurs when the LLM fails to detect the target vulnerability in the unobfuscated code but successfully identifies it after obfuscation.

Specifically, we evaluate the LLM's performance in the vulnerability detection task using a 1--4 scoring scale. 
Formally, let a vulnerable file $A$ contain a ground-truth vulnerability $a$, and let the LLM's detection result be a set of reported vulnerabilities $S$. 
The score is defined as follows:
\begin{itemize}
    \item {Score 1}: $S = \varnothing$.
    \item {Score 2}: $S \neq \varnothing \land a \notin S$ (e.g., $S=\{b,c\}$).
    \item {Score 3}: $S = \{a\}$.
    \item {Score 4}: $a \in S \land S \neq \{a\}$ (e.g., $S=\{a,b,c\}$).
\end{itemize}

We define the criterion for successful LLM-based vulnerability detection as the audit report both identifying the existence of a vulnerability and correctly classifying its type in accordance with the ground truth (corresponding to a score of 3 or 4). Detection is considered a failure if the LLM either reports no vulnerabilities or reports multiple vulnerabilities without including the true positive (corresponding to a score of 1 or 2).

Accordingly, we define a \downgrade as a case where an initial score of 3/4 is reduced to 1/2 after code obfuscation, and an \upgrade as the opposite case, where an initial score of 1/2 is increased to 3/4 after code obfuscation.

\vspace{2mm}
\noindent\textbf{Environment.} Experiments on qn-14b and qn-32b were performed on a high-performance server equipped with 8 NVIDIA GeForce RTX 4090 GPUs (24 GB, PCIe) running CUDA version 12.4 and an Intel Xeon Gold 6426Y CPU. Experiments on the remaining 13 models were performed via the OpenRouter\cite{OpenRouter} API. Experiments on GitHub Copilot and Codex were conducted through official CLI tools and vscode extensions.

\subsection{Experiment Results}


\subsubsection{Implications of Obfuscation Methods (RQ1)}

To systematically examine how obfuscation impacts LLM-based vulnerability detection, we conducted large-scale experiments applying a taxonomy of obfuscation transformations, including layout, data-flow, control-flow, and multi-language, across multiple datasets (\reposvulcpp, \primevulc, \smartbugs, \reposvulpy) and diverse LLM families of varying sizes. For each obfuscated sample, we compared the model’s detection results against its unobfuscated counterpart to quantify how obfuscation alters performance.
\autoref{tab:detection-result-all-datasets-type-evaluation} shows that code obfuscation can trigger two opposing effects on vulnerability detection, which we term \upgrade and \downgrade. {Downgrade} arises when obfuscation reduces detection success compared to the original code. This is the expected and dominant outcome across most datasets, especially \reposvulcpp\ and \primevulc, where layout obfuscation (e.g., L8) and control-flow perturbations (e.g., C3) often lead to sharp performance drops for qwen and llama models. Such cases show how surface-level distortions can obscure vulnerable semantics and mislead detection.

What is more striking, however, is the presence of {upgrade}: cases where obfuscation unexpectedly boosts detection accuracy. On \smartbugs, for instance, data-flow and control-flow manipulations (e.g., D4+D5+D6+D7, C3) sometimes outperform the unobfuscated baseline for qn-32b and lm-70b. Similarly, on \reposvulpy, layout variants L2 and L3 yield modest yet consistent improvements for DeepSeek and OPENAI models. These results suggest that certain perturbations may filter out misleading surface cues and instead drive models to focus more directly on vulnerability-related patterns. 
Finally, model capacity plays a decisive role. Proprietary models such as gpt-4o, o3-mini, and ds-v3 not only reach higher baselines but also display smaller downgrade gaps and more frequent upgrades, while smaller open-source models are more susceptible to obfuscation. Overall, these findings reveal that obfuscation does not merely erode detection but can paradoxically enhance it,  a counterintuitive dual effect that motivates our deeper exploration of the mechanisms behind \upgrade and \downgrade.

\findbox{After code obfuscation, the overall numbers of successful and failed detections remain largely unchanged. Yet, a surprising pattern emerges: beyond expected \downgrade cases (where previously detected vulnerabilities become hidden), we also observe clear \upgrade cases, in which obfuscation unexpectedly enables detection of originally missed vulnerabilities.}
\vspace{-2mm}

\begin{figure*}[t] 
    \centering
    \begin{subfigure}[t]{0.48\textwidth}
        \centering
        \includegraphics[width=\textwidth]{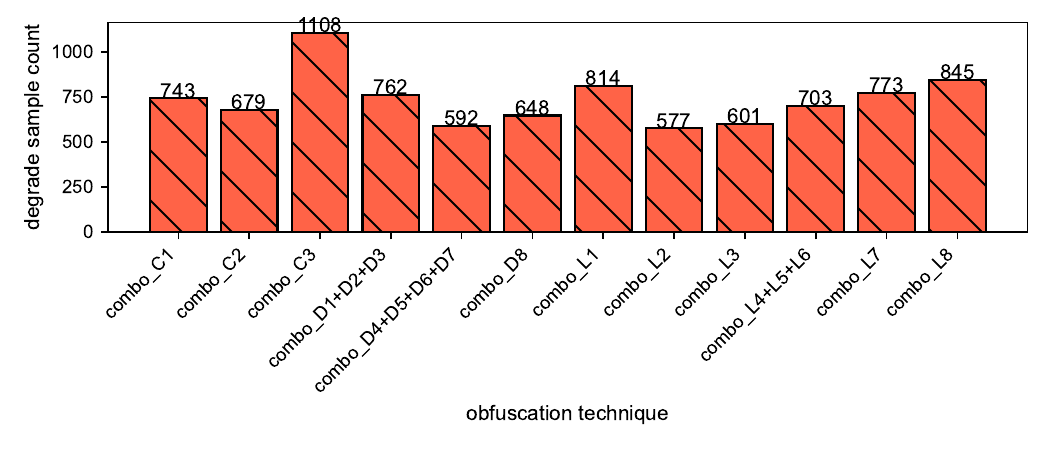}
        \caption{Distribution of downgrade samples on obfuscation technique.}
        \label{fig:upgrade_by_combo}
    \end{subfigure}
    \hfill
    \begin{subfigure}[t]{0.48\textwidth}
        \centering
        \includegraphics[width=\textwidth]{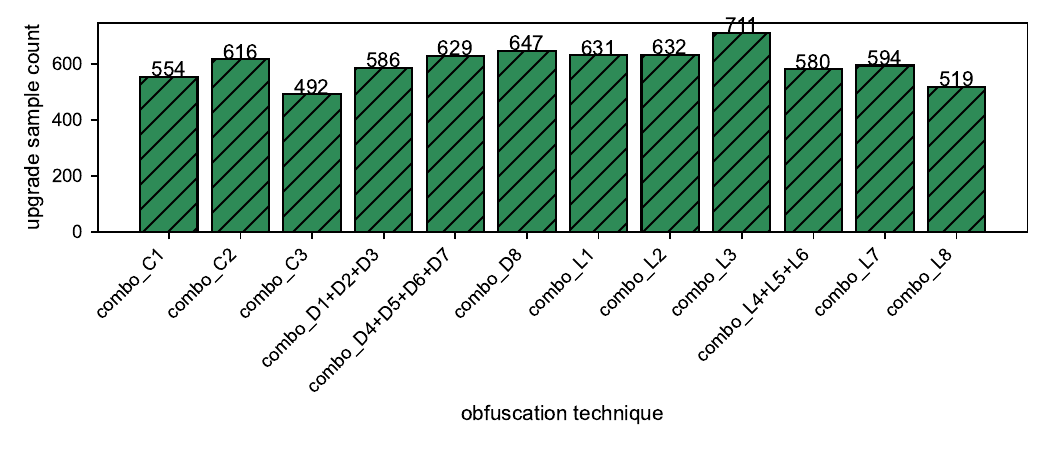}
        \caption{Distribution of upgrade samples on obfuscation technique.}
        \label{fig:downgrade_by_combo}
    \end{subfigure}
    \caption{Distribution of downgrade and upgrade samples on obfuscation technique.}
    \vspace{-3mm}
    \label{fig:upgrade_downgrade_by_combo}
\end{figure*}

\begin{figure*}[t] 
    \centering
    \begin{subfigure}[t]{0.48\textwidth}
        \centering
        \includegraphics[width=\textwidth]{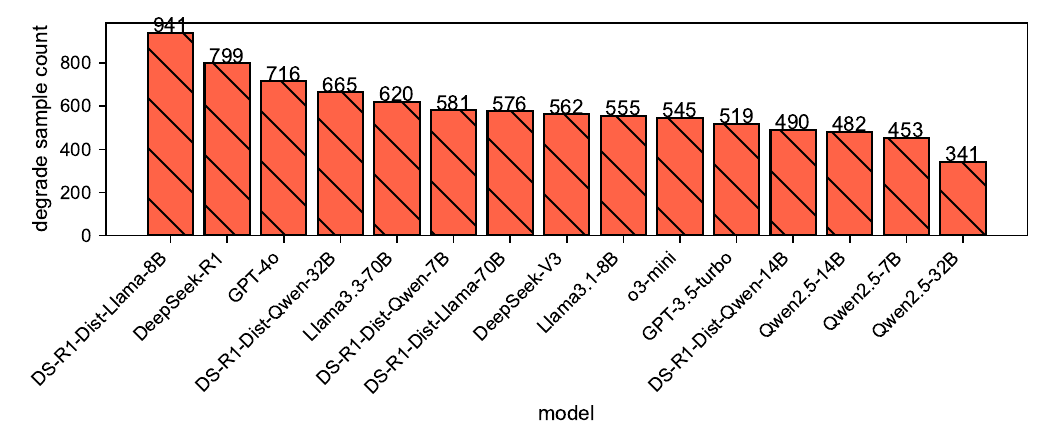}
        \caption{Distribution of downgrade samples on model.}
        \label{fig:upgrade_by_model}
    \end{subfigure}
    \hfill
    \begin{subfigure}[t]{0.48\textwidth}
        \centering
        \includegraphics[width=\textwidth]{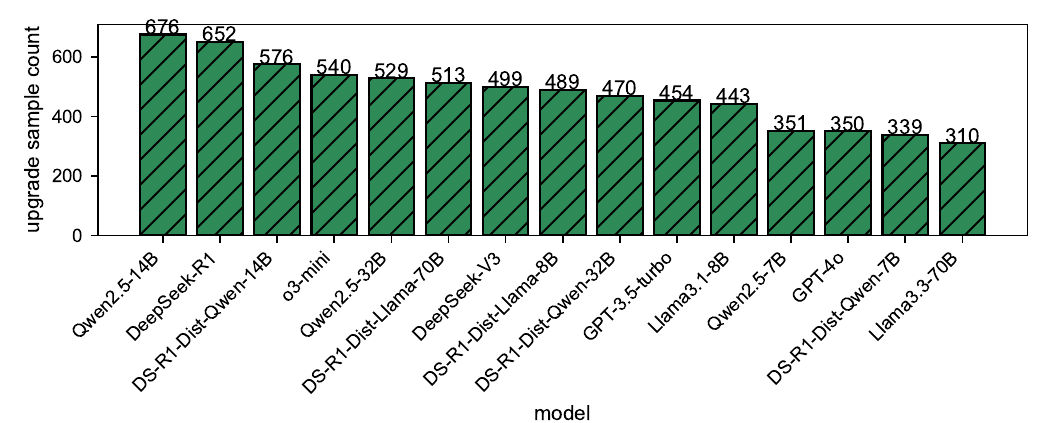}
        \caption{Distribution of upgrade samples on model.}
        \label{fig:downgrade_by_model}
    \end{subfigure}
    \caption{Distribution of downgrade and upgrade samples on model.}
    \vspace{-5mm}
    \label{fig:upgrade_downgrade_by_model}
\end{figure*}

\begin{figure}
    \centering
    \includegraphics[width=0.48\textwidth]{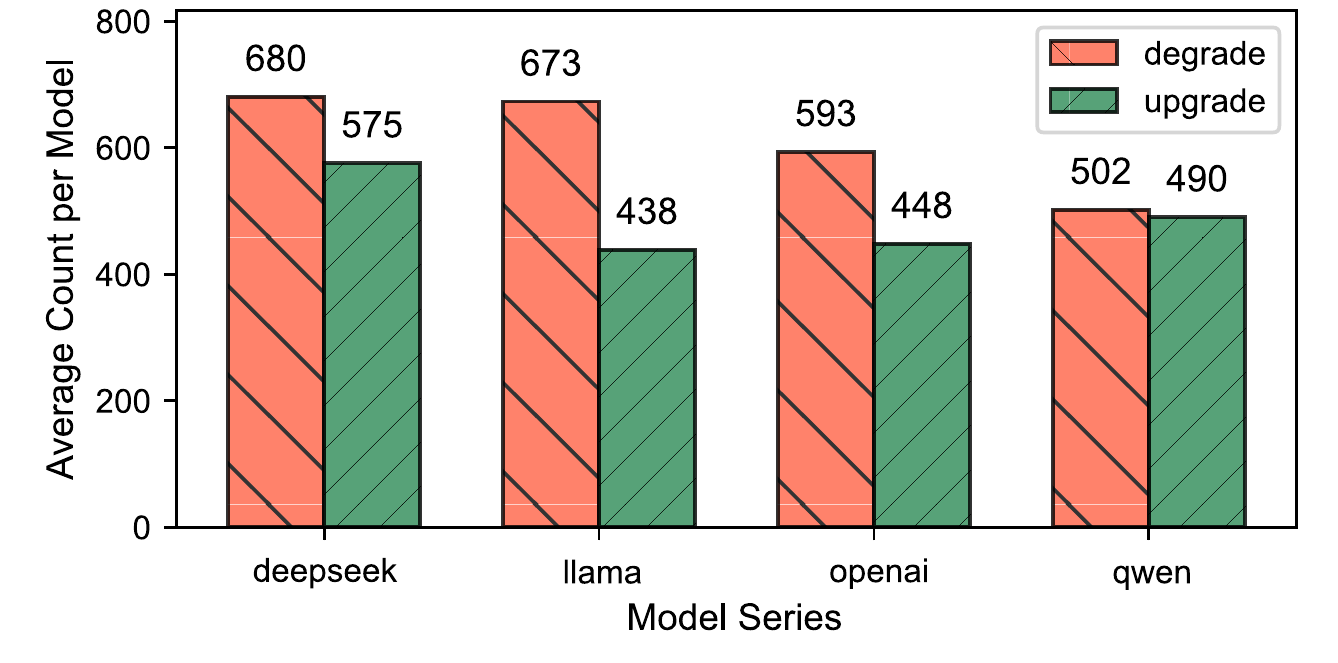}
    \caption{Average Upgrade/Degrade per Model (by Series).}
    \vspace{-5mm}
    
    \label{fig:degrade_upgrade_by_model_series}
\end{figure}

To further analyze the effect of different obfuscation combinations on the upgrade and downgrade phenomena, we adopt the same experimental setup as in RQ1 and count \downgrade and \upgrade samples for each obfuscation type. Results show that the influence of obfuscation is not uniform: some methods mainly introduce noise that degrades detection, while others can more systematically break model reasoning. As shown in ~\autoref{fig:upgrade_downgrade_by_combo} {virtualization}-based obfuscation exhibits the most pronounced impact. On the C/C++ datasets, its attack success rate reaches about $_{\widetilde{~}}$80\%, indicating that transforming original control-flow and execution semantics into virtualized instruction sets creates a representation gap that current LLMs struggle to bridge. The second most effective method is \textit{mixed-programming-language} obfuscation, with an attack success rate of around $_{\widetilde{~}}$75\%. By inserting heterogeneous constructs and blending language syntaxes, it disrupts token patterns and semantics, confusing even large models. In comparison, layout- or identifier-level perturbations (e.g., L1, L8) generally yield weaker downgrade effects and occasionally even lead to upgrades, as models refocus on semantic cues rather than superficial code structures.


\findbox{
Among all source code obfuscation methods, \emph{virtualization}-based obfuscation has the most significant impact on the vulnerability detection accuracy of LLMs, achieving an attack success rate of up to approximately 80\% on the C/C++ datasets. 
The second most effective method is \emph{mixed-programming-language} obfuscation, with an attack success rate of around 75\%.
}

\subsubsection{Implication of Models (RQ2)}

\begin{figure}[t]
    \hspace*{2mm}
    \begin{subfigure}[t]{0.48\textwidth}
        \centering
        \includegraphics[width=\textwidth]{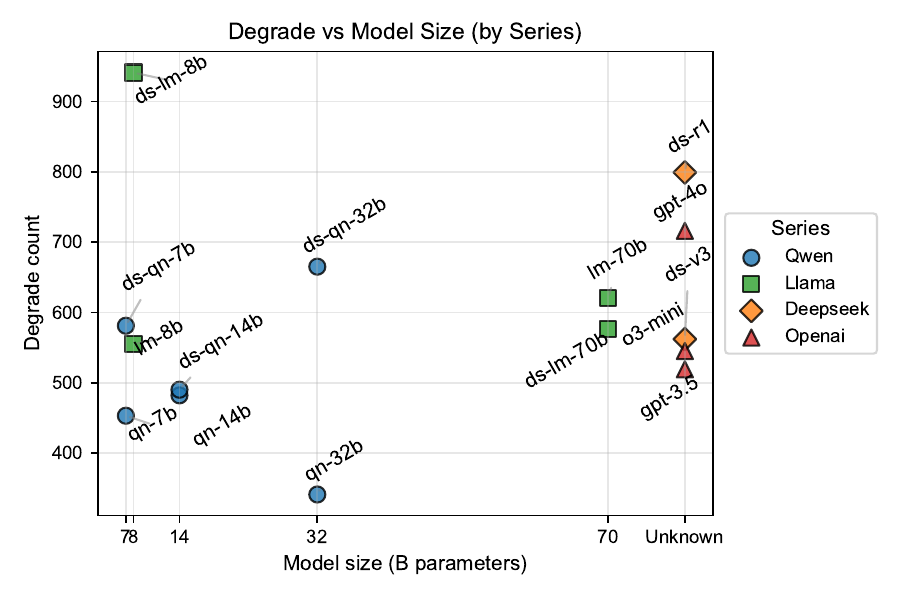}
        \caption{Distribution of downgrade samples on model size.}
        \label{fig:upgrade_by_model_size}
    \end{subfigure}

    \vspace{1em} 
    \hspace*{1mm}
    \begin{subfigure}[t]{0.48\textwidth}
        \centering
        \includegraphics[width=\textwidth]{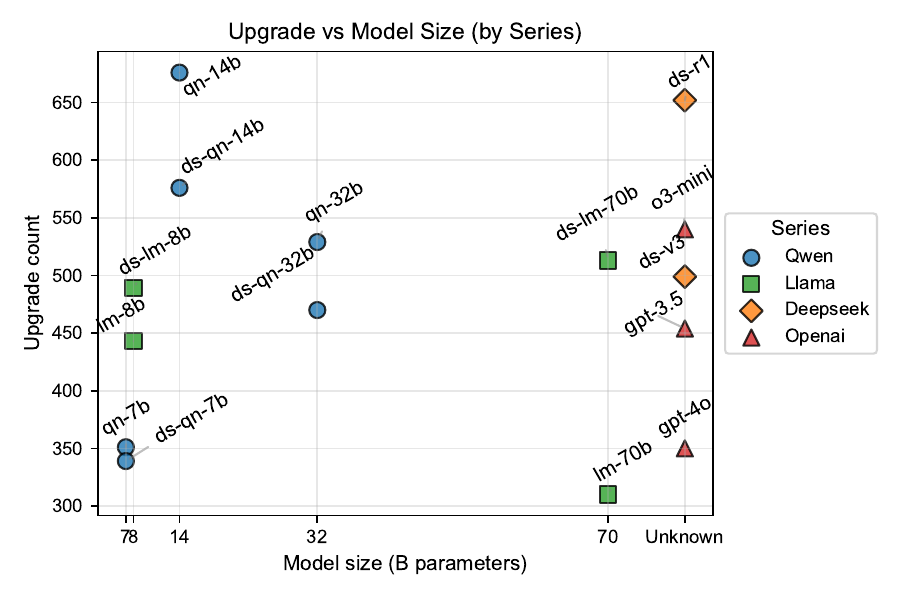}
        \caption{Distribution of upgrade samples on model size.}
        \label{fig:downgrade_by_model_size}
    \end{subfigure}

    \caption{Distribution of downgrade and upgrade samples on model size.}
    \vspace{-5mm}
    \label{fig:upgrade_downgrade_by_model_size}
\end{figure}

To understand how variations in LLM families (e.g., OpenAI, LLaMA, Qwen, DeepSeek), parameter sizes and reasoning capabilities impact their robustness against obfuscation, we use the sample experiment settings in RQ1, and filter \downgrade and \upgrade samples by model families, parameter sizes and reasoning capabilities. As shown in ~\autoref{fig:degrade_upgrade_by_model_series}, in the perspective of  {model series}, we observe clear differences in robustness against obfuscation. \textsc{DeepSeek} demonstrates the lowest stability, followed by \textsc{LLaMA}, \textsc{OpenAI}, and finally \textsc{Qwen}. This ranking indicates that proprietary and carefully optimized models tend to maintain more consistent performance, whereas open-source families, particularly \textsc{Qwen}, are more susceptible to obfuscation-induced shifts.

As shown in ~\autoref{fig:upgrade_downgrade_by_model_size}, in terms of  {model size}, our results reveal a distinct boundary at the 8B parameter scale. Models smaller than 8B are significantly more vulnerable to downgrade, while models larger than 8B achieve noticeably higher robustness under obfuscation. Nevertheless, once the threshold of 8B is crossed, additional scaling yields only marginal improvements, suggesting that sheer model size is not sufficient to guarantee resilience.
Finally, when comparing  {reasoning vs. non-reasoning models}, as illustrated in \autoref{fig:upgrade_downgrade_by_model_size}, we find that reasoning-augmented models generally perform better on unobfuscated code, achieving higher detection success rates. However, in the \textsc{Qwen} and \textsc{LLaMA} families, reasoning models are paradoxically more prone to downgrade once obfuscation is applied. This indicates a trade-off: reasoning enhances \emph{endorsement ability} under clean conditions, but at the cost of reduced \emph{generalization ability} under obfuscation.

\findbox{
Our results reveal a counterintuitive trade-off between detection strength and robustness under obfuscation. 
While larger models ($\geq$8B) and reasoning-oriented variants typically achieve higher detection success on unobfuscated inputs, 
they often suffer from greater instability once obfuscation is applied, exhibiting higher downgrade probabilities 
(and weaker upgrade consistency) than their smaller or non-reasoning counterparts. 
This pattern is especially pronounced in the \textsc{Qwen} and \textsc{LLaMA} families, where reasoning models display 
stronger \emph{endorsement ability} but reduced \emph{generalization ability} under perturbations. 
 
}

\subsubsection{Detection Granularity (RQ3)}

A key limitation of prior work on LLM-based vulnerability detection is its narrow focus on whether a vulnerability simply \emph{exists}, without verifying whether the model can also provide a correct explanation of the vulnerability’s \emph{type} and precise \emph{location}.  
To address this gap, our RQ1 and RQ2 experiments adopted a more fine-grained evaluation: we scored each prediction from 1–4, where \textit{score 3/4} indicates that both the vulnerability type and location are correctly identified (counted as \emph{positive}), and \textit{score 1/2} indicates incomplete or incorrect explanation (counted as \emph{negative}).  
We refer to this as the \emph{type evaluation}.

For RQ3, we additionally replicate the \emph{binary existence} setting commonly used in previous studies by relaxing the criterion:  
\emph{scores 2/3/4} are treated as \emph{positive} and only \textbf{score 1} as \emph{negative}, which we call the \emph{existence evaluation}.  
As shown in \autoref{tab:detection-result-all-datasets-existence-evaluation} in the appendix, this change produces a striking effect.  
Across all datasets and models, the detection success rate increases by roughly \textit{40\%} on average compared with our type evaluation (cf. \autoref{tab:detection-result-all-datasets-type-evaluation}), for both original and obfuscated code.
When we compare original and obfuscated code under the existence evaluation, we observe that obfuscation rarely reduces detection success.  
In fact, most models experience noticeable \emph{improvements}.  
For example, on the \primevulc dataset the \textsc{Qwen-14B} model’s detection success jumps from \emph{0.39} on the original code to \emph{0.80} after applying L1 obfuscation.  
By contrast, under the stricter type evaluation, the same obfuscations produce mixed outcomes (both upgrades and downgrades) across all models.

\findbox{While LLMs can often confirm the \emph{existence} of a vulnerability even after heavy code obfuscation, they struggle to correctly identify the \emph{type} and \emph{location}.  
In other words, obfuscation does not significantly hinder binary yes/no judgments but substantially degrades the fine-grained reasoning required for precise vulnerability characterization.  
This underscores the importance of moving beyond simple existence checks toward evaluations that capture the full explanatory capability of LLM-based vulnerability detection.}

\subsubsection{Implication of Dataset Diversity (RQ4)}

To better understand how the dataset influences the ability of LLMs to resist obfuscation, we analyzed dataset-related attributes under the same experimental setup as RQ1, including vulnerability type, lines of code (LOC), code complexity and programming language.

\begin{figure*}[htbp] 
    \centering
    \begin{subfigure}[t]{0.48\textwidth}
        \centering
        \includegraphics[width=\textwidth]{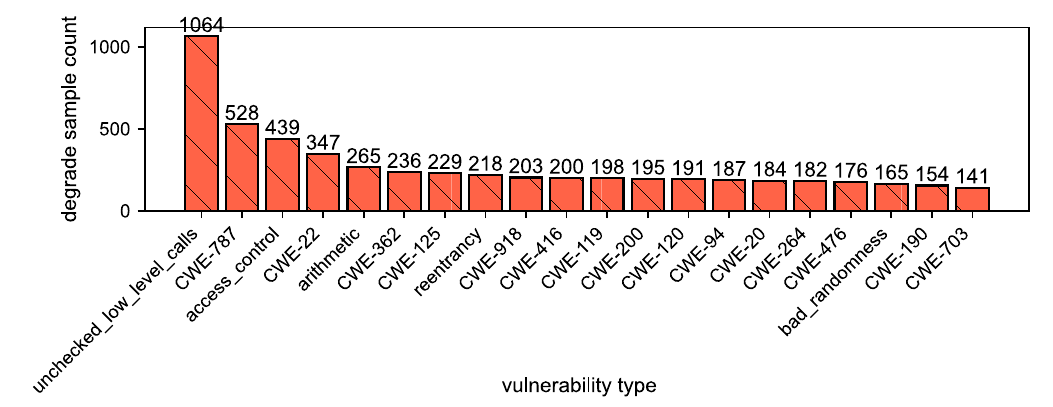}
        \caption{Distribution of downgrade samples on vulnerability type.}
        \label{fig:upgrade_by_vuln_type}
    \end{subfigure}
    \hfill
    \begin{subfigure}[t]{0.48\textwidth}
        \centering
        \includegraphics[width=\textwidth]{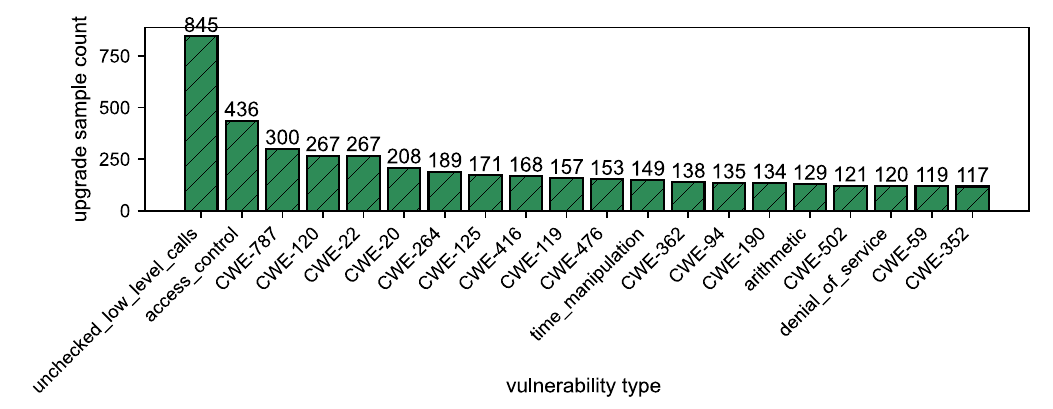}
        \caption{Distribution of upgrade samples on vulnerability type.}
        \label{fig:downgrade_by_vuln_type}
    \end{subfigure}
    \caption{Distribution of downgrade and upgrade samples on vulnerability type.}
    \vspace{-10pt}
    \label{fig:upgrade_downgrade_by_vuln_type}
\end{figure*}

\vspace{2mm}
\noindent\textbf{Impact of Vulnerability Metric.} We further analyze the vulnerability types most frequently associated with downgrade and upgrade cases (~\autoref{fig:upgrade_by_vuln_type} and~\autoref{fig:downgrade_by_vuln_type}). Interestingly, both distributions overlap considerably: categories such as \textit{unchecked low-level calls}, \textit{access control}, and memory-safety related CWEs (e.g., CWE-787, CWE-125, CWE-119, CWE-120, CWE-416, CWE-476) appear in the top ranks of both lists. This indicates that the same vulnerability families can be either suppressed or amplified by obfuscation depending on the obfuscation strategy and the model involved.

A closer inspection reveals common characteristics. First, many of these vulnerabilities are \textit{low-level semantic in nature}, involving pointer/memory errors, unchecked external calls, or improper access control. Such vulnerabilities are tightly coupled with subtle code patterns (e.g., boundary checks, privilege conditions), which makes detection sensitive to syntactic and structural transformations. Second, several vulnerability types (e.g., reentrancy, arithmetic, randomness flaws) rely on \textit{contextual reasoning across function boundaries}, meaning that both obfuscation-induced noise and redundancy may either obscure critical cues (leading to downgrade) or highlight essential dependencies (leading to upgrade). Finally, vulnerabilities like time manipulation and denial-of-service show that \textit{environmental or behavioral properties} are also affected: obfuscation may reduce misleading surface cues and force models to rely on temporal or resource-related semantics, explaining upgrade cases. 
Overall, the vulnerabilities most prone to change share the trait of being \textit{semantically entangled and context-dependent}. Their detection is fragile against superficial obfuscation but can also benefit when obfuscation removes irrelevant signals. This duality reinforces the need to view obfuscation not only as a degradation tool but also as a potential enhancer of model generalization.

\begin{figure}[htbp] 
    \centering
    \begin{subfigure}[t]{0.48\columnwidth}
        \centering
        \includegraphics[width=\textwidth]{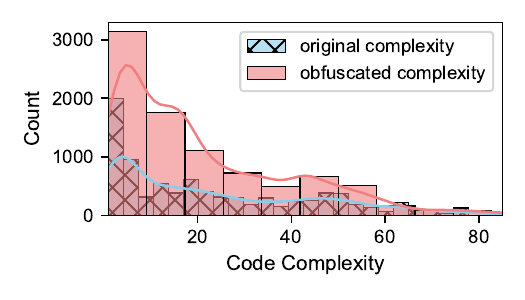}
        \caption{Distribution of downgrade samples on code complexity.}
        \label{fig:downgrade_by_complexity}
    \end{subfigure}
    \hfill
    \begin{subfigure}[t]{0.48\columnwidth}
        \centering
        \includegraphics[width=\textwidth]{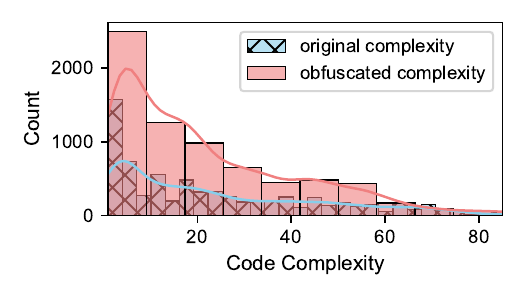}
        \caption{Distribution of upgrade samples on code complexity.}
        \label{fig:upgrade_by_complexity}
    \end{subfigure}
    \caption{Distribution of downgrade and upgrade samples on code complexity.}
    \label{fig:upgrade_downgrade_by_complexity}
    \vspace{-10pt}
\end{figure}

\begin{figure}[htbp] 
    \centering
    \begin{subfigure}[t]{0.48\columnwidth}
        \centering
        \includegraphics[width=\textwidth]{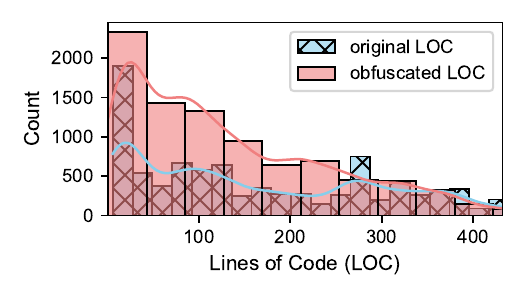}
        \caption{Distribution of downgrade samples on LOC.}
        \label{fig:downgrade_by_loc}
    \end{subfigure}
    \hfill
    \begin{subfigure}[t]{0.48\columnwidth}
        \centering
        \includegraphics[width=\textwidth]{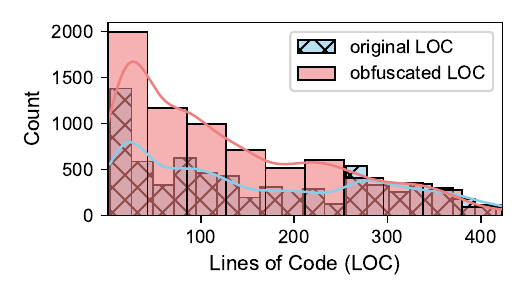}
        \caption{Distribution of upgrade samples on LOC.}
        \label{fig:upgrade_by_loc}
    \end{subfigure}
    \caption{Distribution of downgrade and upgrade samples on LOC.}
    \label{fig:upgrade_downgrade_by_loc}
    \vspace{-3mm}
\end{figure}

\begin{figure}[h] 
    \centering
    \begin{subfigure}[t]{0.48\columnwidth}
        \centering
        \includegraphics[width=\textwidth]{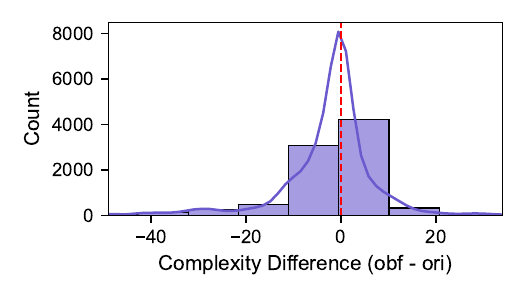}
        \caption{Distribution of downgrade samples on complexity difference.}
        \label{fig:downgrade_by_complexity_diff}
    \end{subfigure}
    \hfill
    \begin{subfigure}[t]{0.48\columnwidth}
        \centering
        \includegraphics[width=\textwidth]{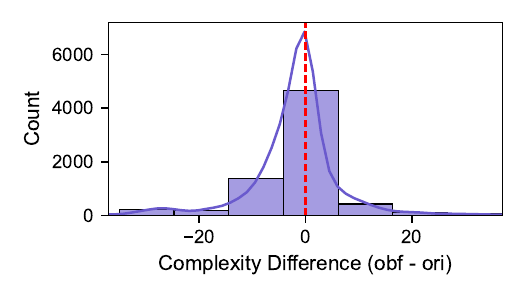}
        \caption{Distribution of upgrade samples on complexity difference.}
        \label{fig:upgrade_by_complexity_diff}
    \end{subfigure}
    \caption{Distribution of downgrade and upgrade samples on code complexity difference.}
    \label{fig:upgrade_downgrade_by_complexity_diff}
\end{figure}

\begin{figure}[htbp] 
    \centering
    \begin{subfigure}[t]{0.48\columnwidth}
        \centering
        \includegraphics[width=\textwidth]{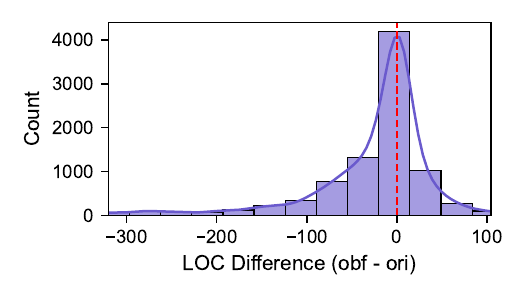}
        \caption{Distribution of downgrade samples on LOC difference.}
        \label{fig:downgrade_by_loc_diff}
    \end{subfigure}
    \hfill
    \begin{subfigure}[t]{0.48\columnwidth}
        \centering
        \includegraphics[width=\textwidth]{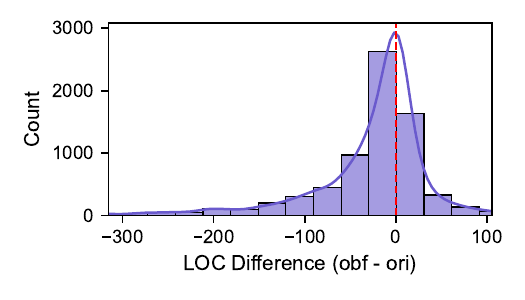}
        \caption{Distribution of upgrade samples on LOC difference.}
        \label{fig:upgrade_by_loc_diff}
    \end{subfigure}
    \caption{Distribution of downgrade and upgrade samples on LOC difference.}
    \vspace{-3mm}
    \label{fig:upgrade_downgrade_by_loc_diff}
\end{figure}

\vspace{2mm}
\noindent\textbf{LOC and Complexity.} We also examine how code length (LOC) and structural complexity correlate with the upgrade and downgrade effects. As shown in ~\autoref{fig:upgrade_downgrade_by_complexity} - ~\autoref{fig:upgrade_downgrade_by_loc_diff}, the results suggest a non-linear relationship: vulnerability detection can fail when the code is either \emph{overly complex} or \emph{overly simple}. In highly complex cases, obfuscation introduces additional noise or intricate control/data-flow dependencies that overwhelm the model’s reasoning ability, leading to downgrade. However, a more striking observation is that the probability of detection failure is actually higher when the code becomes simpler. 
This counterintuitive phenomenon indicates that obfuscation strategies which \textit{reduce code complexity}: for example, by flattening control flow, removing contextual redundancy, or abstracting semantics, tend to strip away auxiliary cues that LLMs often rely on. As a result, the simplified code obscures critical vulnerability patterns and undermines detection accuracy. In contrast, when complexity increases, while detection is challenged, models may still capture latent signals if sufficient semantic redundancy remains. 
In summary, obfuscation that decreases code size or reduces complexity poses a greater threat to LLM-based vulnerability detection than complexity-increasing transformations. This highlights a key weakness of current models: their reliance on superficial structural cues rather than robust semantic understanding.

\begin{figure*}[htbp] 
    \centering
    \begin{subfigure}[t]{0.48\textwidth}
        \centering
        \includegraphics[width=\textwidth]{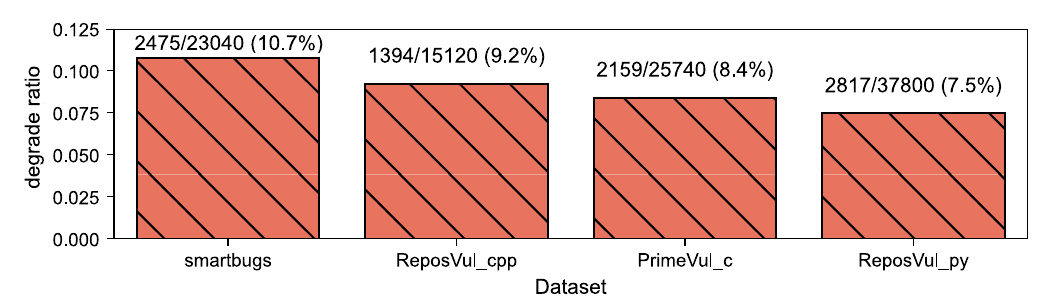}
        \caption{Distribution of downgrade samples on dataset.}
        \vspace{-2mm}
        \label{fig:upgrade_by_dataset}
    \end{subfigure}
    \hfill
    \begin{subfigure}[t]{0.48\textwidth}
        \centering
        \includegraphics[width=\textwidth]{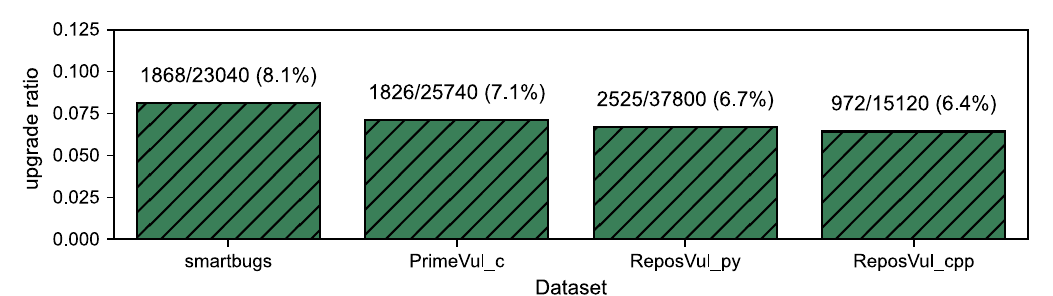}
        \caption{Distribution of upgrade samples on dataset.}
        \vspace{-2mm}
        \label{fig:downgrade_by_dataset}
    \end{subfigure}
    \caption{Distribution of downgrade and upgrade samples on dataset.}
    \vspace{-4mm}\label{fig:upgrade_downgrade_by_dataset}
\end{figure*}

\noindent\textbf{Promgramming Language.} We further compare the distribution of upgrade and downgrade cases across different datasets and programming languages. As illustrated in ~\autoref{fig:upgrade_downgrade_by_dataset}, for downgrade, the highest rate is observed in \smartbugs\ (Solidity), with 2475 out of 23040 cases (10.7\%), followed by \reposvulcpp\ (C++) at 9.2\%, \primevulc\ (C) at 8.4\%, and \reposvulpy\ (Python) at 7.5\%. Similarly, upgrade cases are most frequent in \smartbugs\ (8.1\%), while \primevulc\ (7.1\%), \reposvulpy\ (6.7\%), and \reposvulcpp\ (6.4\%) follow behind.
These results indicate that the likelihood of both upgrading and degrading is strongly dataset-dependent, reflecting differences in language characteristics and vulnerability patterns. Solidity (\smartbugs) shows the greatest sensitivity to obfuscation, possibly due to its contract-oriented programming paradigm and the prevalence of security-critical patterns that are easily disrupted. Python (\reposvulpy) ranks second, where dynamic typing and concise syntax reduce robustness under obfuscation. C (\primevulc) and C++ (\reposvulcpp), despite having complex low-level semantics, appear comparatively more stable, showing fewer upgrade and downgrade shifts.

\findbox{The effect of obfuscation on LLM-based vulnerability detection emerges from the interplay among dataset composition, code complexity, and semantic representation.
Datasets with distinctive or context-dependent patterns are most affected as obfuscation perturbs the lexical and structural cues LLMs use for reasoning.
When obfuscation simplifies or fragments code, it often removes contextual redundancy, causing semantic drift in model interpretation.
Overall, varying robustness across datasets reflects the tension between the model’s level of semantic abstraction and the heterogeneity of real-world code: a gap that obfuscation can easily amplify.}
\vspace{-3mm}

\subsubsection{Effect on Coding Agents (RQ5)}

Previous studies have mainly examined general-purpose LLMs under self-defined prompts, without assessing the effect of code obfuscation within a coding-agent setting, where models are optimized for code tasks and often leverage black-box prompts along with auxiliary toolchains. To address this gap, our study evaluates the effect of code obfuscation on two coding agent, GitHub Copilot and Codex.

Especially, we construct four evaluation subsets, one from each dataset (\smartbugs, \reposvulcpp, \reposvulpy, and \primevulc). For each dataset, we select five code samples that led to the most significant detection \downgrade and five that lead to the most significant \upgrade in RQ1. Each selected sample includes its unobfuscated version and 13 obfuscated variants. We then conduct the same vulnerability detection experiments in RQ1 on two coding agents.

We analyze the vulnerability detection success rate across the four subsets. As shown in \autoref{tab:detection-result-all-datasets-downgrade_top20} and \autoref{tab:detection-result-all-datasets-upgrade_top20} in the appendix, on the \smartbugs, \reposvulcpp, and \primevulc datasets, coding agents consistently outperform general-purpose LLMs on both \textit{downgrade top20} and \textit{upgrade top20}, regardless of code obfuscation. This demonstrates that agents maintain strong detection capability under perturbations that considerably affect general LLMs. On \reposvulpy, the pattern differs slightly: in the \textit{downgrade top20}, agents perform below the DeepSeek series but surpass other models, while in the \textit{upgrade top20}, they are outperformed by lm-8b and ds-lm-8b yet remain stronger than others. Overall, these results suggest that coding agents exhibit higher vulnerability detection success rates compared to general-purpose LLMs, highlighting their robustness against obfuscation-induced performance degradation.

We further evaluate coding agents’ robustness to obfuscation by measuring the \downgrade and \upgrade rates on the selected top20 subsets. As illustrated in \autoref{tab:downgrade-rate-all-datasets-downgrade_top20} in the appendix, on the \textit{downgrade top20} subset, both Copilot and Codex exhibit very low \downgrade rates across nearly all datasets, demonstrating strong resilience to obfuscation-induced performance degradation. Notable exceptions occur on specific obfuscation-dataset combinations, including L7 on SmartBugs and C3 on \reposvulcpp and \primevulc datasets, where agents experience relatively higher \downgrade. As illustrated in \autoref{tab:upgrade-rate-all-datasets-upgrade_top20} in the appendix, on the \textit{upgrade top20} subset, Copilot shows no \upgrade on \smartbugs, as its original samples already achieve 100\% detection success, while in other datasets, its \upgrade rate is moderate relative to other models. Codex exhibits moderate \upgrade on \smartbugs but maintains relatively low \upgrade rates on the remaining datasets. These observations suggest that the most adversarial obfuscations for general LLMs---particularly L7 (inline assembly) and C3 (virtulization)---remain effective against coding agents, though overall the agents are substantially more robust than general-purpose LLMs.

\findbox{
Coding agents achieve higher vulnerability detection accuracy than general-purpose LLMs both before and after obfuscation.
However, the extent to which code obfuscation affects the vulnerability detection performance of coding agents (in terms of downgrade and upgrade rates) is almost identical to that observed in general-purpose LLMs.
}
\vspace{-2mm}

In the Python dataset, we observed that ds-v3 slightly outperformed copilot-gpt-5, which contradicts the trend seen in Solidity, C++ and C datasets. A natural question arises: \textbf{Is this anomaly caused by the underlying model or by the agent framework?}  
To disentangle model effects from agent-framework effects, we introduce a new evaluation setting: \texttt{copilot-ds-v3}, which keeps the Copilot agent framework unchanged but replaces its backbone model with DeepSeek-V3. We then evaluate three models (i.e., ds-v3, copilot-ds-v3, copilot-gpt-5) on the Python dataset.
As shown in the Python portion of \autoref{tab:detection-result-ReposVul_py_downgrade_top20_on_3_model} in the appendix, except for the C3 obfuscation, the overall trend is: \texttt{ds-v3} $\approx$ \texttt{copilot-ds-v3} $>$ \texttt{copilot-gpt-5}. This demonstrates that the inferior performance of \texttt{copilot-gpt-5} on Python is caused by the underlying gpt-5 model. The Copilot agent framework itself is not the root cause.

What's more, across all Python obfuscations, \texttt{ds-v3} and \texttt{copilot-ds-v3} show a mixed pattern (4 wins, 3 losses, 6 ties in favor of \texttt{ds-v3}), indicating no clear dominance; each wins on some obfuscation types. Further questions arise: \textit{Where does \texttt{copilot-ds-v3} lose to \texttt{ds-v3}? Is the lose case also losed by \texttt{copilot-gpt-5}?} 
In the Python dataset, its 20\% drop (relative to \texttt{ds-v3}) under L3, L4+L5+L6, D1+D2+D3, and C1 obfuscations all originate from a single file, \texttt{165\_admin.py}. Taking the L4+L5+L6 setting as an example: \texttt{ds-v3} not only recognizes the transformed layout (loop-to-recursion) but also correctly identifies the CSV injection vulnerability; \texttt{copilot-ds-v3} notices the obfuscation but fails to report the vulnerability; meanwhile, \texttt{copilot-gpt-5} behaves similarly to native \texttt{ds-v3} and successfully detects both. 
This suggests \texttt{copilot-ds-v3}, as an unofficial hybrid, occasionally shows “half-understanding” during knowledge transfer, whereas \texttt{copilot-gpt-5}, as an official hybrid, performs comparably to native \texttt{ds-v3}.
Repeating the analysis on the C++ dataset reveals the same pattern. As shown in the \reposvulcpp portion of \autoref{tab:detection-result-ReposVul_py_downgrade_top20_on_3_model}, \texttt{copilot-ds-v3} again falls behind \texttt{ds-v3} under L3, L7, D1+D2+D3, and C2 obfuscations, concentrated on a single case, \texttt{17\_comment.cpp}. For example, under L7, both \texttt{ds-v3} and \texttt{copilot-gpt-5} correctly detect the out-of-bound read despite inline assembly noise, whereas \texttt{copilot-ds-v3} triggers several false alarms and misses the issue. These cross-language results support our hypothesis: simply hot-plugging a general-purpose LLM into a coding-agent framework does not ensure full capability transfer. Without vendor-side adaptation, hybrids such as \texttt{copilot-ds-v3} tend to show partial understanding, exhibiting solid structural detection but incomplete vulnerability reasoning, whereas officially integrated models avoid this.

\findbox{
Although coding agents generally exhibit stronger anti-obfuscation capability than general-purpose LLMs in vulnerability detection, they can also inherit the weaknesses of their underlying models (e.g., the performance drop of \texttt{copilot-gpt-5} on Python) and may suffer from incomplete capability transfer when hot-plugged into an agent framework without official support (e.g., the \texttt{ds-v3}~$\times$~copilot configuration). Therefore, the choice between agentic and non-agentic tools should be made with caution.
}

\section{Promising Directions for Future Research}\label{sec:future}

\vspace{-0.5mm}
\noindent\textbf{Direction I: Building obfuscation-aware vulnerability detection frameworks that can both leverage and defend against the effects of code transformation (i.e., using the ``upgrade'' phenomena).} 
An important insight from our study is that code obfuscation can both degrade and \emph{improve} the performance of LLM-based vulnerability detection (Finding~\#1). This dual effect suggests two promising directions: on the one hand, the phenomenon of \upgrade\ can be leveraged to strengthen detection models, for example by using obfuscated variants of code to enhance training robustness or to highlight vulnerability-related patterns that are otherwise overlooked. On the other hand, \downgrade\ highlights the potential of obfuscation as an effective attack strategy, capable of systematically impairing the accuracy of LLM-based vulnerability detectors. Understanding how to harness \upgrade\ and how to defend against \downgrade\ therefore represents a crucial challenge for future research.


\vspace{1mm}
\noindent\textbf{Direction II: Developing semantics-preserving, cross-layer obfuscation techniques that can simultaneously disrupt both statistical inference and symbolic reasoning while maintaining program functionality.} 
Our findings reveal several key insights for adversaries seeking to exploit weaknesses in LLM-based vulnerability detectors.
The most effective obfuscation exploits cross-layer semantic mismatches (e.g., Finding \#2, \#5), forcing models to align inconsistent program representations.
Equally important, simplifying code (by flattening control flow or removing contextual redundancy) can be more disruptive than increasing complexity, as it strips away weak contextual cues that LLMs rely on for reasoning (e.g., Finding \#3).
However, attackers must also recognize the upgrade paradox: certain layout or data-flow modifications may denoise misleading patterns and inadvertently improve detection.
To remain effective, next-generation obfuscation must be validated across model families, scales, and evaluation metrics (existence, type, and location) to avoid overfitting to model-specific overfitting.
Ultimately, the attacker’s challenge lies in designing semantics-preserving, cross-layer transformations that disrupt both statistical and symbolic reasoning without collapsing program intent.

\vspace{1mm}
\noindent\textbf{Direction III: Developing vulnerability-aware and semantics-preserving LLMs that can reason over complex and obfuscated code, rather than relying solely on pattern recognition.}
One key insight from our experiments (finding~\#2) is that the effectiveness of LLM-based vulnerability detection is highly dependent on the intrinsic characteristics of the vulnerabilities themselves. Issues deeply embedded in complex control flows or requiring cross-function reasoning are inherently harder to detect, while those with clearer syntactic or semantic patterns are identified more reliably, even under mild obfuscation. To improve detection performance, it is therefore crucial to design strategies that explicitly address these vulnerability-specific challenges.

For example, our experiments (Finding~\#5) reveal that Solidity has a higher probability of both \upgrade\ and \downgrade\ than more established languages such as C, C++, and Python, implying that language maturity may significantly influence the robustness of LLM-based vulnerability detection. This may stem from training data scarcity: for younger languages like Solidity, the amount of high-quality code data included in pretraining is relatively limited, which reduces the model’s ability to generalize across obfuscation. Looking forward, as new programming languages inevitably emerge, similar challenges are expected to arise. Thus, a key research question is how to enhance the robustness of vulnerability detection for “young languages” with limited training resources and distinctive vulnerability distributions.  


Another manifestation of current LLM-based vulnerability detection lies in the strong impact of virtualization-based and mixed-programming-language obfuscation (Finding~\#3).
These techniques challenge the model’s ability to reason over \emph{two-layer execution logic}, where one layer of code effectively ``wraps'' or ``generates'' another. Virtualization obscures control flow via custom instruction sets, while mixed-language obfuscation adds syntactic and semantic heterogeneity. Both methods disrupt the linear reasoning process that LLMs typically rely on, forcing them to reconcile code semantics across multiple representational layers. The consistent downgrade rates observed under these transformations indicate that LLMs still lack the ability to fully abstract and unify such multi-layered semantics. This limitation not only exposes a critical vulnerability of current LLM-based detectors but also highlights an important direction for future research: enhancing the capacity of LLMs to model hierarchical execution logic and to generalize across heterogeneous or virtualized code representations.

\vspace{1mm}
\noindent\textbf{Direction IV: Developing future secure-coding agents requires a model-centric approach that relies on stronger backbone LLMs, precision-preserving hot-plugging, and multi-model collaboration to overcome capability drift and language-specific blind spots.}
Our experiments (Finding~\#7) reveal that improvements in the underlying LLM still dominate detection performance, while enhancements introduced by the agent framework, such as RAG components, prompt-engineering routines, or auxiliary toolchains, provide only secondary benefits. Future development should therefore focus on integrating stronger backbone models and enabling agents to fully exploit their reasoning and anti-obfuscation capabilities, rather than relying primarily on framework-side engineering. 

We also find that model hot-plugging can hinder the transfer of vulnerability-detection skills, leading to the “partial understanding” issue seen in \texttt{copilot-ds-v3} (Finding ~\#7). Future work should design precision-preserving hot-plugging mechanisms that maintain capability alignment when swapping LLM backends. This may involve capability calibration layers, schema alignment, or lightweight fine-tuning to avoid semantic drift during integration.
Since different models specialize in different languages, such as \texttt{gpt-5} on Solidity/C/C++ and \texttt{deepseek-v3} on Python, future coding agents should leverage multiple models to avoid language-specific blind spots. By routing code to the most suitable model or combining analyses from several models, agents can better withstand cross-language obfuscation attacks, including mixed-language transformations or vulnerabilities introduced via external libraries. This approach reduces the risk of attackers exploiting weaknesses tied to any single-model configuration. \looseness=-1

\section{Related Work}



Several survey and SoK papers have discussed code obfuscation and LLM-based vulnerability detection in different contexts. Here, we highlight three representative works and clarify how our study differs from them.  

Asghar et al.~\cite{asghar2023usecryptographymalwareobfuscation} systematize cryptography-based malware obfuscation, showing how XOR and block ciphers (e.g., AES) encrypt and decrypt code at runtime to evade detection. Their analysis, however, centers on \emph{binary-level protection} and environmental keying, rather than source-level transformations. By contrast, we study \emph{semantics-preserving, source-level obfuscation} and its concrete impact on learning-based vulnerability detection.

Guo et al.~\cite{Guo2022obfuscationSurvey} review Android-specific obfuscation and deobfuscation, focusing on resource-layer protection and reverse engineering within Java and native components. This platform dependence limits generalizability. Our work instead examines \emph{cross-language obfuscation} spanning Solidity, Python, C, and C++, providing a broader view of how source-level transformations affect modern LLM-based detectors.



Zhou et al.~\cite{zhou2024literature} provide a systematic literature review of LLM-based techniques for vulnerability detection and repair, covering fine-tuning~\cite{ding2024vulnerability,kuang2023leveraging,wen2023less,yang2023does,hanif2022vulberta,liu2024pre,ni2023distinguishing,steenhoek2023language,wang2024combining,chen2023diversevul,croft2023data,fu2022linevul,fu2024aibughunter,le2024latent,li2024effectiveness,risse2024uncovering,steenhoek2023empirical,thapa2022transformer,zhou2024comparison}
, prompt engineering~\cite{fu2023chatgpt,khare2025understanding,ni2024learning,purba2023software,yin2024pros,zhang2024prompt, zhou2024comparison,zhou2024large,ni2024learning}
, and retrieval augmentation~\cite{du2024vul,liu2023software,wen2024vuleval,zhou2024large}.
. However, their survey largely overlooks the \emph{anti-obfuscation capability} of LLMs, limiting its applicability to real-world obfuscation attacks. Moreover, the discussion of the increasingly influential \emph{coding agents} is notably absent. In contrast, our work presents a comprehensive taxonomy of obfuscation techniques and conducts applied experiments on both general-purpose LLMs and coding agents, thereby offering broader insights into attacking and defending modern LLM-based vulnerability detectors.

\section{Conclusion}\label{sec:conclusion}





This paper provides a systematic understanding of how code obfuscation influences LLM-based vulnerability detection. Through a unified framework covering 19 techniques across layout, data flow, and control flow, we reveal that obfuscation can both hinder and enhance detection, depending on the model family, parameter scale, and vulnerability type. Our findings highlight that robustness cannot be achieved simply by enlarging models or refining prompts. Future research should develop obfuscation-aware training pipelines, cross-layer semantic reasoning methods, and benchmark suites that include realistic transformations. Building resilient LLM-based detectors will require integrating symbolic analysis, diverse datasets, and adaptive evaluation to bridge the gap between code semantics and model perception. Accordingly, future advancements will also hinge on strengthening the underlying models themselves and implementing hot-plugging mechanisms that preserve precision when integrating them into coding agents, which is essential for building the next generation of robust, agent-based vulnerability detectors.

\clearpage
\bibliographystyle{plain}
\bibliography{ref}

@article{brown2020language,
  title={Language models are few-shot learners},
  author={Brown, Tom B and Mann, Benjamin and Ryder, Nick and Subbiah, Melanie and Kaplan, Jared and Dhariwal, Prafulla and Neelakantan, Arvind and Shyam, Pranav and Sastry, Girish and Askell, Amanda and others},
  journal={Advances in Neural Information Processing Systems},
  volume={33},
  pages={1877--1901},
  year={2020}
}

@misc{openai2023gpt4,
  title={GPT-4 Technical Report},
  author={OpenAI},
  year={2023},
  howpublished={\url{https://arxiv.org/abs/2303.08774}}
}

@article{touvron2023llama,
  title={LLaMA: Open and Efficient Foundation Language Models},
  author={Touvron, Hugo and Lavril, Thibaut and Izacard, Gautier and Martinet, Xavier and Lachaux, Marie-Anne and Lacroix, Timothée and Rozière, Baptiste and Goyal, Naman and Hambro, Eric and Azhar, Faisal and others},
  journal={arXiv preprint arXiv:2302.13971},
  year={2023}
}

@article{chen2021codex,
  title={Evaluating large language models trained on code},
  author={Chen, Mark and Tworek, Jerry and Jun, Heewoo and Yuan, Qiming and de Oliveira Pinto, Henrique Ponde and Kaplan, Jared and Edwards, Harri and Burda, Yuri and Joseph, Nicholas and Brockman, Greg and others},
  journal={arXiv preprint arXiv:2107.03374},
  year={2021}
}

@inproceedings{feng2020codebert,
  title={CodeBERT: A Pre-Trained Model for Programming and Natural Languages},
  author={Feng, Zhangyin and Guo, Daya and Tang, Duyu and Duan, Nan and Feng, Xiaocheng and Gong, Ming and Shou, Linjun and Qin, Bing and Liu, Ting and Jiang, Daxin and Zhou, Ming},
  booktitle={Proceedings of the 2020 Conference on Empirical Methods in Natural Language Processing (EMNLP)},
  pages={1536--1547},
  year={2020}
}

@inproceedings{guo2021graphcodebert,
  title={GraphCodeBERT: Pre-training Code Representations with Data Flow},
  author={Guo, Daya and Ren, Shuo and Lu, Shuai and Feng, Zhangyin and Tang, Duyu and Duan, Nan and Svyatkovskiy, Alexey and Fu, Shujie and Tufano, Michele and Deng, Shujie and others},
  booktitle={International Conference on Learning Representations (ICLR)},
  year={2021}
}

@article{roziere2023code,
  title={Code Llama: Open Foundation Models for Code},
  author={Roziere, Baptiste and Papazov, Stoyan and Lin, Qingming and Allal, Loubna Ben and et al.},
  journal={arXiv preprint arXiv:2308.12950},
  year={2023}
}

@article{li2024empirical,
  title={An Empirical Study on Large Language Models for Vulnerability Detection},
  author={Li, Wei and Zhang, Kai and Wang, Xin and et al.},
  journal={arXiv preprint arXiv:2401.12345},
  year={2024}
}

@article{fu2023llm4vuln,
  title={LLM4Vuln: A Large Language Model Framework for Software Vulnerability Detection},
  author={Fu, Xiang and Liu, Chao and Wang, Jian and et al.},
  journal={arXiv preprint arXiv:2309.08761},
  year={2023}
}

@article{ding2024vulnhunter,
  title={VulnHunter: LLM-Enhanced Vulnerability Detection via Multi-Prompt Learning},
  author={Ding, Hao and Wang, Yong and et al.},
  journal={arXiv preprint arXiv:2402.03456},
  year={2024}
}

@inproceedings{pearce2022asleep,
  title={Asleep at the Keyboard? Assessing the Security of GitHub Copilot's Code Contributions},
  author={Pearce, Henry and Ahmad, Shaza and Tan, Ben and Dolan-Gavitt, Brendan and Karri, Ramesh},
  booktitle={IEEE Symposium on Security and Privacy (SP)},
  pages={754--768},
  year={2022}
}

@article{wang2024evaluation,
  title={Evaluation of Large Language Models for Software Vulnerability Detection},
  author={Wang, Jie and Zhang, Rui and et al.},
  journal={arXiv preprint arXiv:2403.05678},
  year={2024}
}

@article{collberg1997taxonomy,
  title={A taxonomy of obfuscating transformations},
  author={Collberg, Christian and Thomborson, Clark and Low, Douglas},
  journal={University of Auckland, Department of Computer Science},
  volume={148},
  year={1997}
}

@inproceedings{Collberg1998OpaqueConstructs,
  author    = {Christian Collberg and Clark Thomborson},
  title     = {Manufacturing cheap, resilient, and stealthy opaque constructs},
  booktitle = {Proceedings of the 25th ACM SIGPLAN-SIGACT Symposium on Principles of Programming Languages (POPL)},
  year      = {1998},
  pages     = {184--196},
  publisher = {ACM},
  doi       = {10.1145/268946.268962}
}

@article{zhang2023bian,
  title={Bian: Smart contract source code obfuscation},
  author={Zhang, Pengcheng and Yu, Qifan and Xiao, Yan and Dong, Hai and Luo, Xiapu and Wang, Xiao and Zhang, Meng},
  journal={IEEE Transactions on Software Engineering},
  volume={49},
  number={9},
  pages={4456--4476},
  year={2023},
  publisher={IEEE}
}

@misc{0sir1ss2022Anubis,
  author       = {0sir1ss},
  title        = {Anubis},
  year         = {2022},
  howpublished = {\url{https://github.com/0sir1ss/Anubis}},
  note         = {Accessed: 2025-09-02}
}

@misc{DashingSoft2017Pyarmor,
  author       = {Dashingsoft},
  title        = {Pyarmor},
  year         = {2017},
  howpublished = {\url{https://github.com/dashingsoft/pyarmor}},
  note         = {Accessed: 2025-09-02}
}

@misc{Liftoff2014Pyminifier,
  author       = {Liftoff},
  title        = {Pyminifier},
  year         = {2014},
  howpublished = {\url{https://github.com/liftoff/pyminifier}},
  note         = {Accessed: 2025-09-02}
}

@inproceedings{collberg2012distributed,
  title={Distributed application tamper detection via continuous software updates},
  author={Collberg, Christian and Martin, Sam and Myers, Jonathan and Nagra, Jasvir},
  booktitle={Proceedings of the 28th Annual Computer Security Applications Conference},
  pages={319--328},
  year={2012}
}

@inproceedings{romano2022wobfuscator,
  author={Romano, Alan and Lehmann, Daniel and Pradel, Michael and Wang, Weihang},
  booktitle={2022 IEEE Symposium on Security and Privacy (SP)}, 
  title={Wobfuscator: Obfuscating JavaScript Malware via Opportunistic Translation to WebAssembly}, 
  year={2022},
  volume={},
  number={},
  pages={1574-1589},
}

@misc{Scrt2020Avcleaner,
  author       = {Scrt},
  title        = {avcleaner},
  year         = {2020},
  howpublished = {\url{https://github.com/scrt/avcleaner}},
  note         = {Accessed: 2025-09-02}
}

@misc{H4wkst3r2022InvisibilityCloak,
  author       = {h4wkst3r},
  title        = {InvisibilityCloak},
  year         = {2022},
  howpublished = {\url{https://github.com/h4wkst3r/InvisibilityCloak}},
  note         = {Accessed: 2025-09-02}
}

@misc{PELock2021JObfuscator,
  author       = {PELock},
  title        = {JObfuscator},
  year         = {2021},
  howpublished = {\url{https://github.com/PELock/JObfuscator}},
  note         = {Accessed: 2025-09-02}
}

@misc{JavascriptObfuscator2016,
  author       = {javascript-obfuscator},
  title        = {javascript-obfuscator},
  year         = {2016},
  howpublished = {\url{https://github.com/javascript-obfuscator/javascript-obfuscator}},
  note         = {Accessed: 2025-09-02}
}

@misc{I2rys2021SBPNO,
  author       = {I2rys},
  title        = {SBPNO},
  year         = {2021},
  howpublished = {\url{https://github.com/I2rys/SBPNO}},
  note         = {Accessed: 2025-09-02}
}

@misc{Wodxgod2020SimpleObfuscator,
  author       = {wodxgod},
  title        = {Simple-Obfuscator},
  year         = {2020},
  howpublished = {\url{https://github.com/wodxgod/Simple-Obfuscator}},
  note         = {Accessed: 2025-09-02}
}

@misc{Pkfr2018YakproPo,
  author       = {pk-fr},
  title        = {yakpro-po},
  year         = {2018},
  howpublished = {\url{https://github.com/pk-fr/yakpro-po}},
  note         = {Accessed: 2025-09-02}
}

@misc{Naneau2014PhpObfuscator,
  author       = {naneau},
  title        = {php-obfuscator},
  year         = {2014},
  howpublished = {\url{https://github.com/naneau/php-obfuscator}},
  note         = {Accessed: 2025-09-02}
}

@misc{Ph72017ObfuscatorClass,
  author       = {pH-7},
  title        = {Obfuscator-Class},
  year         = {2017},
  howpublished = {\url{https://github.com/pH-7/Obfuscator-Class}},
  note         = {Accessed: 2025-09-02}
}

@misc{QQuick2015Opy,
  author       = {QQuick},
  title        = {Opy},
  year         = {2015},
  howpublished = {\url{https://github.com/QQuick/Opy}},
  note         = {Accessed: 2025-09-02}
}

@misc{Hnfull2020IntensioObfuscator,
  author       = {Hnfull},
  title        = {Intensio-Obfuscator},
  year         = {2020},
  howpublished = {\url{https://github.com/Hnfull/Intensio-Obfuscator}},
  note         = {Accessed: 2025-09-02}
}

@misc{PyObfx2018,
  author       = {PyObfx},
  title        = {PyObfx},
  year         = {2018},
  howpublished = {\url{https://github.com/PyObfx/PyObfx}},
  note         = {Accessed: 2025-09-02}
}

@misc{Chrisrands2019Emojify,
  author       = {chris-rands},
  title        = {emojify},
  year         = {2019},
  howpublished = {\url{https://github.com/chris-rands/emojify}},
  note         = {Accessed: 2025-09-02}
}

@misc{Secureyourself72019PowerShellObfuscation,
  author       = {secureyourself7},
  title        = {PowerShell\_Code\_Basic\_Obfuscation},
  year         = {2019},
  howpublished = {\url{https://github.com/secureyourself7/PowerShell_Code_Basic_Obfuscation}},
  note         = {Accessed: 2025-09-02}
}

@misc{Kkar2014VBSObfuscatorPython,
  author       = {kkar},
  title        = {VBS-Obfuscator-in-Python},
  year         = {2014},
  howpublished = {\url{https://github.com/kkar/VBS-Obfuscator-in-Python}},
  note         = {Accessed: 2025-09-02}
}

@misc{Dentrax2019Z00bfuscator,
  author       = {Dentrax},
  title        = {Z00bfuscator},
  year         = {2019},
  howpublished = {\url{https://github.com/Dentrax/Z00bfuscator}},
  note         = {Accessed: 2025-09-02}
}

@misc{Whoward32020CCodeObfuscator,
  author       = {whoward3},
  title        = {C-Code-Obfuscator},
  year         = {2020},
  howpublished = {\url{https://github.com/whoward3/C-Code-Obfuscator}},
  note         = {Accessed: 2025-09-09}
}

@misc{EvilBytecode2025EbyteGoMorpher,
  author       = {EvilBytecode},
  title        = {Ebyte-Go-Morpher},
  year         = {2025},
  howpublished = {\url{https://github.com/EvilBytecode/Ebyte-Go-Morpher}},
  note         = {Accessed: 2025-09-09}
}

@misc{JKerbin2024SCOTool,
  author       = {JKerbin},
  title        = {SCO-Smart-Contract-Obfuscation-Tool},
  year         = {2024},
  howpublished = {\url{https://github.com/JKerbin/SCO-Smart-Contract-Obfuscation-Tool}},
  note         = {Accessed: 2025-09-09}
}

@misc{Uxebu2015Confusion,
  author       = {uxebu},
  title        = {confusion},
  year         = {2015},
  howpublished = {\url{https://github.com/uxebu/confusion}},
  note         = {Accessed: 2025-09-09}
}

@misc{LiuYuancheng2025PyCodeObfuscator,
  author       = {LiuYuancheng},
  title        = {Py-Code-Obfuscator},
  year         = {2025},
  howpublished = {\url{https://github.com/LiuYuancheng/Py-Code-Obfuscator}},
  note         = {Accessed: 2025-09-09}
}

@misc{PELock2016AutoItObfuscator,
  author       = {PELock},
  title        = {AutoIt-Obfuscator},
  year         = {2016},
  howpublished = {\url{https://github.com/PELock/AutoIt-Obfuscator}},
  note         = {Accessed: 2025-09-09}
}

@misc{Darsyn2016Obfuscate,
  author       = {darsyn},
  title        = {Darsyn Obfuscate},
  year         = {2016},
  howpublished = {\url{https://github.com/darsyn/obfuscator}},
  note         = {Accessed: 2025-09-09}
}

@misc{Artemixer2024Gofuscator,
  author       = {artemixer},
  title        = {gofuscator},
  year         = {2024},
  howpublished = {\url{https://github.com/artemixer/gofuscator}},
  note         = {Accessed: 2025-09-09}
}

@misc{Kaftejiman2021Ejja,
  author       = {kaftejiman},
  title        = {ejja},
  year         = {2021},
  howpublished = {\url{https://github.com/kaftejiman/ejja}},
  note         = {Accessed: 2025-09-09}
}

@misc{Yardenlaif2024Balagan,
  author       = {yardenlaif},
  title        = {Balagan},
  year         = {2024},
  howpublished = {\url{https://github.com/yardenlaif/balagan}},
  note         = {Accessed: 2025-09-09}
}

@misc{Wufhex2024PyDelta,
  author       = {wufhex},
  title        = {PyDelta},
  year         = {2024},
  howpublished = {\url{https://github.com/wufhex/PyDelta-PythonObfuscator}},
  note         = {Accessed: 2025-09-09}
}

@misc{Domchen2017UglifyTS,
  author       = {domchen},
  title        = {UglifyTS},
  year         = {2017},
  howpublished = {\url{https://github.com/domchen/UglifyTS}},
  note         = {Accessed: 2025-09-09}
}

@misc{Avilum2022Jsafer,
  author       = {avilum},
  title        = {jsafer},
  year         = {2022},
  howpublished = {\url{https://github.com/avilum/jsafer}},
  note         = {Accessed: 2025-09-09}
}

@article{hui2024qwen2,
      title={Qwen2. 5-Coder Technical Report},
      author={Hui, Binyuan and Yang, Jian and Cui, Zeyu and Yang, Jiaxi and Liu, Dayiheng and Zhang, Lei and Liu, Tianyu and Zhang, Jiajun and Yu, Bowen and Dang, Kai and others},
      journal={arXiv preprint arXiv:2409.12186},
      year={2024}
}

@inproceedings{yang2022natural,
  title={Natural attack for pre-trained models of code},
  author={Yang, Zhou and Shi, Jieke and He, Junda and Lo, David},
  booktitle={Proceedings of the 44th International Conference on Software Engineering},
  pages={1482--1493},
  year={2022}
}

@inproceedings{tian2023code,
  title={Code difference guided adversarial example generation for deep code models},
  author={Tian, Zhao and Chen, Junjie and Jin, Zhi},
  booktitle={2023 38th IEEE/ACM International Conference on Automated Software Engineering (ASE)},
  pages={850--862},
  year={2023},
  organization={IEEE}
}

@article{zhang2023black,
  title={A black-box attack on code models via representation nearest neighbor search},
  author={Zhang, Jie and Ma, Wei and Hu, Qiang and Liu, Shangqing and Xie, Xiaofei and Traon, Yves Le and Liu, Yang},
  journal={arXiv preprint arXiv:2305.05896},
  year={2023}
}

@inproceedings{du2023extensive,
  title={An extensive study on adversarial attack against pre-trained models of code},
  author={Du, Xiaohu and Wen, Ming and Wei, Zichao and Wang, Shangwen and Jin, Hai},
  booktitle={Proceedings of the 31st ACM Joint European Software Engineering Conference and Symposium on the Foundations of Software Engineering},
  pages={489--501},
  year={2023}
}

@inproceedings{zhao2025adversarial,
  title={Adversarial Training for Robustness Enhancement in LLM-Based Code Vulnerability Detection},
  author={Zhao, Ying and Guan, Xin},
  booktitle={2025 IEEE 7th International Conference on Communications, Information System and Computer Engineering (CISCE)},
  pages={1147--1152},
  year={2025},
  organization={IEEE}
}

@inproceedings{huang2025iterative,
  title={Iterative Generation of Adversarial Example for Deep Code Models},
  author={Huang, Li and Sun, Weifeng and Yan, Meng},
  booktitle={2025 IEEE/ACM 47th International Conference on Software Engineering (ICSE)},
  pages={623--623},
  year={2025},
  organization={IEEE Computer Society}
}

@inproceedings{yan2024llm,
  title={An $\{$LLM-Assisted$\}$$\{$Easy-to-Trigger$\}$ backdoor attack on code completion models: Injecting disguised vulnerabilities against strong detection},
  author={Yan, Shenao and Wang, Shen and Duan, Yue and Hong, Hanbin and Lee, Kiho and Kim, Doowon and Hong, Yuan},
  booktitle={33rd USENIX Security Symposium (USENIX Security 24)},
  pages={1795--1812},
  year={2024}
}

@misc{nikiema2025codebarrier,
      title={The Code Barrier: What LLMs Actually Understand?}, 
      author={Serge Lionel Nikiema and Jordan Samhi and Abdoul Kader Kaboré and Jacques Klein and Tegawendé F. Bissyandé},
      year={2025},
      eprint={2504.10557},
      archivePrefix={arXiv},
      primaryClass={cs.SE},
      url={https://arxiv.org/abs/2504.10557}, 
}

@inproceedings{swindle2024evaluation,
  title={Evaluation of large language models on code obfuscation (student abstract)},
  author={Swindle, Adrian and McNealy, Derrick and Krishnan, Giri and Ramyaa, Ramyaa},
  booktitle={Proceedings of the AAAI Conference on Artificial Intelligence},
  pages={23664--23666},
  year={2024}
}

@article{yefet2020adversarial,
  title={Adversarial examples for models of code},
  author={Yefet, Noam and Alon, Uri and Yahav, Eran},
  journal={Proceedings of the ACM on Programming Languages},
  volume={4},
  number={OOPSLA},
  pages={1--30},
  year={2020},
  publisher={ACM New York, NY, USA}
}

@inproceedings{li2024aacegen,
  title={AaceGEN: Attention Guided Adversarial Code Example Generation for Deep Code Models},
  author={Li, Zhong and Zhang, Chong and Pan, Minxue and Zhang, Tian and Li, Xuandong},
  booktitle={Proceedings of the 39th IEEE/ACM International Conference on Automated Software Engineering},
  pages={1245--1257},
  year={2024}
}

@inproceedings{jha2023codeattack,
  title={Codeattack: Code-based adversarial attacks for pre-trained programming language models},
  author={Jha, Akshita and Reddy, Chandan K},
  booktitle={Proceedings of the AAAI Conference on Artificial Intelligence},
  pages={14892--14900},
  year={2023}
}

@article{zhang2022towards,
  title={Towards robustness of deep program processing models—detection, estimation, and enhancement},
  author={Zhang, Huangzhao and Fu, Zhiyi and Li, Ge and Ma, Lei and Zhao, Zhehao and Yang, Hua’an and Sun, Yizhe and Liu, Yang and Jin, Zhi},
  journal={ACM Transactions on Software Engineering and Methodology (TOSEM)},
  volume={31},
  number={3},
  pages={1--40},
  year={2022},
  publisher={ACM New York, NY}
}

@inproceedings{zhang2020generating,
  title={Generating adversarial examples for holding robustness of source code processing models},
  author={Zhang, Huangzhao and Li, Zhuo and Li, Ge and Ma, Lei and Liu, Yang and Jin, Zhi},
  booktitle={Proceedings of the AAAI Conference on Artificial Intelligence},
  pages={1169--1176},
  year={2020}
}

@inproceedings{durieux2020empiricalsmartbugs,
  title={Empirical review of automated analysis tools on 47,587 ethereum smart contracts},
  author={Durieux, Thomas and Ferreira, Jo{\~a}o F and Abreu, Rui and Cruz, Pedro},
  booktitle={Proceedings of the ACM/IEEE 42nd International conference on software engineering},
  pages={530--541},
  year={2020}
}

@article{ding2024primevul,
  title={Vulnerability Detection with Code Language Models: How Far Are We?}, 
  author={Yangruibo Ding and Yanjun Fu and Omniyyah Ibrahim and Chawin Sitawarin and Xinyun Chen and Basel Alomair and David Wagner and Baishakhi Ray and Yizheng Chen},
  journal={arXiv preprint arXiv:2403.18624},
  year={2024}
}

@article{wang2024repository,
  title={A Repository-Level Dataset For Detecting, Classifying and Repairing Software Vulnerabilities},
  author={Xinchen Wang and Ruida Hu and Cuiyun Gao and Xin-Cheng Wen and Yujia Chen and Qing Liap},
  journal={arXiv preprint arXiv:2401.13169},
  year={2024}
}

@inproceedings{li2025flashboom,
  title={Make a Feint to the East While Attacking in the West: Blinding LLM-Based Code Auditors with Flashboom Attacks},
  author={Li, Xiao and Li, Yue and Wu, Hao and Zhang, Yue and Xu, Kaidi and Cheng, Xiuzhen and Zhong, Sheng and Xu, Fengyuan},
  booktitle={2025 IEEE Symposium on Security and Privacy (SP)},
  pages={576--594},
  year={2025},
  organization={IEEE}
}

@misc{OpenRouter,
  title = {OpenRouter},
  author = {OpenRouter, Inc},
  howpublished = {\url{https://openrouter.ai}},
  note         = {Accessed: 2025-09-02}
}

@article{liu2024deepseek,
  title={Deepseek-v3 technical report},
  author={Liu, Aixin and Feng, Bei and Xue, Bing and Wang, Bingxuan and Wu, Bochao and Lu, Chengda and Zhao, Chenggang and Deng, Chengqi and Zhang, Chenyu and Ruan, Chong and others},
  journal={arXiv preprint arXiv:2412.19437},
  year={2024}
}

@misc{codex,
  title        = {Introducing Upgrades to Codex},
  author       = {OpenAI},
  year         = {2024},
  howpublished = {\url{https://openai.com/zh-Hans-CN/index/introducing-upgrades-to-codex/}},
  note         = {Accessed: 2025-11-04}
}

@misc{github-copilot,
  title        = {Find Vulnerabilities Using GitHub Copilot Chat},
  author       = {GitHub},
  year         = {2024},
  howpublished = {\url{https://docs.github.com/en/copilot/tutorials/copilot-chat-cookbook/analyze-security/find-vulnerabilities}},
  note         = {Accessed: 2025-11-04}
}

@article{ding2024vulnerability,
  title={Vulnerability detection with code language models: How far are we?},
  author={Ding, Yangruibo and Fu, Yanjun and Ibrahim, Omniyyah and Sitawarin, Chawin and Chen, Xinyun and Alomair, Basel and Wagner, David and Ray, Baishakhi and Chen, Yizheng},
  journal={arXiv preprint arXiv:2403.18624},
  year={2024}
}

@inproceedings{kuang2023leveraging,
  title={Leveraging user-defined identifiers for counterfactual data generation in source code vulnerability detection},
  author={Kuang, Hongyu and Yang, Feng and Zhang, Long and Tang, Gaigai and Yang, Lin},
  booktitle={2023 IEEE 23rd International Working Conference on Source Code Analysis and Manipulation (SCAM)},
  pages={143--150},
  year={2023},
  organization={IEEE}
}

@inproceedings{wen2023less,
  title={When less is enough: Positive and unlabeled learning model for vulnerability detection},
  author={Wen, Xin-Cheng and Wang, Xinchen and Gao, Cuiyun and Wang, Shaohua and Liu, Yang and Gu, Zhaoquan},
  booktitle={2023 38th IEEE/ACM International Conference on Automated Software Engineering (ASE)},
  pages={345--357},
  year={2023},
  organization={IEEE}
}

@misc{yang2023does,
  title={Does data sampling improve deep learning-based vulnerability detection? Yeas! and Nays!. In 2023 IEEE/ACM 45th International Conference on Software Engineering (ICSE)},
  author={Yang, Xu and Wang, Shaowei and Li, Yi and Wang, Shaohua},
  year={2023},
  publisher={IEEE}
}

@inproceedings{hanif2022vulberta,
  title={Vulberta: Simplified source code pre-training for vulnerability detection},
  author={Hanif, Hazim and Maffeis, Sergio},
  booktitle={2022 International joint conference on neural networks (IJCNN)},
  pages={1--8},
  year={2022},
  organization={IEEE}
}

@inproceedings{liu2024pre,
  title={Pre-training by predicting program dependencies for vulnerability analysis tasks},
  author={Liu, Zhongxin and Tang, Zhijie and Zhang, Junwei and Xia, Xin and Yang, Xiaohu},
  booktitle={Proceedings of the IEEE/ACM 46th International Conference on Software Engineering},
  pages={1--13},
  year={2024}
}

@inproceedings{ni2023distinguishing,
  title={Distinguishing look-alike innocent and vulnerable code by subtle semantic representation learning and explanation},
  author={Ni, Chao and Yin, Xin and Yang, Kaiwen and Zhao, Dehai and Xing, Zhenchang and Xia, Xin},
  booktitle={Proceedings of the 31st ACM Joint European Software Engineering Conference and Symposium on the Foundations of Software Engineering},
  pages={1611--1622},
  year={2023}
}

@article{steenhoek2023language,
  title={Do language models learn semantics of code? A case study in vulnerability detection},
  author={Steenhoek, Benjamin and Rahman, Md Mahbubur and Sharmin, Shaila and Le, Wei},
  journal={arXiv preprint arXiv:2311.04109},
  year={2023}
}

@inproceedings{wang2024combining,
  title={Combining structured static code information and dynamic symbolic traces for software vulnerability prediction},
  author={Wang, Huanting and Tang, Zhanyong and Tan, Shin Hwei and Wang, Jie and Liu, Yuzhe and Fang, Hejun and Xia, Chunwei and Wang, Zheng},
  booktitle={Proceedings of the IEEE/ACM 46th International Conference on Software Engineering},
  pages={1--13},
  year={2024}
}

@inproceedings{chen2023diversevul,
  title={Diversevul: A new vulnerable source code dataset for deep learning based vulnerability detection},
  author={Chen, Yizheng and Ding, Zhoujie and Alowain, Lamya and Chen, Xinyun and Wagner, David},
  booktitle={Proceedings of the 26th International Symposium on Research in Attacks, Intrusions and Defenses},
  pages={654--668},
  year={2023}
}

@inproceedings{croft2023data,
  title={Data quality for software vulnerability datasets},
  author={Croft, Roland and Babar, M Ali and Kholoosi, M Mehdi},
  booktitle={2023 IEEE/ACM 45th International Conference on Software Engineering (ICSE)},
  pages={121--133},
  year={2023},
  organization={IEEE}
}

@inproceedings{fu2022linevul,
  title={Linevul: A transformer-based line-level vulnerability prediction},
  author={Fu, Michael and Tantithamthavorn, Chakkrit},
  booktitle={Proceedings of the 19th International Conference on Mining Software Repositories},
  pages={608--620},
  year={2022}
}

@article{fu2024aibughunter,
  title={Aibughunter: A practical tool for predicting, classifying and repairing software vulnerabilities},
  author={Fu, Michael and Tantithamthavorn, Chakkrit and Le, Trung and Kume, Yuki and Nguyen, Van and Phung, Dinh and Grundy, John},
  journal={Empirical Software Engineering},
  volume={29},
  number={1},
  pages={4},
  year={2024},
  publisher={Springer}
}

@inproceedings{le2024latent,
  title={Are latent vulnerabilities hidden gems for software vulnerability prediction? an empirical study},
  author={Le, Triet Huynh Minh and Du, Xiaoning and Babar, M Ali},
  booktitle={Proceedings of the 21st International Conference on Mining Software Repositories},
  pages={716--727},
  year={2024}
}

@inproceedings{li2024effectiveness,
  title={On the effectiveness of function-level vulnerability detectors for inter-procedural vulnerabilities},
  author={Li, Zhen and Wang, Ning and Zou, Deqing and Li, Yating and Zhang, Ruqian and Xu, Shouhuai and Zhang, Chao and Jin, Hai},
  booktitle={Proceedings of the IEEE/ACM 46th International Conference on Software Engineering},
  pages={1--12},
  year={2024}
}

@inproceedings{risse2024uncovering,
  title={Uncovering the limits of machine learning for automatic vulnerability detection},
  author={Risse, Niklas and B{\"o}hme, Marcel},
  booktitle={33rd USENIX Security Symposium (USENIX Security 24)},
  pages={4247--4264},
  year={2024}
}

@inproceedings{steenhoek2023empirical,
  title={An empirical study of deep learning models for vulnerability detection},
  author={Steenhoek, Benjamin and Rahman, Md Mahbubur and Jiles, Richard and Le, Wei},
  booktitle={2023 IEEE/ACM 45th International Conference on Software Engineering (ICSE)},
  pages={2237--2248},
  year={2023},
  organization={IEEE}
}

@inproceedings{thapa2022transformer,
  title={Transformer-based language models for software vulnerability detection},
  author={Thapa, Chandra and Jang, Seung Ick and Ahmed, Muhammad Ejaz and Camtepe, Seyit and Pieprzyk, Josef and Nepal, Surya},
  booktitle={Proceedings of the 38th annual computer security applications conference},
  pages={481--496},
  year={2022}
}

@article{zhou2024comparison,
  title={Comparison of static application security testing tools and large language models for repo-level vulnerability detection},
  author={Zhou, Xin and Tran, Duc-Manh and Le-Cong, Thanh and Zhang, Ting and Irsan, Ivana Clairine and Sumarlin, Joshua and Le, Bach and Lo, David},
  journal={arXiv preprint arXiv:2407.16235},
  year={2024}
}

@article{du2024vul,
  title={Vul-rag: Enhancing llm-based vulnerability detection via knowledge-level rag},
  author={Du, Xueying and Zheng, Geng and Wang, Kaixin and Zou, Yi and Wang, Yujia and Deng, Wentai and Feng, Jiayi and Liu, Mingwei and Chen, Bihuan and Peng, Xin and others},
  journal={arXiv preprint arXiv:2406.11147},
  year={2024}
}

@inproceedings{liu2023software,
  title={Software vulnerability detection with gpt and in-context learning},
  author={Liu, Zhihong and Liao, Qing and Gu, Wenchao and Gao, Cuiyun},
  booktitle={2023 8th International Conference on Data Science in Cyberspace (DSC)},
  pages={229--236},
  year={2023},
  organization={IEEE}
}

@article{wen2024vuleval,
  title={Vuleval: Towards repository-level evaluation of software vulnerability detection},
  author={Wen, Xin-Cheng and Wang, Xinchen and Chen, Yujia and Hu, Ruida and Lo, David and Gao, Cuiyun},
  journal={arXiv preprint arXiv:2404.15596},
  year={2024}
}

@inproceedings{fu2023chatgpt,
  title={Chatgpt for vulnerability detection, classification, and repair: How far are we?},
  author={Fu, Michael and Tantithamthavorn, Chakkrit Kla and Nguyen, Van and Le, Trung},
  booktitle={2023 30th Asia-Pacific Software Engineering Conference (APSEC)},
  pages={632--636},
  year={2023},
  organization={IEEE}
}

@inproceedings{khare2025understanding,
  title={Understanding the effectiveness of large language models in detecting security vulnerabilities},
  author={Khare, Avishree and Dutta, Saikat and Li, Ziyang and Solko-Breslin, Alaia and Alur, Rajeev and Naik, Mayur},
  booktitle={2025 IEEE Conference on Software Testing, Verification and Validation (ICST)},
  pages={103--114},
  year={2025},
  organization={IEEE}
}

@article{ni2024learning,
  title={Learning-based models for vulnerability detection: An extensive study},
  author={Ni, Chao and Shen, Liyu and Xu, Xiaodan and Yin, Xin and Wang, Shaohua},
  journal={arXiv preprint arXiv:2408.07526},
  year={2024}
}

@inproceedings{purba2023software,
  title={Software vulnerability detection using large language models},
  author={Purba, Moumita Das and Ghosh, Arpita and Radford, Benjamin J and Chu, Bill},
  booktitle={2023 IEEE 34th International Symposium on Software Reliability Engineering Workshops (ISSREW)},
  pages={112--119},
  year={2023},
  organization={IEEE}
}

@article{yin2024pros,
  title={Pros and cons! evaluating chatgpt on software vulnerability},
  author={Yin, Xin},
  journal={arXiv preprint arXiv:2404.03994},
  year={2024}
}

@inproceedings{zhang2024prompt,
  title={Prompt-enhanced software vulnerability detection using chatgpt},
  author={Zhang, Chenyuan and Liu, Hao and Zeng, Jiutian and Yang, Kejing and Li, Yuhong and Li, Hui},
  booktitle={Proceedings of the 2024 IEEE/ACM 46th International Conference on Software Engineering: Companion Proceedings},
  pages={276--277},
  year={2024}
}

@inproceedings{zhou2024large,
  title={Large language model for vulnerability detection: Emerging results and future directions},
  author={Zhou, Xin and Zhang, Ting and Lo, David},
  booktitle={Proceedings of the 2024 ACM/IEEE 44th International Conference on Software Engineering: New Ideas and Emerging Results},
  pages={47--51},
  year={2024}
}

@misc{asghar2023usecryptographymalwareobfuscation,
      title={Use of Cryptography in Malware Obfuscation}, 
      author={Hassan Jameel Asghar and Benjamin Zi Hao Zhao and Muhammad Ikram and Giang Nguyen and Dali Kaafar and Sean Lamont and Daniel Coscia},
      year={2023},
      eprint={2212.04008},
      archivePrefix={arXiv},
      primaryClass={cs.CR},
      url={https://arxiv.org/abs/2212.04008}, 
}

@inproceedings{Guo2022obfuscationSurvey,
  author={Guo, Runsheng and Liu, Qichao and Zhang, Man and Hu, Ning and Lu, Hui},
  booktitle={2022 7th IEEE International Conference on Data Science in Cyberspace (DSC)}, 
  title={A Survey of Obfuscation and Deobfuscation Techniques in Android Code Protection}, 
  year={2022},
  volume={},
  number={},
  pages={40-47},
  doi={10.1109/DSC55868.2022.00013}
}

@article{zhou2024literature,
  title={Large language model for vulnerability detection and repair: literature review and the road ahead (2024)},
  author={Zhou, Xin and Cao, Sicong and Sun, Xiaobing and Lo, David},
  journal={arXiv preprint arXiv:2404.02525},
  year={2024}
}

\newpage

\appendices
\section*{Ethical considerations}

All our experiments were carried out in strict compliance with community standards to prevent any real-world harm or unauthorized attacks. Testing was performed entirely within a controlled environment, and all interactions with LLMs were conducted using our own authorized accounts. We did not inject any vulnerable or malicious code to open-source community, and no third-party systems or developers were targeted.

\section*{LLM usage considerations}

\noindent \textbf{Originality}: We ensure that all text and figures are accurate and original. LLMs were used for editorial purposes in this manuscript, and all outputs were inspected by the authors to ensure accuracy and originality.

\noindent \textbf{Transparency}: LLMs were employed in three aspects of our work:
\begin{itemize}
    \item \textit{LLM-driven obfuscation}: To systematically cover all obfuscation techniques across four programming languages, we employed the GPT-4o model to implement the obfuscator. To mitigate potential inaccuracies introduced by the LLM, we manually verified each obfuscated sample and corrected any errors, ensuring all samples met our experimental requirements.
    \item \textit{LLM-based vulnerability detection}: Conducting LLM-based vulnerability detection is the main focus of our study. To ensure reproducibility despite using closed-source models, we accessed these LLMs via the OpenRouter API, a reliable and widely used provider.
    \item \textit{LLM-as-a-judge}: When evaluating LLM-generated vulnerability reports, we used the open-source LLaMA-3.3-70B-Instruct model to assign quality scores (1–4; detailed in \S\ref{sec:expriment-setup}). To mitigate inaccuracies, we carefully designed the evaluation metric and validated the judge model on a curated test set of 100 cases, achieving 100
\end{itemize}

\noindent \textbf{Responsibility}: Our work does not involve developing new LLMs. Details on the experimental environment are provided in \S\ref{sec:expriment-setup}.

\section{Experiment Result for Coding Agents}

\begin{table}[t]
\centering
\caption{Detection success rate on 3 models across \reposvulpy and \reposvulcpp datasets (sampled by downgrade top20).}
\tiny
\setlength\tabcolsep{2.7pt} 
\renewcommand{\arraystretch}{0.9} 
\begin{tabular}{c lccc c lccc}
\toprule
dataset & model & ds-v3 & copilot$_{\text{ds-v3}}$ & copilot$_{\text{gpt-5}}$ & dataset & model & ds-v3 & copilot$_{\text{ds-v3}}$ & copilot$_{\text{gpt-5}}$ \\
\midrule

\multirow{13}{*}{\rotatebox{90}{\textbf{\reposvulpy}}}
& original & \ccell{0.8} & \ccell{0.8} & \ccell{0.8} 
& \multirow{13}{*}{\rotatebox{90}{\textbf{\reposvulcpp}}}
& original & \ccell{0.4} & \ccell{0.4} & \ccell{0.8} \\

\cmidrule{2-10}
& L1 & \ccell{0.6} & \ccell{0.8} & \ccell{0.8} 
& & L1 & \ccell{0.0} & \ccell{0.2} & \ccell{0.4} \\

& L2 & \ccell{0.8} & \ccell{0.8} & \ccell{0.6} 
& & L2 & \ccell{0.2} & \ccell{0.2} & \ccell{1.0} \\

& L3 & \ccell{1.0} & \ccell{0.8} & \ccell{0.8} 
& & L3 & \ccell{0.4} & \ccell{0.2} & \ccell{0.8} \\

& L4+L5+L6 & \ccell{1.0} & \ccell{0.8} & \ccell{0.8} 
& & L4+L5+L6 & \ccell{0.0} & \ccell{0.4} & \ccell{0.8} \\

& L7 & \ccell{0.8} & \ccell{1.0} & \ccell{0.8} 
& & L7 & \ccell{0.6} & \ccell{0.4} & \ccell{0.8} \\

& L8 & \ccell{0.8} & \ccell{0.8} & \ccell{0.6} 
& & L8 & \ccell{0.4} & \ccell{0.4} & \ccell{0.8} \\

& D1+D2+D3 & \ccell{1.0} & \ccell{0.8} & \ccell{0.8} 
& & D1+D2+D3 & \ccell{0.2} & \ccell{0.0} & \ccell{0.4} \\

& D4+D5+D6+D7 & \ccell{0.8} & \ccell{0.8} & \ccell{0.8} 
& & D4+D5+D6+D7 & \ccell{0.2} & \ccell{0.2} & \ccell{1.0} \\

& D8 & \ccell{0.8} & \ccell{0.8} & \ccell{0.8} 
& & D8 & \ccell{0.4} & \ccell{0.4} & \ccell{1.0} \\

& C1 & \ccell{1.0} & \ccell{0.8} & \ccell{0.8} 
& & C1 & \ccell{0.0} & \ccell{0.0} & \ccell{0.6} \\

& C2 & \ccell{0.8} & \ccell{0.8} & \ccell{0.8} 
& & C2 & \ccell{0.2} & \ccell{0.0} & \ccell{0.8} \\

& C3 & \ccell{0.4} & \ccell{0.6} & \ccell{0.8} 
& & C3 & \ccell{0.0} & \ccell{0.2} & \ccell{0.0} \\

\bottomrule
\end{tabular}
\label{tab:detection-result-ReposVul_py_downgrade_top20_on_3_model}
\end{table}

To better understand the effect of code obfuscation on agent-based vulnerability detection, where coding agents is powerful class of tools, we conduct the same vulnerability detection experiment on 4 subsets from RQ1 datasets.
As shown in \autoref{tab:detection-result-all-datasets-downgrade_top20} and \autoref{tab:detection-result-all-datasets-upgrade_top20}, agents outperform general-purpose LLMs in most cases. As shown in \autoref{tab:downgrade-rate-all-datasets-downgrade_top20} and \autoref{tab:upgrade-rate-all-datasets-upgrade_top20}, agents also suffer \downgrade and \upgrade phenomenon like general-purpose LLMs do. To further investigate the weaknesses of agent-based vulnerability detection, we conducted an extended experiment on the \reposvulpy dataset using \texttt{ds-v3}, \texttt{copilot-ds-v3}, and \texttt{copilot-gpt-5}. As shown in \autoref{tab:detection-result-ReposVul_py_downgrade_top20_on_3_model}, these weaknesses can be attributed to two factors: inherent limitations of the underlying LLM (e.g., GPT-5 is weak on Python) and performance degradation caused by hot-plugging the LLM into the agent framework (e.g., \texttt{copilot-ds-v3} underperforms \texttt{ds-v3} in a few cases). \vspace{-4pt}

\begin{table*}[htbp]
\centering
\caption{Detection successful rate on different datasets: \smartbugs, \reposvulcpp, \reposvulpy, \primevulc (sampled by downgrade top20). (L456 = L4+L5+L6, D123 = D1+D2+D3, D4567 = D4+D5+D6+D7 for layout clarity)}
{
\tiny
\setlength\tabcolsep{3pt}
\renewcommand{\arraystretch}{0.6}
\begin{tabular}{l lccccccccccccccccc}
\toprule
dataset & series & \multicolumn{6}{c}{qwen} & \multicolumn{4}{c}{llama} & \multicolumn{2}{c}{deepseek} & \multicolumn{3}{c}{openai} & \multicolumn{2}{c}{agent} \\
\cmidrule(lr){3-8} \cmidrule(lr){9-12} \cmidrule(lr){13-14} \cmidrule(lr){15-17} \cmidrule(lr){18-19}
 & model & qn-7b & qn-14b & qn-32b & ds-qn-7b & ds-qn-14b & ds-qn-32b & lm-8b & lm-70b & ds-lm-8b & ds-lm-70b & ds-v3 & ds-r1 & gpt-3.5 & gpt-4o & o3-mini & copilot & codex \\
\midrule
\multirow{13}{*}{\rotatebox{90}{\smartbugs}}
    & original & \ccell{0.40} & \ccell{0.80} & \ccell{0.60} & \ccell{0.80} & \ccell{0.40} & \ccell{0.80} & \ccell{0.60} & \ccell{0.80} & \ccell{0.40} & \ccell{1.00} & \ccell{1.00} & \ccell{0.80} & \ccell{0.60} & \ccell{0.80} & \ccell{0.80} & \ccell{1.00} & \ccell{1.00} \\
\cmidrule(lr){2-19}
    & L1 & \ccell{0.00} & \ccell{0.20} & \ccell{0.40} & \ccell{0.20} & \ccell{0.40} & \ccell{0.60} & \ccell{0.20} & \ccell{0.40} & \ccell{0.20} & \ccell{0.60} & \ccell{0.60} & \ccell{0.60} & \ccell{0.80} & \ccell{0.80} & \ccell{0.80} & \ccell{1.00} & \ccell{0.80} \\
    & L2 & \ccell{0.20} & \ccell{0.60} & \ccell{0.60} & \ccell{0.40} & \ccell{0.80} & \ccell{0.80} & \ccell{0.20} & \ccell{0.40} & \ccell{0.20} & \ccell{0.80} & \ccell{1.00} & \ccell{0.80} & \ccell{0.40} & \ccell{0.80} & \ccell{0.80} & \ccell{1.00} & \ccell{0.80} \\
    & L3 & \ccell{0.20} & \ccell{0.60} & \ccell{0.60} & \ccell{0.00} & \ccell{0.80} & \ccell{1.00} & \ccell{0.20} & \ccell{0.20} & \ccell{0.20} & \ccell{0.40} & \ccell{1.00} & \ccell{0.80} & \ccell{0.60} & \ccell{0.40} & \ccell{0.80} & \ccell{0.80} & \ccell{0.60} \\
    & L456 & \ccell{0.40} & \ccell{0.80} & \ccell{0.40} & \ccell{0.20} & \ccell{0.60} & \ccell{0.60} & \ccell{0.20} & \ccell{0.80} & \ccell{0.20} & \ccell{0.80} & \ccell{0.80} & \ccell{0.80} & \ccell{0.80} & \ccell{1.00} & \ccell{0.80} & \ccell{1.00} & \ccell{0.60} \\
    & L7 & \ccell{0.00} & \ccell{0.60} & \ccell{0.40} & \ccell{0.20} & \ccell{0.60} & \ccell{0.40} & \ccell{0.20} & \ccell{0.60} & \ccell{0.20} & \ccell{0.20} & \ccell{0.60} & \ccell{0.60} & \ccell{0.40} & \ccell{0.40} & \ccell{0.60} & \ccell{0.60} & \ccell{0.40} \\
    & L8 & \ccell{0.40} & \ccell{0.40} & \ccell{0.00} & \ccell{0.20} & \ccell{0.20} & \ccell{0.40} & \ccell{0.40} & \ccell{0.20} & \ccell{0.20} & \ccell{0.60} & \ccell{1.00} & \ccell{1.00} & \ccell{0.20} & \ccell{0.60} & \ccell{0.80} & \ccell{1.00} & \ccell{1.00} \\
    & D123 & \ccell{0.20} & \ccell{0.60} & \ccell{0.60} & \ccell{0.00} & \ccell{0.20} & \ccell{0.80} & \ccell{0.20} & \ccell{0.40} & \ccell{0.00} & \ccell{0.80} & \ccell{0.60} & \ccell{0.60} & \ccell{0.60} & \ccell{0.80} & \ccell{0.80} & \ccell{1.00} & \ccell{1.00} \\
    & D4567 & \ccell{0.40} & \ccell{0.80} & \ccell{0.20} & \ccell{0.20} & \ccell{0.60} & \ccell{0.80} & \ccell{0.00} & \ccell{0.20} & \ccell{0.00} & \ccell{0.80} & \ccell{1.00} & \ccell{0.60} & \ccell{0.60} & \ccell{0.80} & \ccell{0.60} & \ccell{1.00} & \ccell{0.60} \\
    & D8 & \ccell{0.00} & \ccell{0.60} & \ccell{0.40} & \ccell{0.40} & \ccell{0.40} & \ccell{0.80} & \ccell{0.20} & \ccell{0.40} & \ccell{0.20} & \ccell{0.60} & \ccell{0.80} & \ccell{0.80} & \ccell{0.60} & \ccell{0.60} & \ccell{0.60} & \ccell{1.00} & \ccell{0.80} \\
    & C1 & \ccell{0.40} & \ccell{0.40} & \ccell{0.20} & \ccell{0.40} & \ccell{0.20} & \ccell{0.80} & \ccell{0.20} & \ccell{0.40} & \ccell{0.20} & \ccell{0.60} & \ccell{0.80} & \ccell{0.60} & \ccell{0.60} & \ccell{0.20} & \ccell{1.00} & \ccell{1.00} & \ccell{1.00} \\
    & C2 & \ccell{0.60} & \ccell{0.80} & \ccell{0.20} & \ccell{0.20} & \ccell{0.80} & \ccell{0.80} & \ccell{0.20} & \ccell{0.40} & \ccell{0.20} & \ccell{0.60} & \ccell{0.80} & \ccell{0.60} & \ccell{0.20} & \ccell{0.00} & \ccell{0.40} & \ccell{1.00} & \ccell{0.80} \\
    & C3 & \ccell{0.60} & \ccell{0.80} & \ccell{0.20} & \ccell{0.00} & \ccell{0.20} & \ccell{0.60} & \ccell{0.00} & \ccell{0.40} & \ccell{0.00} & \ccell{0.60} & \ccell{0.80} & \ccell{1.00} & \ccell{0.60} & \ccell{0.80} & \ccell{0.80} & \ccell{1.00} & \ccell{1.00} \\
 
\midrule
\multirow{13}{*}{\rotatebox{90}{\reposvulcpp}}
    & original & \ccell{0.20} & \ccell{0.60} & \ccell{0.20} & \ccell{0.60} & \ccell{0.60} & \ccell{0.80} & \ccell{0.40} & \ccell{0.40} & \ccell{0.60} & \ccell{0.40} & \ccell{0.40} & \ccell{0.80} & \ccell{0.00} & \ccell{0.60} & \ccell{0.80} & \ccell{0.80} & \ccell{0.60} \\
\cmidrule(lr){2-19}
    & L1 & \ccell{0.20} & \ccell{0.00} & \ccell{0.20} & \ccell{0.60} & \ccell{0.20} & \ccell{0.40} & \ccell{0.20} & \ccell{0.00} & \ccell{0.00} & \ccell{0.00} & \ccell{0.00} & \ccell{0.20} & \ccell{0.00} & \ccell{0.00} & \ccell{0.80} & \ccell{0.40} & \ccell{0.40} \\
    & L2 & \ccell{0.00} & \ccell{0.40} & \ccell{0.60} & \ccell{0.40} & \ccell{0.00} & \ccell{0.80} & \ccell{0.40} & \ccell{0.00} & \ccell{0.00} & \ccell{0.40} & \ccell{0.20} & \ccell{0.40} & \ccell{0.00} & \ccell{0.20} & \ccell{0.80} & \ccell{1.00} & \ccell{0.60} \\
    & L3 & \ccell{0.20} & \ccell{0.40} & \ccell{0.20} & \ccell{0.60} & \ccell{0.20} & \ccell{0.20} & \ccell{0.40} & \ccell{0.20} & \ccell{0.60} & \ccell{0.20} & \ccell{0.40} & \ccell{0.80} & \ccell{0.40} & \ccell{0.00} & \ccell{0.60} & \ccell{0.80} & \ccell{0.40} \\
    & L456 & \ccell{0.00} & \ccell{0.60} & \ccell{0.20} & \ccell{0.20} & \ccell{0.20} & \ccell{0.20} & \ccell{0.40} & \ccell{0.00} & \ccell{0.40} & \ccell{0.20} & \ccell{0.00} & \ccell{0.00} & \ccell{0.20} & \ccell{0.40} & \ccell{0.20} & \ccell{0.80} & \ccell{0.80} \\
    & L7 & \ccell{0.20} & \ccell{0.20} & \ccell{0.40} & \ccell{0.40} & \ccell{0.20} & \ccell{0.40} & \ccell{0.20} & \ccell{0.00} & \ccell{0.20} & \ccell{0.00} & \ccell{0.60} & \ccell{0.20} & \ccell{0.20} & \ccell{0.20} & \ccell{0.40} & \ccell{0.80} & \ccell{0.40} \\
    & L8 & \ccell{0.20} & \ccell{0.40} & \ccell{0.20} & \ccell{0.00} & \ccell{0.40} & \ccell{0.20} & \ccell{0.20} & \ccell{0.20} & \ccell{0.40} & \ccell{0.40} & \ccell{0.40} & \ccell{0.20} & \ccell{0.00} & \ccell{0.00} & \ccell{0.40} & \ccell{0.80} & \ccell{0.60} \\
    & D123 & \ccell{0.00} & \ccell{0.20} & \ccell{0.00} & \ccell{0.20} & \ccell{0.60} & \ccell{0.40} & \ccell{0.20} & \ccell{0.20} & \ccell{0.20} & \ccell{0.00} & \ccell{0.20} & \ccell{0.40} & \ccell{0.00} & \ccell{0.60} & \ccell{0.60} & \ccell{0.40} & \ccell{0.80} \\
    & D4567 & \ccell{0.00} & \ccell{0.40} & \ccell{0.60} & \ccell{0.60} & \ccell{0.20} & \ccell{0.20} & \ccell{0.20} & \ccell{0.00} & \ccell{0.20} & \ccell{0.40} & \ccell{0.20} & \ccell{0.60} & \ccell{0.60} & \ccell{0.20} & \ccell{0.60} & \ccell{1.00} & \ccell{0.60} \\
    & D8 & \ccell{0.00} & \ccell{0.60} & \ccell{0.40} & \ccell{0.20} & \ccell{0.20} & \ccell{0.60} & \ccell{0.20} & \ccell{0.20} & \ccell{0.40} & \ccell{0.20} & \ccell{0.40} & \ccell{0.40} & \ccell{0.20} & \ccell{0.20} & \ccell{0.60} & \ccell{1.00} & \ccell{0.60} \\
    & C1 & \ccell{0.20} & \ccell{0.40} & \ccell{0.60} & \ccell{0.20} & \ccell{0.00} & \ccell{0.20} & \ccell{0.00} & \ccell{0.00} & \ccell{0.00} & \ccell{0.20} & \ccell{0.00} & \ccell{0.20} & \ccell{0.20} & \ccell{0.40} & \ccell{0.80} & \ccell{0.60} & \ccell{0.80} \\
    & C2 & \ccell{0.00} & \ccell{0.40} & \ccell{0.40} & \ccell{0.40} & \ccell{0.40} & \ccell{0.40} & \ccell{0.20} & \ccell{0.00} & \ccell{0.40} & \ccell{0.00} & \ccell{0.20} & \ccell{0.40} & \ccell{0.20} & \ccell{0.20} & \ccell{0.80} & \ccell{0.80} & \ccell{0.60} \\
    & C3 & \ccell{0.00} & \ccell{0.40} & \ccell{0.20} & \ccell{0.00} & \ccell{0.00} & \ccell{0.00} & \ccell{0.00} & \ccell{0.00} & \ccell{0.00} & \ccell{0.00} & \ccell{0.00} & \ccell{0.00} & \ccell{0.20} & \ccell{0.20} & \ccell{0.00} & \ccell{0.00} & \ccell{0.20} \\
 
\midrule
\multirow{13}{*}{\rotatebox{90}{\reposvulpy}}
    & original & \ccell{0.40} & \ccell{0.60} & \ccell{0.40} & \ccell{0.20} & \ccell{0.40} & \ccell{0.40} & \ccell{0.80} & \ccell{1.00} & \ccell{1.00} & \ccell{0.80} & \ccell{0.80} & \ccell{0.80} & \ccell{0.40} & \ccell{1.00} & \ccell{0.60} & \ccell{0.80} & \ccell{0.40} \\
\cmidrule(lr){2-19}
    & L1 & \ccell{0.00} & \ccell{0.00} & \ccell{0.20} & \ccell{0.00} & \ccell{0.00} & \ccell{0.00} & \ccell{0.00} & \ccell{0.40} & \ccell{0.00} & \ccell{0.40} & \ccell{0.60} & \ccell{0.40} & \ccell{0.00} & \ccell{0.20} & \ccell{0.40} & \ccell{0.80} & \ccell{0.80} \\
    & L2 & \ccell{0.00} & \ccell{0.00} & \ccell{0.20} & \ccell{0.00} & \ccell{0.00} & \ccell{0.20} & \ccell{0.40} & \ccell{0.60} & \ccell{0.00} & \ccell{0.60} & \ccell{0.80} & \ccell{0.80} & \ccell{0.00} & \ccell{0.60} & \ccell{0.40} & \ccell{0.60} & \ccell{0.40} \\
    & L3 & \ccell{0.40} & \ccell{0.20} & \ccell{0.20} & \ccell{0.00} & \ccell{0.40} & \ccell{0.20} & \ccell{0.20} & \ccell{0.60} & \ccell{0.40} & \ccell{0.40} & \ccell{1.00} & \ccell{0.80} & \ccell{0.40} & \ccell{0.60} & \ccell{0.40} & \ccell{0.80} & \ccell{0.60} \\
    & L456 & \ccell{0.00} & \ccell{0.00} & \ccell{0.40} & \ccell{0.20} & \ccell{0.20} & \ccell{0.40} & \ccell{0.20} & \ccell{0.40} & \ccell{0.20} & \ccell{0.80} & \ccell{1.00} & \ccell{0.60} & \ccell{0.20} & \ccell{0.60} & \ccell{0.40} & \ccell{0.80} & \ccell{0.20} \\
    & L7 & \ccell{0.00} & \ccell{0.20} & \ccell{0.20} & \ccell{0.20} & \ccell{0.00} & \ccell{0.00} & \ccell{0.00} & \ccell{0.20} & \ccell{0.00} & \ccell{0.20} & \ccell{0.80} & \ccell{0.60} & \ccell{0.20} & \ccell{0.20} & \ccell{0.20} & \ccell{0.80} & \ccell{0.40} \\
    & L8 & \ccell{0.20} & \ccell{0.20} & \ccell{0.40} & \ccell{0.20} & \ccell{0.00} & \ccell{0.00} & \ccell{0.40} & \ccell{0.60} & \ccell{0.40} & \ccell{0.60} & \ccell{0.80} & \ccell{0.60} & \ccell{0.20} & \ccell{0.40} & \ccell{0.20} & \ccell{0.60} & \ccell{0.20} \\
    & D123 & \ccell{0.20} & \ccell{0.00} & \ccell{0.20} & \ccell{0.20} & \ccell{0.60} & \ccell{0.20} & \ccell{0.20} & \ccell{0.20} & \ccell{0.40} & \ccell{0.20} & \ccell{1.00} & \ccell{0.80} & \ccell{0.20} & \ccell{0.40} & \ccell{1.00} & \ccell{0.80} & \ccell{0.60} \\
    & D4567 & \ccell{0.20} & \ccell{0.40} & \ccell{0.20} & \ccell{0.00} & \ccell{0.40} & \ccell{0.20} & \ccell{0.00} & \ccell{0.00} & \ccell{0.20} & \ccell{0.80} & \ccell{0.80} & \ccell{0.40} & \ccell{0.40} & \ccell{0.60} & \ccell{0.40} & \ccell{0.80} & \ccell{0.40} \\
    & D8 & \ccell{0.20} & \ccell{0.20} & \ccell{0.20} & \ccell{0.20} & \ccell{0.40} & \ccell{0.20} & \ccell{0.20} & \ccell{0.20} & \ccell{0.20} & \ccell{0.00} & \ccell{0.80} & \ccell{0.40} & \ccell{0.40} & \ccell{0.60} & \ccell{0.20} & \ccell{0.80} & \ccell{0.40} \\
    & C1 & \ccell{0.20} & \ccell{0.00} & \ccell{0.20} & \ccell{0.00} & \ccell{0.60} & \ccell{0.40} & \ccell{0.00} & \ccell{0.20} & \ccell{0.20} & \ccell{1.00} & \ccell{1.00} & \ccell{0.80} & \ccell{0.20} & \ccell{0.40} & \ccell{0.40} & \ccell{0.80} & \ccell{0.60} \\
    & C2 & \ccell{0.20} & \ccell{0.20} & \ccell{0.20} & \ccell{0.40} & \ccell{0.00} & \ccell{0.20} & \ccell{0.20} & \ccell{0.60} & \ccell{0.40} & \ccell{0.60} & \ccell{0.80} & \ccell{0.40} & \ccell{0.40} & \ccell{0.60} & \ccell{0.40} & \ccell{0.80} & \ccell{0.40} \\
    & C3 & \ccell{0.00} & \ccell{0.00} & \ccell{0.20} & \ccell{0.40} & \ccell{0.00} & \ccell{0.00} & \ccell{0.00} & \ccell{0.40} & \ccell{0.20} & \ccell{0.20} & \ccell{0.40} & \ccell{0.40} & \ccell{0.00} & \ccell{0.40} & \ccell{0.40} & \ccell{0.80} & \ccell{0.40} \\
 
\midrule
\multirow{13}{*}{\rotatebox{90}{\primevulc}}
    & original & \ccell{0.40} & \ccell{0.40} & \ccell{0.60} & \ccell{0.60} & \ccell{0.40} & \ccell{1.00} & \ccell{0.40} & \ccell{0.40} & \ccell{0.60} & \ccell{0.00} & \ccell{0.60} & \ccell{1.00} & \ccell{0.60} & \ccell{0.60} & \ccell{0.80} & \ccell{1.00} & \ccell{1.00} \\
\cmidrule(lr){2-19}
    & L1 & \ccell{0.40} & \ccell{0.60} & \ccell{0.60} & \ccell{0.00} & \ccell{0.00} & \ccell{0.20} & \ccell{0.00} & \ccell{0.20} & \ccell{0.40} & \ccell{0.60} & \ccell{0.40} & \ccell{0.60} & \ccell{0.20} & \ccell{0.00} & \ccell{0.60} & \ccell{1.00} & \ccell{1.00} \\
    & L2 & \ccell{0.40} & \ccell{0.80} & \ccell{0.80} & \ccell{0.40} & \ccell{0.40} & \ccell{0.20} & \ccell{0.20} & \ccell{0.40} & \ccell{0.40} & \ccell{0.60} & \ccell{1.00} & \ccell{0.80} & \ccell{0.40} & \ccell{0.00} & \ccell{0.80} & \ccell{1.00} & \ccell{0.80} \\
    & L3 & \ccell{0.40} & \ccell{0.60} & \ccell{0.60} & \ccell{0.20} & \ccell{0.40} & \ccell{0.60} & \ccell{0.40} & \ccell{0.20} & \ccell{0.40} & \ccell{0.20} & \ccell{0.60} & \ccell{0.40} & \ccell{0.40} & \ccell{0.20} & \ccell{0.60} & \ccell{1.00} & \ccell{1.00} \\
    & L456 & \ccell{0.40} & \ccell{0.60} & \ccell{0.60} & \ccell{0.20} & \ccell{0.40} & \ccell{0.20} & \ccell{0.20} & \ccell{0.20} & \ccell{0.40} & \ccell{0.20} & \ccell{0.40} & \ccell{0.60} & \ccell{0.20} & \ccell{0.20} & \ccell{0.60} & \ccell{1.00} & \ccell{1.00} \\
    & L7 & \ccell{0.20} & \ccell{0.80} & \ccell{0.60} & \ccell{0.20} & \ccell{0.60} & \ccell{0.60} & \ccell{0.20} & \ccell{0.40} & \ccell{0.60} & \ccell{0.60} & \ccell{0.80} & \ccell{0.80} & \ccell{0.40} & \ccell{0.40} & \ccell{0.60} & \ccell{1.00} & \ccell{1.00} \\
    & L8 & \ccell{0.20} & \ccell{0.40} & \ccell{0.80} & \ccell{0.40} & \ccell{0.40} & \ccell{0.40} & \ccell{0.40} & \ccell{0.20} & \ccell{0.60} & \ccell{0.00} & \ccell{0.60} & \ccell{0.60} & \ccell{0.60} & \ccell{0.00} & \ccell{0.40} & \ccell{0.60} & \ccell{0.80} \\
    & D123 & \ccell{0.20} & \ccell{0.40} & \ccell{0.40} & \ccell{0.60} & \ccell{0.60} & \ccell{0.40} & \ccell{0.20} & \ccell{0.40} & \ccell{0.20} & \ccell{0.80} & \ccell{0.80} & \ccell{0.40} & \ccell{0.00} & \ccell{0.40} & \ccell{1.00} & \ccell{1.00} & \ccell{1.00} \\
    & D4567 & \ccell{0.20} & \ccell{0.40} & \ccell{0.60} & \ccell{0.20} & \ccell{0.80} & \ccell{0.60} & \ccell{0.40} & \ccell{0.60} & \ccell{0.20} & \ccell{0.40} & \ccell{0.80} & \ccell{0.80} & \ccell{0.60} & \ccell{0.20} & \ccell{0.60} & \ccell{1.00} & \ccell{1.00} \\
    & D8 & \ccell{0.20} & \ccell{0.60} & \ccell{0.80} & \ccell{0.40} & \ccell{0.60} & \ccell{0.20} & \ccell{0.60} & \ccell{0.20} & \ccell{0.40} & \ccell{0.40} & \ccell{0.60} & \ccell{0.60} & \ccell{0.00} & \ccell{0.00} & \ccell{0.60} & \ccell{1.00} & \ccell{1.00} \\
    & C1 & \ccell{0.20} & \ccell{0.80} & \ccell{0.60} & \ccell{0.20} & \ccell{0.80} & \ccell{0.60} & \ccell{0.00} & \ccell{0.20} & \ccell{0.40} & \ccell{0.40} & \ccell{0.80} & \ccell{0.60} & \ccell{0.00} & \ccell{0.40} & \ccell{0.60} & \ccell{0.80} & \ccell{1.00} \\
    & C2 & \ccell{0.60} & \ccell{0.60} & \ccell{0.60} & \ccell{0.20} & \ccell{0.40} & \ccell{0.80} & \ccell{0.40} & \ccell{0.20} & \ccell{0.60} & \ccell{0.60} & \ccell{0.80} & \ccell{0.60} & \ccell{0.60} & \ccell{0.20} & \ccell{0.80} & \ccell{1.00} & \ccell{1.00} \\
    & C3 & \ccell{0.00} & \ccell{0.00} & \ccell{0.20} & \ccell{0.00} & \ccell{0.00} & \ccell{0.00} & \ccell{0.00} & \ccell{0.00} & \ccell{0.00} & \ccell{0.00} & \ccell{0.00} & \ccell{0.00} & \ccell{0.20} & \ccell{0.00} & \ccell{0.00} & \ccell{0.20} & \ccell{0.00} \\
 
\bottomrule
\end{tabular}
}
\label{tab:detection-result-all-datasets-downgrade_top20}
\end{table*}

\begin{table*}[htbp]
\centering
\caption{Downgrade rate on different datasets: \smartbugs, \reposvulcpp, \reposvulpy, \primevulc (sampled by downgrade top20). (L456 = L4+L5+L6, D123 = D1+D2+D3, D4567 = D4+D5+D6+D7 for layout clarity)}
{
\tiny
\setlength\tabcolsep{3pt}
\renewcommand{\arraystretch}{0.6}
\begin{tabular}{l lccccccccccccccccc}
\toprule
dataset & series & \multicolumn{6}{c}{qwen} & \multicolumn{4}{c}{llama} & \multicolumn{2}{c}{deepseek} & \multicolumn{3}{c}{openai} & \multicolumn{2}{c}{agent} \\
\cmidrule(lr){3-8} \cmidrule(lr){9-12} \cmidrule(lr){13-14} \cmidrule(lr){15-17} \cmidrule(lr){18-19}
 & model & qn-7b & qn-14b & qn-32b & ds-qn-7b & ds-qn-14b & ds-qn-32b & lm-8b & lm-70b & ds-lm-8b & ds-lm-70b & ds-v3 & ds-r1 & gpt-3.5 & gpt-4o & o3-mini & copilot & codex \\
\midrule
\multirow{12}{*}{\rotatebox{90}{\smartbugs}}
    & L1 & \ccell{0.40} & \ccell{0.60} & \ccell{0.40} & \ccell{0.60} & \ccell{0.20} & \ccell{0.20} & \ccell{0.40} & \ccell{0.60} & \ccell{0.40} & \ccell{0.40} & \ccell{0.40} & \ccell{0.20} & \ccell{0.00} & \ccell{0.20} & \ccell{0.00} & \ccell{0.00} & \ccell{0.20} \\
    & L2 & \ccell{0.20} & \ccell{0.40} & \ccell{0.40} & \ccell{0.40} & \ccell{0.00} & \ccell{0.20} & \ccell{0.40} & \ccell{0.40} & \ccell{0.40} & \ccell{0.20} & \ccell{0.00} & \ccell{0.00} & \ccell{0.20} & \ccell{0.20} & \ccell{0.00} & \ccell{0.00} & \ccell{0.20} \\
    & L3 & \ccell{0.20} & \ccell{0.40} & \ccell{0.20} & \ccell{0.80} & \ccell{0.00} & \ccell{0.00} & \ccell{0.40} & \ccell{0.60} & \ccell{0.40} & \ccell{0.60} & \ccell{0.00} & \ccell{0.00} & \ccell{0.00} & \ccell{0.40} & \ccell{0.00} & \ccell{0.20} & \ccell{0.40} \\
    & L456 & \ccell{0.20} & \ccell{0.20} & \ccell{0.20} & \ccell{0.60} & \ccell{0.20} & \ccell{0.20} & \ccell{0.40} & \ccell{0.00} & \ccell{0.40} & \ccell{0.20} & \ccell{0.20} & \ccell{0.20} & \ccell{0.00} & \ccell{0.00} & \ccell{0.00} & \ccell{0.00} & \ccell{0.40} \\
    & L7 & \ccell{0.40} & \ccell{0.20} & \ccell{0.40} & \ccell{0.60} & \ccell{0.20} & \ccell{0.40} & \ccell{0.40} & \ccell{0.20} & \ccell{0.40} & \ccell{0.80} & \ccell{0.40} & \ccell{0.40} & \ccell{0.20} & \ccell{0.40} & \ccell{0.20} & \ccell{0.40} & \ccell{0.60} \\
    & L8 & \ccell{0.40} & \ccell{0.60} & \ccell{0.60} & \ccell{0.60} & \ccell{0.40} & \ccell{0.60} & \ccell{0.40} & \ccell{0.60} & \ccell{0.40} & \ccell{0.40} & \ccell{0.00} & \ccell{0.00} & \ccell{0.40} & \ccell{0.40} & \ccell{0.20} & \ccell{0.00} & \ccell{0.00} \\
    & D123 & \ccell{0.20} & \ccell{0.20} & \ccell{0.20} & \ccell{0.80} & \ccell{0.40} & \ccell{0.20} & \ccell{0.40} & \ccell{0.60} & \ccell{0.40} & \ccell{0.20} & \ccell{0.40} & \ccell{0.20} & \ccell{0.00} & \ccell{0.20} & \ccell{0.20} & \ccell{0.00} & \ccell{0.00} \\
    & D4567 & \ccell{0.00} & \ccell{0.20} & \ccell{0.60} & \ccell{0.60} & \ccell{0.40} & \ccell{0.20} & \ccell{0.60} & \ccell{0.60} & \ccell{0.40} & \ccell{0.20} & \ccell{0.00} & \ccell{0.20} & \ccell{0.00} & \ccell{0.20} & \ccell{0.20} & \ccell{0.00} & \ccell{0.40} \\
    & D8 & \ccell{0.40} & \ccell{0.40} & \ccell{0.40} & \ccell{0.40} & \ccell{0.20} & \ccell{0.20} & \ccell{0.40} & \ccell{0.40} & \ccell{0.40} & \ccell{0.40} & \ccell{0.20} & \ccell{0.00} & \ccell{0.20} & \ccell{0.20} & \ccell{0.20} & \ccell{0.00} & \ccell{0.20} \\
    & C1 & \ccell{0.20} & \ccell{0.40} & \ccell{0.40} & \ccell{0.40} & \ccell{0.20} & \ccell{0.00} & \ccell{0.60} & \ccell{0.40} & \ccell{0.40} & \ccell{0.40} & \ccell{0.20} & \ccell{0.20} & \ccell{0.20} & \ccell{0.60} & \ccell{0.00} & \ccell{0.00} & \ccell{0.00} \\
    & C2 & \ccell{0.20} & \ccell{0.20} & \ccell{0.60} & \ccell{0.60} & \ccell{0.00} & \ccell{0.20} & \ccell{0.40} & \ccell{0.40} & \ccell{0.40} & \ccell{0.40} & \ccell{0.20} & \ccell{0.20} & \ccell{0.60} & \ccell{0.80} & \ccell{0.40} & \ccell{0.00} & \ccell{0.20} \\
    & C3 & \ccell{0.20} & \ccell{0.00} & \ccell{0.40} & \ccell{0.80} & \ccell{0.20} & \ccell{0.40} & \ccell{0.60} & \ccell{0.40} & \ccell{0.40} & \ccell{0.40} & \ccell{0.20} & \ccell{0.00} & \ccell{0.20} & \ccell{0.20} & \ccell{0.00} & \ccell{0.00} & \ccell{0.00} \\
 
\midrule
\multirow{12}{*}{\rotatebox{90}{\reposvulcpp}}
    & L1 & \ccell{0.20} & \ccell{0.60} & \ccell{0.20} & \ccell{0.20} & \ccell{0.40} & \ccell{0.40} & \ccell{0.20} & \ccell{0.40} & \ccell{0.60} & \ccell{0.40} & \ccell{0.40} & \ccell{0.60} & \ccell{0.00} & \ccell{0.60} & \ccell{0.00} & \ccell{0.40} & \ccell{0.20} \\
    & L2 & \ccell{0.20} & \ccell{0.20} & \ccell{0.00} & \ccell{0.20} & \ccell{0.60} & \ccell{0.20} & \ccell{0.20} & \ccell{0.40} & \ccell{0.60} & \ccell{0.20} & \ccell{0.20} & \ccell{0.40} & \ccell{0.00} & \ccell{0.40} & \ccell{0.00} & \ccell{0.00} & \ccell{0.00} \\
    & L3 & \ccell{0.20} & \ccell{0.40} & \ccell{0.20} & \ccell{0.00} & \ccell{0.40} & \ccell{0.60} & \ccell{0.00} & \ccell{0.40} & \ccell{0.20} & \ccell{0.40} & \ccell{0.20} & \ccell{0.20} & \ccell{0.00} & \ccell{0.60} & \ccell{0.20} & \ccell{0.20} & \ccell{0.20} \\
    & L456 & \ccell{0.20} & \ccell{0.40} & \ccell{0.20} & \ccell{0.40} & \ccell{0.40} & \ccell{0.60} & \ccell{0.00} & \ccell{0.40} & \ccell{0.60} & \ccell{0.20} & \ccell{0.40} & \ccell{0.80} & \ccell{0.00} & \ccell{0.20} & \ccell{0.60} & \ccell{0.20} & \ccell{0.00} \\
    & L7 & \ccell{0.20} & \ccell{0.40} & \ccell{0.20} & \ccell{0.40} & \ccell{0.60} & \ccell{0.40} & \ccell{0.20} & \ccell{0.40} & \ccell{0.40} & \ccell{0.40} & \ccell{0.00} & \ccell{0.60} & \ccell{0.00} & \ccell{0.40} & \ccell{0.40} & \ccell{0.00} & \ccell{0.20} \\
    & L8 & \ccell{0.20} & \ccell{0.60} & \ccell{0.20} & \ccell{0.60} & \ccell{0.40} & \ccell{0.80} & \ccell{0.20} & \ccell{0.20} & \ccell{0.40} & \ccell{0.20} & \ccell{0.00} & \ccell{0.60} & \ccell{0.00} & \ccell{0.60} & \ccell{0.40} & \ccell{0.00} & \ccell{0.00} \\
    & D123 & \ccell{0.20} & \ccell{0.60} & \ccell{0.20} & \ccell{0.40} & \ccell{0.20} & \ccell{0.60} & \ccell{0.20} & \ccell{0.20} & \ccell{0.60} & \ccell{0.40} & \ccell{0.20} & \ccell{0.60} & \ccell{0.00} & \ccell{0.00} & \ccell{0.20} & \ccell{0.40} & \ccell{0.00} \\
    & D4567 & \ccell{0.20} & \ccell{0.20} & \ccell{0.00} & \ccell{0.00} & \ccell{0.60} & \ccell{0.60} & \ccell{0.20} & \ccell{0.40} & \ccell{0.60} & \ccell{0.20} & \ccell{0.20} & \ccell{0.20} & \ccell{0.00} & \ccell{0.40} & \ccell{0.20} & \ccell{0.00} & \ccell{0.00} \\
    & D8 & \ccell{0.20} & \ccell{0.20} & \ccell{0.00} & \ccell{0.40} & \ccell{0.60} & \ccell{0.40} & \ccell{0.20} & \ccell{0.20} & \ccell{0.60} & \ccell{0.40} & \ccell{0.40} & \ccell{0.40} & \ccell{0.00} & \ccell{0.40} & \ccell{0.20} & \ccell{0.00} & \ccell{0.00} \\
    & C1 & \ccell{0.20} & \ccell{0.20} & \ccell{0.00} & \ccell{0.40} & \ccell{0.60} & \ccell{0.60} & \ccell{0.40} & \ccell{0.40} & \ccell{0.60} & \ccell{0.20} & \ccell{0.40} & \ccell{0.60} & \ccell{0.00} & \ccell{0.20} & \ccell{0.00} & \ccell{0.40} & \ccell{0.00} \\
    & C2 & \ccell{0.20} & \ccell{0.40} & \ccell{0.20} & \ccell{0.40} & \ccell{0.40} & \ccell{0.40} & \ccell{0.20} & \ccell{0.40} & \ccell{0.40} & \ccell{0.40} & \ccell{0.20} & \ccell{0.40} & \ccell{0.00} & \ccell{0.40} & \ccell{0.00} & \ccell{0.00} & \ccell{0.00} \\
    & C3 & \ccell{0.20} & \ccell{0.40} & \ccell{0.20} & \ccell{0.60} & \ccell{0.60} & \ccell{0.80} & \ccell{0.40} & \ccell{0.40} & \ccell{0.60} & \ccell{0.40} & \ccell{0.40} & \ccell{0.80} & \ccell{0.00} & \ccell{0.40} & \ccell{0.80} & \ccell{0.80} & \ccell{0.40} \\
 
\midrule
\multirow{12}{*}{\rotatebox{90}{\reposvulpy}}
    & L1 & \ccell{0.40} & \ccell{0.60} & \ccell{0.20} & \ccell{0.00} & \ccell{0.40} & \ccell{0.40} & \ccell{0.80} & \ccell{0.60} & \ccell{1.00} & \ccell{0.60} & \ccell{0.20} & \ccell{0.40} & \ccell{0.40} & \ccell{0.80} & \ccell{0.20} & \ccell{0.00} & \ccell{0.00} \\
    & L2 & \ccell{0.40} & \ccell{0.60} & \ccell{0.20} & \ccell{0.20} & \ccell{0.40} & \ccell{0.20} & \ccell{0.40} & \ccell{0.40} & \ccell{1.00} & \ccell{0.20} & \ccell{0.20} & \ccell{0.20} & \ccell{0.40} & \ccell{0.40} & \ccell{0.20} & \ccell{0.20} & \ccell{0.00} \\
    & L3 & \ccell{0.20} & \ccell{0.40} & \ccell{0.20} & \ccell{0.20} & \ccell{0.00} & \ccell{0.40} & \ccell{0.60} & \ccell{0.40} & \ccell{0.60} & \ccell{0.40} & \ccell{0.00} & \ccell{0.00} & \ccell{0.00} & \ccell{0.40} & \ccell{0.20} & \ccell{0.00} & \ccell{0.00} \\
    & L456 & \ccell{0.40} & \ccell{0.60} & \ccell{0.00} & \ccell{0.20} & \ccell{0.20} & \ccell{0.00} & \ccell{0.80} & \ccell{0.60} & \ccell{0.80} & \ccell{0.20} & \ccell{0.00} & \ccell{0.20} & \ccell{0.20} & \ccell{0.40} & \ccell{0.20} & \ccell{0.00} & \ccell{0.20} \\
    & L7 & \ccell{0.40} & \ccell{0.40} & \ccell{0.20} & \ccell{0.00} & \ccell{0.40} & \ccell{0.40} & \ccell{0.80} & \ccell{0.80} & \ccell{1.00} & \ccell{0.80} & \ccell{0.00} & \ccell{0.20} & \ccell{0.20} & \ccell{0.80} & \ccell{0.40} & \ccell{0.00} & \ccell{0.00} \\
    & L8 & \ccell{0.40} & \ccell{0.40} & \ccell{0.20} & \ccell{0.20} & \ccell{0.40} & \ccell{0.40} & \ccell{0.40} & \ccell{0.40} & \ccell{0.60} & \ccell{0.40} & \ccell{0.00} & \ccell{0.20} & \ccell{0.20} & \ccell{0.60} & \ccell{0.40} & \ccell{0.20} & \ccell{0.20} \\
    & D123 & \ccell{0.20} & \ccell{0.60} & \ccell{0.20} & \ccell{0.20} & \ccell{0.00} & \ccell{0.20} & \ccell{0.60} & \ccell{0.80} & \ccell{0.60} & \ccell{0.60} & \ccell{0.00} & \ccell{0.00} & \ccell{0.20} & \ccell{0.60} & \ccell{0.00} & \ccell{0.00} & \ccell{0.00} \\
    & D4567 & \ccell{0.20} & \ccell{0.40} & \ccell{0.20} & \ccell{0.20} & \ccell{0.00} & \ccell{0.20} & \ccell{0.80} & \ccell{1.00} & \ccell{0.80} & \ccell{0.20} & \ccell{0.00} & \ccell{0.40} & \ccell{0.00} & \ccell{0.40} & \ccell{0.20} & \ccell{0.00} & \ccell{0.00} \\
    & D8 & \ccell{0.20} & \ccell{0.60} & \ccell{0.20} & \ccell{0.00} & \ccell{0.00} & \ccell{0.20} & \ccell{0.80} & \ccell{0.80} & \ccell{0.80} & \ccell{0.80} & \ccell{0.20} & \ccell{0.40} & \ccell{0.00} & \ccell{0.40} & \ccell{0.40} & \ccell{0.00} & \ccell{0.00} \\
    & C1 & \ccell{0.20} & \ccell{0.60} & \ccell{0.20} & \ccell{0.20} & \ccell{0.00} & \ccell{0.20} & \ccell{0.80} & \ccell{0.80} & \ccell{0.80} & \ccell{0.00} & \ccell{0.00} & \ccell{0.20} & \ccell{0.20} & \ccell{0.60} & \ccell{0.20} & \ccell{0.00} & \ccell{0.00} \\
    & C2 & \ccell{0.20} & \ccell{0.40} & \ccell{0.20} & \ccell{0.20} & \ccell{0.40} & \ccell{0.20} & \ccell{0.60} & \ccell{0.40} & \ccell{0.60} & \ccell{0.40} & \ccell{0.00} & \ccell{0.40} & \ccell{0.00} & \ccell{0.40} & \ccell{0.20} & \ccell{0.00} & \ccell{0.00} \\
    & C3 & \ccell{0.40} & \ccell{0.60} & \ccell{0.20} & \ccell{0.00} & \ccell{0.40} & \ccell{0.40} & \ccell{0.80} & \ccell{0.60} & \ccell{0.80} & \ccell{0.80} & \ccell{0.40} & \ccell{0.40} & \ccell{0.40} & \ccell{0.60} & \ccell{0.40} & \ccell{0.00} & \ccell{0.20} \\
 
\midrule
\multirow{12}{*}{\rotatebox{90}{\primevulc}}
    & L1 & \ccell{0.00} & \ccell{0.20} & \ccell{0.20} & \ccell{0.60} & \ccell{0.40} & \ccell{0.80} & \ccell{0.40} & \ccell{0.20} & \ccell{0.40} & \ccell{0.00} & \ccell{0.20} & \ccell{0.40} & \ccell{0.40} & \ccell{0.60} & \ccell{0.40} & \ccell{0.00} & \ccell{0.00} \\
    & L2 & \ccell{0.20} & \ccell{0.00} & \ccell{0.00} & \ccell{0.60} & \ccell{0.20} & \ccell{0.80} & \ccell{0.40} & \ccell{0.00} & \ccell{0.40} & \ccell{0.00} & \ccell{0.00} & \ccell{0.20} & \ccell{0.40} & \ccell{0.60} & \ccell{0.20} & \ccell{0.00} & \ccell{0.20} \\
    & L3 & \ccell{0.00} & \ccell{0.00} & \ccell{0.00} & \ccell{0.40} & \ccell{0.20} & \ccell{0.40} & \ccell{0.20} & \ccell{0.20} & \ccell{0.40} & \ccell{0.00} & \ccell{0.00} & \ccell{0.60} & \ccell{0.20} & \ccell{0.40} & \ccell{0.20} & \ccell{0.00} & \ccell{0.00} \\
    & L456 & \ccell{0.00} & \ccell{0.00} & \ccell{0.20} & \ccell{0.40} & \ccell{0.20} & \ccell{0.80} & \ccell{0.20} & \ccell{0.20} & \ccell{0.20} & \ccell{0.00} & \ccell{0.20} & \ccell{0.40} & \ccell{0.40} & \ccell{0.60} & \ccell{0.40} & \ccell{0.00} & \ccell{0.00} \\
    & L7 & \ccell{0.20} & \ccell{0.00} & \ccell{0.00} & \ccell{0.60} & \ccell{0.20} & \ccell{0.40} & \ccell{0.20} & \ccell{0.00} & \ccell{0.20} & \ccell{0.00} & \ccell{0.00} & \ccell{0.20} & \ccell{0.40} & \ccell{0.20} & \ccell{0.40} & \ccell{0.00} & \ccell{0.00} \\
    & L8 & \ccell{0.20} & \ccell{0.20} & \ccell{0.00} & \ccell{0.60} & \ccell{0.40} & \ccell{0.60} & \ccell{0.20} & \ccell{0.20} & \ccell{0.20} & \ccell{0.00} & \ccell{0.00} & \ccell{0.40} & \ccell{0.00} & \ccell{0.60} & \ccell{0.60} & \ccell{0.40} & \ccell{0.20} \\
    & D123 & \ccell{0.40} & \ccell{0.00} & \ccell{0.20} & \ccell{0.00} & \ccell{0.20} & \ccell{0.60} & \ccell{0.40} & \ccell{0.20} & \ccell{0.60} & \ccell{0.00} & \ccell{0.00} & \ccell{0.60} & \ccell{0.60} & \ccell{0.40} & \ccell{0.00} & \ccell{0.00} & \ccell{0.00} \\
    & D4567 & \ccell{0.20} & \ccell{0.00} & \ccell{0.00} & \ccell{0.60} & \ccell{0.00} & \ccell{0.40} & \ccell{0.40} & \ccell{0.00} & \ccell{0.40} & \ccell{0.00} & \ccell{0.00} & \ccell{0.20} & \ccell{0.00} & \ccell{0.40} & \ccell{0.40} & \ccell{0.00} & \ccell{0.00} \\
    & D8 & \ccell{0.20} & \ccell{0.00} & \ccell{0.00} & \ccell{0.40} & \ccell{0.00} & \ccell{0.80} & \ccell{0.20} & \ccell{0.20} & \ccell{0.60} & \ccell{0.00} & \ccell{0.20} & \ccell{0.40} & \ccell{0.60} & \ccell{0.60} & \ccell{0.40} & \ccell{0.00} & \ccell{0.00} \\
    & C1 & \ccell{0.20} & \ccell{0.00} & \ccell{0.00} & \ccell{0.40} & \ccell{0.20} & \ccell{0.40} & \ccell{0.40} & \ccell{0.20} & \ccell{0.40} & \ccell{0.00} & \ccell{0.00} & \ccell{0.40} & \ccell{0.60} & \ccell{0.20} & \ccell{0.40} & \ccell{0.20} & \ccell{0.00} \\
    & C2 & \ccell{0.00} & \ccell{0.00} & \ccell{0.00} & \ccell{0.40} & \ccell{0.20} & \ccell{0.20} & \ccell{0.20} & \ccell{0.20} & \ccell{0.20} & \ccell{0.00} & \ccell{0.20} & \ccell{0.40} & \ccell{0.20} & \ccell{0.60} & \ccell{0.20} & \ccell{0.00} & \ccell{0.00} \\
    & C3 & \ccell{0.40} & \ccell{0.40} & \ccell{0.40} & \ccell{0.60} & \ccell{0.40} & \ccell{1.00} & \ccell{0.40} & \ccell{0.40} & \ccell{0.60} & \ccell{0.00} & \ccell{0.60} & \ccell{1.00} & \ccell{0.40} & \ccell{0.60} & \ccell{0.80} & \ccell{0.80} & \ccell{1.00} \\
 
\bottomrule
\end{tabular}
}
\label{tab:downgrade-rate-all-datasets-downgrade_top20}
\end{table*}

\begin{table*}[htbp]
\centering
\caption{Detection successful rate on different datasets: \smartbugs, \reposvulcpp, \reposvulpy, \primevulc (sampled by upgrade top20). ). (L456 = L4+L5+L6, D123 = D1+D2+D3, D4567 = D4+D5+D6+D7 for layout clarity)}
{
\tiny
\setlength\tabcolsep{3pt}
\renewcommand{\arraystretch}{0.6}
\begin{tabular}{l lccccccccccccccccc}
\toprule
dataset & series & \multicolumn{6}{c}{qwen} & \multicolumn{4}{c}{llama} & \multicolumn{2}{c}{deepseek} & \multicolumn{3}{c}{openai} & \multicolumn{2}{c}{agent} \\
\cmidrule(lr){3-8} \cmidrule(lr){9-12} \cmidrule(lr){13-14} \cmidrule(lr){15-17} \cmidrule(lr){18-19}
 & model & qn-7b & qn-14b & qn-32b & ds-qn-7b & ds-qn-14b & ds-qn-32b & lm-8b & lm-70b & ds-lm-8b & ds-lm-70b & ds-v3 & ds-r1 & gpt-3.5 & gpt-4o & o3-mini & copilot & codex \\
\midrule
\multirow{13}{*}{\rotatebox{90}{\smartbugs}}
    & original & \ccell{0.00} & \ccell{0.20} & \ccell{0.00} & \ccell{0.40} & \ccell{0.20} & \ccell{0.40} & \ccell{0.00} & \ccell{0.00} & \ccell{0.20} & \ccell{0.20} & \ccell{0.60} & \ccell{0.60} & \ccell{0.20} & \ccell{0.60} & \ccell{0.80} & \ccell{1.00} & \ccell{0.60} \\
\cmidrule(lr){2-19}
    & L1 & \ccell{0.20} & \ccell{0.40} & \ccell{0.00} & \ccell{0.60} & \ccell{0.20} & \ccell{0.40} & \ccell{0.20} & \ccell{0.20} & \ccell{0.00} & \ccell{0.00} & \ccell{0.60} & \ccell{0.20} & \ccell{0.60} & \ccell{0.80} & \ccell{1.00} & \ccell{1.00} & \ccell{1.00} \\
    & L2 & \ccell{0.20} & \ccell{0.00} & \ccell{0.20} & \ccell{0.60} & \ccell{0.20} & \ccell{0.80} & \ccell{0.40} & \ccell{0.20} & \ccell{0.80} & \ccell{0.40} & \ccell{0.60} & \ccell{0.80} & \ccell{0.20} & \ccell{1.00} & \ccell{0.80} & \ccell{1.00} & \ccell{0.80} \\
    & L3 & \ccell{0.00} & \ccell{0.20} & \ccell{0.00} & \ccell{0.60} & \ccell{0.20} & \ccell{0.80} & \ccell{0.60} & \ccell{0.60} & \ccell{0.20} & \ccell{0.40} & \ccell{0.40} & \ccell{1.00} & \ccell{0.00} & \ccell{0.60} & \ccell{1.00} & \ccell{0.80} & \ccell{1.00} \\
    & L456 & \ccell{0.00} & \ccell{0.40} & \ccell{0.20} & \ccell{0.40} & \ccell{0.20} & \ccell{0.20} & \ccell{0.40} & \ccell{0.20} & \ccell{0.20} & \ccell{0.60} & \ccell{0.40} & \ccell{0.80} & \ccell{0.20} & \ccell{0.60} & \ccell{0.80} & \ccell{0.80} & \ccell{0.80} \\
    & L7 & \ccell{0.00} & \ccell{0.40} & \ccell{0.00} & \ccell{0.60} & \ccell{0.20} & \ccell{0.60} & \ccell{0.40} & \ccell{0.40} & \ccell{0.40} & \ccell{0.20} & \ccell{0.40} & \ccell{1.00} & \ccell{0.40} & \ccell{1.00} & \ccell{0.60} & \ccell{0.80} & \ccell{0.80} \\
    & L8 & \ccell{0.20} & \ccell{0.20} & \ccell{0.20} & \ccell{0.40} & \ccell{0.60} & \ccell{0.40} & \ccell{0.20} & \ccell{0.40} & \ccell{0.00} & \ccell{0.40} & \ccell{0.80} & \ccell{0.60} & \ccell{0.80} & \ccell{0.60} & \ccell{0.80} & \ccell{1.00} & \ccell{1.00} \\
    & D123 & \ccell{0.20} & \ccell{0.20} & \ccell{0.40} & \ccell{0.80} & \ccell{0.80} & \ccell{0.40} & \ccell{0.20} & \ccell{0.20} & \ccell{0.40} & \ccell{0.60} & \ccell{1.00} & \ccell{1.00} & \ccell{0.40} & \ccell{1.00} & \ccell{0.80} & \ccell{1.00} & \ccell{0.80} \\
    & D4567 & \ccell{0.20} & \ccell{0.00} & \ccell{0.60} & \ccell{0.60} & \ccell{0.40} & \ccell{0.40} & \ccell{0.60} & \ccell{0.80} & \ccell{0.60} & \ccell{0.40} & \ccell{0.60} & \ccell{0.80} & \ccell{0.60} & \ccell{0.80} & \ccell{0.80} & \ccell{1.00} & \ccell{1.00} \\
    & D8 & \ccell{0.20} & \ccell{0.60} & \ccell{0.40} & \ccell{0.60} & \ccell{0.20} & \ccell{0.60} & \ccell{0.80} & \ccell{0.20} & \ccell{0.40} & \ccell{0.20} & \ccell{0.60} & \ccell{0.80} & \ccell{0.20} & \ccell{0.40} & \ccell{1.00} & \ccell{1.00} & \ccell{0.80} \\
    & C1 & \ccell{0.00} & \ccell{0.20} & \ccell{0.20} & \ccell{0.20} & \ccell{0.00} & \ccell{0.20} & \ccell{0.20} & \ccell{0.20} & \ccell{0.20} & \ccell{0.20} & \ccell{0.60} & \ccell{0.80} & \ccell{0.00} & \ccell{0.80} & \ccell{1.00} & \ccell{1.00} & \ccell{1.00} \\
    & C2 & \ccell{0.00} & \ccell{0.60} & \ccell{0.20} & \ccell{0.20} & \ccell{0.20} & \ccell{0.20} & \ccell{0.80} & \ccell{0.40} & \ccell{0.40} & \ccell{0.60} & \ccell{0.60} & \ccell{0.80} & \ccell{0.00} & \ccell{0.60} & \ccell{1.00} & \ccell{1.00} & \ccell{0.80} \\
    & C3 & \ccell{0.20} & \ccell{0.40} & \ccell{0.20} & \ccell{0.40} & \ccell{0.40} & \ccell{0.40} & \ccell{0.20} & \ccell{0.40} & \ccell{0.00} & \ccell{0.80} & \ccell{1.00} & \ccell{1.00} & \ccell{0.20} & \ccell{0.80} & \ccell{0.40} & \ccell{0.80} & \ccell{0.80} \\
 
\midrule
\multirow{13}{*}{\rotatebox{90}{\reposvulcpp}}
    & original & \ccell{0.40} & \ccell{0.20} & \ccell{0.20} & \ccell{0.00} & \ccell{0.20} & \ccell{0.20} & \ccell{0.60} & \ccell{0.40} & \ccell{0.20} & \ccell{0.40} & \ccell{0.40} & \ccell{0.20} & \ccell{0.00} & \ccell{0.40} & \ccell{0.60} & \ccell{0.80} & \ccell{0.80} \\
\cmidrule(lr){2-19}
    & L1 & \ccell{0.60} & \ccell{0.60} & \ccell{0.80} & \ccell{0.00} & \ccell{0.40} & \ccell{0.60} & \ccell{0.20} & \ccell{0.20} & \ccell{0.20} & \ccell{0.40} & \ccell{0.80} & \ccell{0.00} & \ccell{0.20} & \ccell{0.40} & \ccell{0.60} & \ccell{0.80} & \ccell{0.80} \\
    & L2 & \ccell{0.20} & \ccell{0.40} & \ccell{0.40} & \ccell{0.20} & \ccell{0.40} & \ccell{0.60} & \ccell{0.80} & \ccell{0.40} & \ccell{0.20} & \ccell{0.20} & \ccell{0.60} & \ccell{0.20} & \ccell{0.40} & \ccell{0.80} & \ccell{0.60} & \ccell{0.80} & \ccell{0.80} \\
    & L3 & \ccell{0.20} & \ccell{0.40} & \ccell{0.60} & \ccell{0.20} & \ccell{0.20} & \ccell{0.40} & \ccell{0.80} & \ccell{0.40} & \ccell{0.40} & \ccell{0.20} & \ccell{0.80} & \ccell{0.40} & \ccell{0.40} & \ccell{0.80} & \ccell{0.20} & \ccell{0.80} & \ccell{0.80} \\
    & L456 & \ccell{0.40} & \ccell{0.40} & \ccell{0.60} & \ccell{0.20} & \ccell{0.20} & \ccell{0.40} & \ccell{0.60} & \ccell{0.40} & \ccell{0.20} & \ccell{0.00} & \ccell{0.80} & \ccell{0.40} & \ccell{0.60} & \ccell{0.60} & \ccell{0.60} & \ccell{0.80} & \ccell{1.00} \\
    & L7 & \ccell{1.00} & \ccell{0.20} & \ccell{0.80} & \ccell{0.00} & \ccell{0.20} & \ccell{0.20} & \ccell{0.40} & \ccell{0.60} & \ccell{0.40} & \ccell{0.20} & \ccell{0.60} & \ccell{0.60} & \ccell{0.20} & \ccell{0.80} & \ccell{0.80} & \ccell{0.60} & \ccell{0.80} \\
    & L8 & \ccell{0.20} & \ccell{0.20} & \ccell{0.40} & \ccell{0.20} & \ccell{0.40} & \ccell{0.20} & \ccell{0.40} & \ccell{0.20} & \ccell{0.60} & \ccell{0.00} & \ccell{0.20} & \ccell{0.20} & \ccell{0.20} & \ccell{0.40} & \ccell{0.60} & \ccell{0.80} & \ccell{0.80} \\
    & D123 & \ccell{0.20} & \ccell{0.60} & \ccell{0.80} & \ccell{0.00} & \ccell{0.40} & \ccell{0.20} & \ccell{0.40} & \ccell{0.60} & \ccell{0.00} & \ccell{0.40} & \ccell{0.60} & \ccell{0.20} & \ccell{0.40} & \ccell{0.40} & \ccell{0.60} & \ccell{0.80} & \ccell{0.80} \\
    & D4567 & \ccell{0.20} & \ccell{0.40} & \ccell{0.60} & \ccell{0.20} & \ccell{0.40} & \ccell{0.40} & \ccell{0.40} & \ccell{0.40} & \ccell{0.20} & \ccell{0.20} & \ccell{0.40} & \ccell{0.40} & \ccell{0.00} & \ccell{0.60} & \ccell{0.40} & \ccell{0.80} & \ccell{0.80} \\
    & D8 & \ccell{0.40} & \ccell{0.40} & \ccell{0.60} & \ccell{0.60} & \ccell{0.00} & \ccell{0.40} & \ccell{0.60} & \ccell{0.60} & \ccell{0.20} & \ccell{0.40} & \ccell{0.80} & \ccell{0.00} & \ccell{0.40} & \ccell{0.60} & \ccell{0.80} & \ccell{0.80} & \ccell{0.80} \\
    & C1 & \ccell{0.60} & \ccell{0.20} & \ccell{0.60} & \ccell{0.00} & \ccell{0.20} & \ccell{0.60} & \ccell{0.40} & \ccell{0.60} & \ccell{0.20} & \ccell{0.20} & \ccell{0.80} & \ccell{0.20} & \ccell{0.00} & \ccell{0.80} & \ccell{0.40} & \ccell{0.80} & \ccell{1.00} \\
    & C2 & \ccell{0.60} & \ccell{0.40} & \ccell{0.60} & \ccell{0.40} & \ccell{0.80} & \ccell{0.40} & \ccell{0.80} & \ccell{0.40} & \ccell{0.20} & \ccell{0.40} & \ccell{0.60} & \ccell{0.40} & \ccell{0.20} & \ccell{0.80} & \ccell{0.40} & \ccell{0.80} & \ccell{0.80} \\
    & C3 & \ccell{0.20} & \ccell{0.60} & \ccell{0.60} & \ccell{0.00} & \ccell{0.20} & \ccell{0.60} & \ccell{0.40} & \ccell{0.20} & \ccell{0.00} & \ccell{0.00} & \ccell{0.20} & \ccell{0.00} & \ccell{0.40} & \ccell{0.00} & \ccell{0.00} & \ccell{0.20} & \ccell{0.40} \\
 
\midrule
\multirow{13}{*}{\rotatebox{90}{\reposvulpy}}
    & original & \ccell{0.60} & \ccell{0.00} & \ccell{0.00} & \ccell{0.20} & \ccell{0.00} & \ccell{0.20} & \ccell{0.60} & \ccell{0.20} & \ccell{0.20} & \ccell{0.00} & \ccell{0.40} & \ccell{0.20} & \ccell{0.20} & \ccell{0.60} & \ccell{0.00} & \ccell{0.20} & \ccell{0.20} \\
\cmidrule(lr){2-19}
    & L1 & \ccell{0.60} & \ccell{0.40} & \ccell{0.60} & \ccell{0.20} & \ccell{0.60} & \ccell{0.40} & \ccell{1.00} & \ccell{0.40} & \ccell{0.60} & \ccell{0.60} & \ccell{0.80} & \ccell{0.20} & \ccell{0.20} & \ccell{0.60} & \ccell{0.60} & \ccell{0.60} & \ccell{0.40} \\
    & L2 & \ccell{0.40} & \ccell{0.60} & \ccell{0.60} & \ccell{0.20} & \ccell{0.20} & \ccell{0.20} & \ccell{0.80} & \ccell{0.60} & \ccell{0.60} & \ccell{0.20} & \ccell{0.60} & \ccell{0.20} & \ccell{0.20} & \ccell{0.60} & \ccell{0.00} & \ccell{0.60} & \ccell{0.40} \\
    & L3 & \ccell{0.40} & \ccell{0.20} & \ccell{0.40} & \ccell{0.40} & \ccell{0.20} & \ccell{0.20} & \ccell{0.80} & \ccell{0.40} & \ccell{0.80} & \ccell{0.20} & \ccell{0.60} & \ccell{0.00} & \ccell{0.80} & \ccell{0.40} & \ccell{0.00} & \ccell{0.60} & \ccell{0.00} \\
    & L456 & \ccell{0.20} & \ccell{0.60} & \ccell{0.20} & \ccell{0.40} & \ccell{0.00} & \ccell{0.20} & \ccell{0.40} & \ccell{0.40} & \ccell{0.60} & \ccell{0.20} & \ccell{0.60} & \ccell{0.20} & \ccell{0.20} & \ccell{0.40} & \ccell{0.20} & \ccell{0.20} & \ccell{0.40} \\
    & L7 & \ccell{0.00} & \ccell{0.20} & \ccell{0.40} & \ccell{0.20} & \ccell{0.40} & \ccell{0.00} & \ccell{0.60} & \ccell{0.80} & \ccell{0.80} & \ccell{0.60} & \ccell{0.60} & \ccell{0.40} & \ccell{0.60} & \ccell{0.20} & \ccell{0.20} & \ccell{0.20} & \ccell{0.00} \\
    & L8 & \ccell{0.40} & \ccell{0.20} & \ccell{0.00} & \ccell{0.40} & \ccell{0.00} & \ccell{0.20} & \ccell{0.60} & \ccell{0.20} & \ccell{0.60} & \ccell{0.20} & \ccell{0.20} & \ccell{0.20} & \ccell{0.20} & \ccell{0.20} & \ccell{0.00} & \ccell{0.40} & \ccell{0.40} \\
    & D123 & \ccell{0.00} & \ccell{0.40} & \ccell{0.60} & \ccell{0.20} & \ccell{0.20} & \ccell{0.40} & \ccell{0.60} & \ccell{0.80} & \ccell{0.60} & \ccell{0.20} & \ccell{0.40} & \ccell{0.80} & \ccell{0.60} & \ccell{0.60} & \ccell{0.00} & \ccell{0.60} & \ccell{0.20} \\
    & D4567 & \ccell{0.40} & \ccell{0.20} & \ccell{0.40} & \ccell{0.20} & \ccell{0.20} & \ccell{0.20} & \ccell{0.60} & \ccell{0.60} & \ccell{0.20} & \ccell{0.60} & \ccell{0.40} & \ccell{0.40} & \ccell{0.60} & \ccell{0.60} & \ccell{0.40} & \ccell{0.40} & \ccell{0.40} \\
    & D8 & \ccell{0.40} & \ccell{0.20} & \ccell{0.20} & \ccell{0.40} & \ccell{0.20} & \ccell{0.20} & \ccell{0.80} & \ccell{0.60} & \ccell{0.80} & \ccell{0.40} & \ccell{0.60} & \ccell{0.40} & \ccell{0.20} & \ccell{0.60} & \ccell{0.20} & \ccell{0.00} & \ccell{0.20} \\
    & C1 & \ccell{0.00} & \ccell{0.40} & \ccell{0.20} & \ccell{0.40} & \ccell{0.40} & \ccell{0.00} & \ccell{0.20} & \ccell{0.60} & \ccell{0.00} & \ccell{0.20} & \ccell{0.60} & \ccell{0.20} & \ccell{0.00} & \ccell{0.40} & \ccell{0.00} & \ccell{0.40} & \ccell{0.20} \\
    & C2 & \ccell{0.20} & \ccell{0.60} & \ccell{0.40} & \ccell{0.20} & \ccell{0.40} & \ccell{0.20} & \ccell{0.80} & \ccell{0.60} & \ccell{0.60} & \ccell{0.20} & \ccell{0.60} & \ccell{0.40} & \ccell{0.40} & \ccell{0.60} & \ccell{0.00} & \ccell{0.40} & \ccell{0.40} \\
    & C3 & \ccell{0.00} & \ccell{0.40} & \ccell{0.40} & \ccell{0.20} & \ccell{0.40} & \ccell{0.20} & \ccell{0.60} & \ccell{0.60} & \ccell{0.40} & \ccell{0.20} & \ccell{0.40} & \ccell{0.60} & \ccell{0.40} & \ccell{0.60} & \ccell{0.40} & \ccell{0.60} & \ccell{0.20} \\
 
\midrule
\multirow{13}{*}{\rotatebox{90}{\primevulc}}
    & original & \ccell{0.20} & \ccell{0.00} & \ccell{0.20} & \ccell{0.20} & \ccell{0.00} & \ccell{0.20} & \ccell{0.40} & \ccell{0.60} & \ccell{0.60} & \ccell{0.20} & \ccell{0.60} & \ccell{0.40} & \ccell{0.00} & \ccell{0.60} & \ccell{0.20} & \ccell{0.60} & \ccell{0.80} \\
\cmidrule(lr){2-19}
    & L1 & \ccell{0.40} & \ccell{0.80} & \ccell{0.80} & \ccell{0.20} & \ccell{0.40} & \ccell{0.20} & \ccell{0.60} & \ccell{0.20} & \ccell{0.40} & \ccell{0.40} & \ccell{0.00} & \ccell{0.80} & \ccell{0.00} & \ccell{0.20} & \ccell{0.60} & \ccell{1.00} & \ccell{0.80} \\
    & L2 & \ccell{0.20} & \ccell{0.60} & \ccell{0.80} & \ccell{0.20} & \ccell{0.20} & \ccell{0.20} & \ccell{0.40} & \ccell{0.60} & \ccell{0.40} & \ccell{0.20} & \ccell{0.60} & \ccell{0.40} & \ccell{0.60} & \ccell{0.60} & \ccell{0.60} & \ccell{0.80} & \ccell{0.60} \\
    & L3 & \ccell{0.20} & \ccell{0.80} & \ccell{0.60} & \ccell{0.40} & \ccell{0.20} & \ccell{0.00} & \ccell{0.40} & \ccell{0.80} & \ccell{0.40} & \ccell{0.40} & \ccell{0.60} & \ccell{0.40} & \ccell{0.00} & \ccell{0.40} & \ccell{0.80} & \ccell{1.00} & \ccell{0.40} \\
    & L456 & \ccell{0.20} & \ccell{0.40} & \ccell{0.80} & \ccell{0.40} & \ccell{0.20} & \ccell{0.20} & \ccell{0.40} & \ccell{0.60} & \ccell{0.40} & \ccell{0.20} & \ccell{0.40} & \ccell{0.40} & \ccell{0.20} & \ccell{0.20} & \ccell{0.60} & \ccell{0.80} & \ccell{0.80} \\
    & L7 & \ccell{0.20} & \ccell{0.60} & \ccell{0.80} & \ccell{0.20} & \ccell{0.60} & \ccell{0.40} & \ccell{0.60} & \ccell{0.60} & \ccell{0.60} & \ccell{0.40} & \ccell{1.00} & \ccell{0.00} & \ccell{0.60} & \ccell{0.60} & \ccell{0.60} & \ccell{0.60} & \ccell{0.60} \\
    & L8 & \ccell{0.40} & \ccell{0.60} & \ccell{0.80} & \ccell{0.20} & \ccell{0.40} & \ccell{0.20} & \ccell{0.40} & \ccell{0.80} & \ccell{0.80} & \ccell{0.60} & \ccell{0.40} & \ccell{0.00} & \ccell{0.00} & \ccell{0.40} & \ccell{0.20} & \ccell{0.80} & \ccell{0.60} \\
    & D123 & \ccell{0.20} & \ccell{0.40} & \ccell{0.40} & \ccell{0.00} & \ccell{0.20} & \ccell{0.40} & \ccell{0.40} & \ccell{0.80} & \ccell{0.20} & \ccell{0.40} & \ccell{0.00} & \ccell{0.60} & \ccell{0.20} & \ccell{0.20} & \ccell{0.80} & \ccell{0.80} & \ccell{1.00} \\
    & D4567 & \ccell{0.20} & \ccell{0.80} & \ccell{0.60} & \ccell{0.20} & \ccell{0.40} & \ccell{0.40} & \ccell{0.40} & \ccell{0.40} & \ccell{0.20} & \ccell{0.20} & \ccell{0.80} & \ccell{0.40} & \ccell{0.40} & \ccell{0.40} & \ccell{0.80} & \ccell{1.00} & \ccell{0.80} \\
    & D8 & \ccell{0.40} & \ccell{0.80} & \ccell{0.80} & \ccell{0.00} & \ccell{0.20} & \ccell{0.20} & \ccell{0.60} & \ccell{0.40} & \ccell{0.20} & \ccell{0.60} & \ccell{0.40} & \ccell{0.40} & \ccell{0.40} & \ccell{0.60} & \ccell{0.80} & \ccell{0.80} & \ccell{0.80} \\
    & C1 & \ccell{0.40} & \ccell{0.40} & \ccell{0.60} & \ccell{0.20} & \ccell{0.40} & \ccell{0.00} & \ccell{0.60} & \ccell{0.60} & \ccell{0.20} & \ccell{0.80} & \ccell{0.40} & \ccell{0.60} & \ccell{0.00} & \ccell{0.40} & \ccell{0.40} & \ccell{0.80} & \ccell{0.80} \\
    & C2 & \ccell{0.20} & \ccell{0.40} & \ccell{0.60} & \ccell{0.40} & \ccell{0.20} & \ccell{0.20} & \ccell{0.60} & \ccell{0.40} & \ccell{0.20} & \ccell{0.40} & \ccell{1.00} & \ccell{0.20} & \ccell{0.20} & \ccell{0.60} & \ccell{0.60} & \ccell{1.00} & \ccell{1.00} \\
    & C3 & \ccell{0.00} & \ccell{0.20} & \ccell{0.20} & \ccell{0.00} & \ccell{0.00} & \ccell{0.00} & \ccell{0.40} & \ccell{0.00} & \ccell{0.40} & \ccell{0.00} & \ccell{0.00} & \ccell{0.20} & \ccell{0.20} & \ccell{0.40} & \ccell{0.20} & \ccell{0.80} & \ccell{0.60} \\
 
\bottomrule
\end{tabular}
}
\label{tab:detection-result-all-datasets-upgrade_top20}
\end{table*}

\begin{table*}[htbp]
\centering
\caption{Upgrade rate on different datasets: \smartbugs, \reposvulcpp, \reposvulpy, \primevulc (sampled by upgrade top20). (L456 = L4+L5+L6, D123 = D1+D2+D3, D4567 = D4+D5+D6+D7 for layout clarity)}
{
\tiny
\setlength\tabcolsep{3pt}
\renewcommand{\arraystretch}{0.6}
\begin{tabular}{l lccccccccccccccccc}
\toprule
dataset & series & \multicolumn{6}{c}{qwen} & \multicolumn{4}{c}{llama} & \multicolumn{2}{c}{deepseek} & \multicolumn{3}{c}{openai} & \multicolumn{2}{c}{agent} \\
\cmidrule(lr){3-8} \cmidrule(lr){9-12} \cmidrule(lr){13-14} \cmidrule(lr){15-17} \cmidrule(lr){18-19}
 & model & qn-7b & qn-14b & qn-32b & ds-qn-7b & ds-qn-14b & ds-qn-32b & lm-8b & lm-70b & ds-lm-8b & ds-lm-70b & ds-v3 & ds-r1 & gpt-3.5 & gpt-4o & o3-mini & copilot & codex \\
\midrule
\multirow{12}{*}{\rotatebox{90}{\smartbugs}}
    & L1 & \ccell{0.20} & \ccell{0.20} & \ccell{0.00} & \ccell{0.40} & \ccell{0.20} & \ccell{0.20} & \ccell{0.20} & \ccell{0.20} & \ccell{0.00} & \ccell{0.00} & \ccell{0.20} & \ccell{0.20} & \ccell{0.60} & \ccell{0.40} & \ccell{0.20} & \ccell{0.00} & \ccell{0.40} \\
    & L2 & \ccell{0.20} & \ccell{0.00} & \ccell{0.20} & \ccell{0.20} & \ccell{0.20} & \ccell{0.40} & \ccell{0.40} & \ccell{0.20} & \ccell{0.60} & \ccell{0.20} & \ccell{0.00} & \ccell{0.40} & \ccell{0.20} & \ccell{0.40} & \ccell{0.00} & \ccell{0.00} & \ccell{0.20} \\
    & L3 & \ccell{0.00} & \ccell{0.20} & \ccell{0.00} & \ccell{0.20} & \ccell{0.20} & \ccell{0.40} & \ccell{0.60} & \ccell{0.60} & \ccell{0.00} & \ccell{0.20} & \ccell{0.00} & \ccell{0.40} & \ccell{0.00} & \ccell{0.40} & \ccell{0.20} & \ccell{0.00} & \ccell{0.40} \\
    & L456 & \ccell{0.00} & \ccell{0.20} & \ccell{0.20} & \ccell{0.20} & \ccell{0.20} & \ccell{0.20} & \ccell{0.40} & \ccell{0.20} & \ccell{0.20} & \ccell{0.40} & \ccell{0.00} & \ccell{0.40} & \ccell{0.00} & \ccell{0.20} & \ccell{0.00} & \ccell{0.00} & \ccell{0.20} \\
    & L7 & \ccell{0.00} & \ccell{0.20} & \ccell{0.00} & \ccell{0.20} & \ccell{0.20} & \ccell{0.40} & \ccell{0.40} & \ccell{0.40} & \ccell{0.40} & \ccell{0.20} & \ccell{0.00} & \ccell{0.40} & \ccell{0.40} & \ccell{0.40} & \ccell{0.00} & \ccell{0.00} & \ccell{0.20} \\
    & L8 & \ccell{0.20} & \ccell{0.00} & \ccell{0.20} & \ccell{0.20} & \ccell{0.40} & \ccell{0.20} & \ccell{0.20} & \ccell{0.40} & \ccell{0.00} & \ccell{0.20} & \ccell{0.20} & \ccell{0.40} & \ccell{0.80} & \ccell{0.40} & \ccell{0.00} & \ccell{0.00} & \ccell{0.40} \\
    & D123 & \ccell{0.20} & \ccell{0.00} & \ccell{0.40} & \ccell{0.40} & \ccell{0.60} & \ccell{0.40} & \ccell{0.20} & \ccell{0.20} & \ccell{0.20} & \ccell{0.40} & \ccell{0.40} & \ccell{0.40} & \ccell{0.40} & \ccell{0.40} & \ccell{0.20} & \ccell{0.00} & \ccell{0.20} \\
    & D4567 & \ccell{0.20} & \ccell{0.00} & \ccell{0.60} & \ccell{0.20} & \ccell{0.20} & \ccell{0.20} & \ccell{0.60} & \ccell{0.80} & \ccell{0.40} & \ccell{0.20} & \ccell{0.00} & \ccell{0.20} & \ccell{0.40} & \ccell{0.40} & \ccell{0.00} & \ccell{0.00} & \ccell{0.40} \\
    & D8 & \ccell{0.20} & \ccell{0.40} & \ccell{0.40} & \ccell{0.20} & \ccell{0.20} & \ccell{0.40} & \ccell{0.80} & \ccell{0.20} & \ccell{0.20} & \ccell{0.20} & \ccell{0.00} & \ccell{0.40} & \ccell{0.20} & \ccell{0.20} & \ccell{0.20} & \ccell{0.00} & \ccell{0.20} \\
    & C1 & \ccell{0.00} & \ccell{0.20} & \ccell{0.20} & \ccell{0.20} & \ccell{0.00} & \ccell{0.00} & \ccell{0.20} & \ccell{0.20} & \ccell{0.00} & \ccell{0.20} & \ccell{0.20} & \ccell{0.40} & \ccell{0.00} & \ccell{0.40} & \ccell{0.20} & \ccell{0.00} & \ccell{0.40} \\
    & C2 & \ccell{0.00} & \ccell{0.40} & \ccell{0.20} & \ccell{0.00} & \ccell{0.00} & \ccell{0.20} & \ccell{0.80} & \ccell{0.40} & \ccell{0.40} & \ccell{0.60} & \ccell{0.20} & \ccell{0.40} & \ccell{0.00} & \ccell{0.20} & \ccell{0.20} & \ccell{0.00} & \ccell{0.20} \\
    & C3 & \ccell{0.20} & \ccell{0.40} & \ccell{0.20} & \ccell{0.20} & \ccell{0.20} & \ccell{0.20} & \ccell{0.20} & \ccell{0.40} & \ccell{0.00} & \ccell{0.60} & \ccell{0.40} & \ccell{0.40} & \ccell{0.20} & \ccell{0.40} & \ccell{0.20} & \ccell{0.00} & \ccell{0.40} \\
 
\midrule
\multirow{12}{*}{\rotatebox{90}{\reposvulcpp}}
    & L1 & \ccell{0.20} & \ccell{0.40} & \ccell{0.60} & \ccell{0.00} & \ccell{0.40} & \ccell{0.40} & \ccell{0.00} & \ccell{0.20} & \ccell{0.20} & \ccell{0.20} & \ccell{0.60} & \ccell{0.00} & \ccell{0.20} & \ccell{0.20} & \ccell{0.20} & \ccell{0.20} & \ccell{0.00} \\
    & L2 & \ccell{0.20} & \ccell{0.40} & \ccell{0.40} & \ccell{0.20} & \ccell{0.20} & \ccell{0.40} & \ccell{0.40} & \ccell{0.20} & \ccell{0.20} & \ccell{0.00} & \ccell{0.20} & \ccell{0.00} & \ccell{0.40} & \ccell{0.40} & \ccell{0.20} & \ccell{0.20} & \ccell{0.00} \\
    & L3 & \ccell{0.20} & \ccell{0.40} & \ccell{0.40} & \ccell{0.20} & \ccell{0.20} & \ccell{0.40} & \ccell{0.40} & \ccell{0.20} & \ccell{0.20} & \ccell{0.00} & \ccell{0.40} & \ccell{0.40} & \ccell{0.40} & \ccell{0.40} & \ccell{0.00} & \ccell{0.20} & \ccell{0.00} \\
    & L456 & \ccell{0.20} & \ccell{0.20} & \ccell{0.60} & \ccell{0.20} & \ccell{0.20} & \ccell{0.20} & \ccell{0.20} & \ccell{0.20} & \ccell{0.20} & \ccell{0.00} & \ccell{0.40} & \ccell{0.20} & \ccell{0.60} & \ccell{0.40} & \ccell{0.20} & \ccell{0.20} & \ccell{0.20} \\
    & L7 & \ccell{0.60} & \ccell{0.20} & \ccell{0.60} & \ccell{0.00} & \ccell{0.20} & \ccell{0.00} & \ccell{0.00} & \ccell{0.40} & \ccell{0.40} & \ccell{0.00} & \ccell{0.20} & \ccell{0.40} & \ccell{0.20} & \ccell{0.60} & \ccell{0.20} & \ccell{0.20} & \ccell{0.00} \\
    & L8 & \ccell{0.20} & \ccell{0.20} & \ccell{0.20} & \ccell{0.20} & \ccell{0.40} & \ccell{0.00} & \ccell{0.20} & \ccell{0.20} & \ccell{0.40} & \ccell{0.00} & \ccell{0.00} & \ccell{0.00} & \ccell{0.20} & \ccell{0.20} & \ccell{0.00} & \ccell{0.20} & \ccell{0.20} \\
    & D123 & \ccell{0.20} & \ccell{0.40} & \ccell{0.60} & \ccell{0.00} & \ccell{0.40} & \ccell{0.00} & \ccell{0.00} & \ccell{0.20} & \ccell{0.00} & \ccell{0.00} & \ccell{0.20} & \ccell{0.20} & \ccell{0.40} & \ccell{0.00} & \ccell{0.20} & \ccell{0.20} & \ccell{0.00} \\
    & D4567 & \ccell{0.20} & \ccell{0.20} & \ccell{0.40} & \ccell{0.20} & \ccell{0.20} & \ccell{0.20} & \ccell{0.20} & \ccell{0.20} & \ccell{0.20} & \ccell{0.00} & \ccell{0.20} & \ccell{0.20} & \ccell{0.00} & \ccell{0.20} & \ccell{0.20} & \ccell{0.20} & \ccell{0.00} \\
    & D8 & \ccell{0.00} & \ccell{0.40} & \ccell{0.40} & \ccell{0.60} & \ccell{0.00} & \ccell{0.20} & \ccell{0.00} & \ccell{0.20} & \ccell{0.00} & \ccell{0.20} & \ccell{0.40} & \ccell{0.00} & \ccell{0.40} & \ccell{0.20} & \ccell{0.20} & \ccell{0.20} & \ccell{0.00} \\
    & C1 & \ccell{0.40} & \ccell{0.20} & \ccell{0.40} & \ccell{0.00} & \ccell{0.20} & \ccell{0.40} & \ccell{0.20} & \ccell{0.20} & \ccell{0.20} & \ccell{0.00} & \ccell{0.40} & \ccell{0.00} & \ccell{0.00} & \ccell{0.40} & \ccell{0.20} & \ccell{0.20} & \ccell{0.20} \\
    & C2 & \ccell{0.20} & \ccell{0.20} & \ccell{0.40} & \ccell{0.40} & \ccell{0.60} & \ccell{0.20} & \ccell{0.40} & \ccell{0.20} & \ccell{0.20} & \ccell{0.20} & \ccell{0.40} & \ccell{0.20} & \ccell{0.20} & \ccell{0.40} & \ccell{0.20} & \ccell{0.20} & \ccell{0.00} \\
    & C3 & \ccell{0.20} & \ccell{0.60} & \ccell{0.40} & \ccell{0.00} & \ccell{0.20} & \ccell{0.40} & \ccell{0.20} & \ccell{0.20} & \ccell{0.00} & \ccell{0.00} & \ccell{0.20} & \ccell{0.00} & \ccell{0.40} & \ccell{0.00} & \ccell{0.00} & \ccell{0.20} & \ccell{0.00} \\
 
\midrule
\multirow{12}{*}{\rotatebox{90}{\reposvulpy}}
    & L1 & \ccell{0.40} & \ccell{0.40} & \ccell{0.60} & \ccell{0.00} & \ccell{0.60} & \ccell{0.20} & \ccell{0.40} & \ccell{0.20} & \ccell{0.40} & \ccell{0.60} & \ccell{0.40} & \ccell{0.00} & \ccell{0.00} & \ccell{0.20} & \ccell{0.60} & \ccell{0.60} & \ccell{0.20} \\
    & L2 & \ccell{0.20} & \ccell{0.60} & \ccell{0.60} & \ccell{0.00} & \ccell{0.20} & \ccell{0.00} & \ccell{0.20} & \ccell{0.40} & \ccell{0.40} & \ccell{0.20} & \ccell{0.20} & \ccell{0.00} & \ccell{0.00} & \ccell{0.00} & \ccell{0.00} & \ccell{0.40} & \ccell{0.20} \\
    & L3 & \ccell{0.00} & \ccell{0.20} & \ccell{0.40} & \ccell{0.40} & \ccell{0.20} & \ccell{0.20} & \ccell{0.20} & \ccell{0.20} & \ccell{0.60} & \ccell{0.20} & \ccell{0.20} & \ccell{0.00} & \ccell{0.60} & \ccell{0.00} & \ccell{0.00} & \ccell{0.40} & \ccell{0.00} \\
    & L456 & \ccell{0.00} & \ccell{0.60} & \ccell{0.20} & \ccell{0.20} & \ccell{0.00} & \ccell{0.20} & \ccell{0.00} & \ccell{0.20} & \ccell{0.40} & \ccell{0.20} & \ccell{0.40} & \ccell{0.00} & \ccell{0.00} & \ccell{0.00} & \ccell{0.20} & \ccell{0.20} & \ccell{0.20} \\
    & L7 & \ccell{0.00} & \ccell{0.20} & \ccell{0.40} & \ccell{0.00} & \ccell{0.40} & \ccell{0.00} & \ccell{0.20} & \ccell{0.60} & \ccell{0.60} & \ccell{0.60} & \ccell{0.20} & \ccell{0.20} & \ccell{0.40} & \ccell{0.00} & \ccell{0.20} & \ccell{0.20} & \ccell{0.00} \\
    & L8 & \ccell{0.00} & \ccell{0.20} & \ccell{0.00} & \ccell{0.00} & \ccell{0.00} & \ccell{0.20} & \ccell{0.20} & \ccell{0.00} & \ccell{0.40} & \ccell{0.20} & \ccell{0.00} & \ccell{0.20} & \ccell{0.20} & \ccell{0.00} & \ccell{0.00} & \ccell{0.40} & \ccell{0.20} \\
    & D123 & \ccell{0.00} & \ccell{0.40} & \ccell{0.60} & \ccell{0.00} & \ccell{0.20} & \ccell{0.20} & \ccell{0.00} & \ccell{0.60} & \ccell{0.40} & \ccell{0.20} & \ccell{0.00} & \ccell{0.60} & \ccell{0.40} & \ccell{0.20} & \ccell{0.00} & \ccell{0.60} & \ccell{0.20} \\
    & D4567 & \ccell{0.00} & \ccell{0.20} & \ccell{0.40} & \ccell{0.20} & \ccell{0.20} & \ccell{0.00} & \ccell{0.00} & \ccell{0.40} & \ccell{0.00} & \ccell{0.60} & \ccell{0.00} & \ccell{0.40} & \ccell{0.40} & \ccell{0.20} & \ccell{0.40} & \ccell{0.20} & \ccell{0.20} \\
    & D8 & \ccell{0.20} & \ccell{0.20} & \ccell{0.20} & \ccell{0.20} & \ccell{0.20} & \ccell{0.20} & \ccell{0.40} & \ccell{0.40} & \ccell{0.60} & \ccell{0.40} & \ccell{0.20} & \ccell{0.20} & \ccell{0.20} & \ccell{0.00} & \ccell{0.20} & \ccell{0.00} & \ccell{0.00} \\
    & C1 & \ccell{0.00} & \ccell{0.40} & \ccell{0.20} & \ccell{0.20} & \ccell{0.40} & \ccell{0.00} & \ccell{0.00} & \ccell{0.40} & \ccell{0.00} & \ccell{0.20} & \ccell{0.20} & \ccell{0.00} & \ccell{0.00} & \ccell{0.00} & \ccell{0.00} & \ccell{0.20} & \ccell{0.00} \\
    & C2 & \ccell{0.00} & \ccell{0.60} & \ccell{0.40} & \ccell{0.00} & \ccell{0.40} & \ccell{0.20} & \ccell{0.20} & \ccell{0.40} & \ccell{0.40} & \ccell{0.20} & \ccell{0.20} & \ccell{0.20} & \ccell{0.20} & \ccell{0.00} & \ccell{0.00} & \ccell{0.40} & \ccell{0.20} \\
    & C3 & \ccell{0.00} & \ccell{0.40} & \ccell{0.40} & \ccell{0.00} & \ccell{0.40} & \ccell{0.20} & \ccell{0.20} & \ccell{0.40} & \ccell{0.20} & \ccell{0.20} & \ccell{0.00} & \ccell{0.40} & \ccell{0.40} & \ccell{0.20} & \ccell{0.40} & \ccell{0.40} & \ccell{0.20} \\
 
\midrule
\multirow{12}{*}{\rotatebox{90}{\primevulc}}
    & L1 & \ccell{0.40} & \ccell{0.80} & \ccell{0.60} & \ccell{0.20} & \ccell{0.40} & \ccell{0.20} & \ccell{0.20} & \ccell{0.20} & \ccell{0.20} & \ccell{0.20} & \ccell{0.00} & \ccell{0.40} & \ccell{0.00} & \ccell{0.00} & \ccell{0.40} & \ccell{0.40} & \ccell{0.20} \\
    & L2 & \ccell{0.20} & \ccell{0.60} & \ccell{0.60} & \ccell{0.00} & \ccell{0.20} & \ccell{0.20} & \ccell{0.00} & \ccell{0.20} & \ccell{0.20} & \ccell{0.20} & \ccell{0.00} & \ccell{0.20} & \ccell{0.60} & \ccell{0.00} & \ccell{0.40} & \ccell{0.20} & \ccell{0.00} \\
    & L3 & \ccell{0.20} & \ccell{0.80} & \ccell{0.60} & \ccell{0.40} & \ccell{0.20} & \ccell{0.00} & \ccell{0.00} & \ccell{0.20} & \ccell{0.00} & \ccell{0.20} & \ccell{0.20} & \ccell{0.20} & \ccell{0.00} & \ccell{0.00} & \ccell{0.60} & \ccell{0.40} & \ccell{0.00} \\
    & L456 & \ccell{0.20} & \ccell{0.40} & \ccell{0.60} & \ccell{0.40} & \ccell{0.20} & \ccell{0.00} & \ccell{0.00} & \ccell{0.20} & \ccell{0.00} & \ccell{0.00} & \ccell{0.00} & \ccell{0.40} & \ccell{0.20} & \ccell{0.00} & \ccell{0.60} & \ccell{0.20} & \ccell{0.00} \\
    & L7 & \ccell{0.20} & \ccell{0.60} & \ccell{0.60} & \ccell{0.20} & \ccell{0.60} & \ccell{0.40} & \ccell{0.20} & \ccell{0.20} & \ccell{0.20} & \ccell{0.40} & \ccell{0.40} & \ccell{0.00} & \ccell{0.60} & \ccell{0.00} & \ccell{0.40} & \ccell{0.00} & \ccell{0.00} \\
    & L8 & \ccell{0.40} & \ccell{0.60} & \ccell{0.80} & \ccell{0.20} & \ccell{0.40} & \ccell{0.00} & \ccell{0.20} & \ccell{0.20} & \ccell{0.40} & \ccell{0.40} & \ccell{0.20} & \ccell{0.00} & \ccell{0.00} & \ccell{0.00} & \ccell{0.20} & \ccell{0.20} & \ccell{0.00} \\
    & D123 & \ccell{0.20} & \ccell{0.40} & \ccell{0.20} & \ccell{0.00} & \ccell{0.20} & \ccell{0.20} & \ccell{0.00} & \ccell{0.40} & \ccell{0.20} & \ccell{0.20} & \ccell{0.00} & \ccell{0.60} & \ccell{0.20} & \ccell{0.00} & \ccell{0.60} & \ccell{0.20} & \ccell{0.20} \\
    & D4567 & \ccell{0.00} & \ccell{0.80} & \ccell{0.40} & \ccell{0.20} & \ccell{0.40} & \ccell{0.20} & \ccell{0.00} & \ccell{0.00} & \ccell{0.20} & \ccell{0.00} & \ccell{0.20} & \ccell{0.20} & \ccell{0.40} & \ccell{0.00} & \ccell{0.60} & \ccell{0.40} & \ccell{0.00} \\
    & D8 & \ccell{0.40} & \ccell{0.80} & \ccell{0.60} & \ccell{0.00} & \ccell{0.20} & \ccell{0.00} & \ccell{0.20} & \ccell{0.00} & \ccell{0.00} & \ccell{0.40} & \ccell{0.00} & \ccell{0.20} & \ccell{0.40} & \ccell{0.20} & \ccell{0.60} & \ccell{0.20} & \ccell{0.00} \\
    & C1 & \ccell{0.20} & \ccell{0.40} & \ccell{0.40} & \ccell{0.20} & \ccell{0.40} & \ccell{0.00} & \ccell{0.20} & \ccell{0.20} & \ccell{0.00} & \ccell{0.60} & \ccell{0.00} & \ccell{0.40} & \ccell{0.00} & \ccell{0.00} & \ccell{0.20} & \ccell{0.20} & \ccell{0.20} \\
    & C2 & \ccell{0.20} & \ccell{0.40} & \ccell{0.60} & \ccell{0.40} & \ccell{0.20} & \ccell{0.20} & \ccell{0.20} & \ccell{0.00} & \ccell{0.00} & \ccell{0.20} & \ccell{0.40} & \ccell{0.00} & \ccell{0.20} & \ccell{0.00} & \ccell{0.40} & \ccell{0.40} & \ccell{0.20} \\
    & C3 & \ccell{0.00} & \ccell{0.20} & \ccell{0.00} & \ccell{0.00} & \ccell{0.00} & \ccell{0.00} & \ccell{0.20} & \ccell{0.00} & \ccell{0.00} & \ccell{0.00} & \ccell{0.00} & \ccell{0.20} & \ccell{0.20} & \ccell{0.00} & \ccell{0.00} & \ccell{0.40} & \ccell{0.20} \\
 
\bottomrule
\end{tabular}
}
\label{tab:upgrade-rate-all-datasets-upgrade_top20}
\end{table*}

\section{Prompt Template for Code Obfuscation}\label{sec:prompt-for-obfuscation}

In our experiment, we utilized an LLM-driven approach to implement the code obfuscator. \autoref{fig:prompt-for-obfuscation} illustrates the prompt used. The contents of the \texttt{instruction\_block} field in the user prompt can be found in \autoref{tab:combo-instruction}.

\begin{figure}[ht]
    \centering
    
    \begin{tcolorbox}[title=\small{\textbf{System prompt}}, colframe=lightblue, colback=white]
    
    \small{You are an expert code obfuscater. Your task is to rewrite the given code by the given instructions, to make it less readable while preserving its functionality. Maintain code correctness and ensure the code can still be compiled and run correctly. Do not add any comments indicating the obfuscation process.}
    
    \end{tcolorbox}

    \begin{tcolorbox}[title=\small{\textbf{User prompt}}, colframe=lightblue, colback=white]
    
    \small{Modify the given code by the following instructions:}

    \small{\textcolor{blue}{\{instruction\_block\}}}
 
    \small{Code:}
    \small{\textcolor{blue}{\{code\}}}

    \small{Your answer must contain only the modified code! Do not explan anything extra! Wrap the output code with \texttt{```}language \texttt{```}, where language is the programming language of the code, like python, c, c++, solidity, etc.}
    
    \end{tcolorbox}
    \caption{Prompt template for code obfuscation.}
    \label{fig:prompt-for-obfuscation}
\end{figure}

\begin{table*}[htbp]
\caption{Mapping between obfuscation combos and instructions.}
\centering
\small
\setlength\tabcolsep{3pt}
\renewcommand{\arraystretch}{0.3}
\begin{tabularx}{\textwidth}{p{0.05\linewidth} p{0.28\linewidth} X}
\toprule
\textbf{Combo} & \textbf{Name} & \textbf{Instruction} \\
\midrule
\texttt{L1} & Identifier Deletion & Delete names of identifiers in the code, including function names, variable names, class names, and method names. Replace them with hashed names, like \texttt{OX7B4DF339}. \\

\texttt{L2} & Comment Deletion & Delete all comments in the code, including docstrings. \\

\texttt{L3} & Formatting Obfuscation & Change code formatting to make the code less readable, e.g., remove or add whitespaces, indentation, line breaks, and line continuations. \\

\texttt{L4} & For-to-While Transformation & Replace for-loops with while-loops or do-while-loops, or replace do-while-loops/while-loops with for-loops. \\

\texttt{L5} & If-to-Switch Transformation & Replace if-else with switch in C++; replace case with if-else in Python. \\

\texttt{L6} & Loop-to-Recursion Transformation & Replace all loops with recursion if possible in C++/Python/Solidity. \\

\texttt{L7} & Mix-Language with Inline Assembly & Replace single programming language with mixed languages, e.g., use C/C++ and inline assembly, or Solidity with inline assembly. \\

\texttt{L8} & Mix-Language with External Calls & Replace single programming language with mixed languages, e.g., use Python with C/C++ by loading C DLL. \\

\texttt{D1} & Arithmetic Substitution & Replace arithmetic constants with equivalent expressions, e.g., \texttt{a=1} $\to$ \texttt{a=(999-900)/99+0*250}. \\

\texttt{D2} & Boolean Substitution & Replace boolean constants with equivalent expressions, e.g., \texttt{a=True} $\to$ \texttt{a=(1==2)||(not False||True||1==1)}. \\

\texttt{D3} & String Substitution & Replace string constants with equivalent concatenations, e.g., \texttt{a='hello'} $\to$ \texttt{a='h'+'e'+'llo'}. \\

\texttt{D4} & Data Aggregation & Aggregate scalars into vectors/structs, e.g., \texttt{a=1,b=2,c=3} $\to$ \texttt{v=[1,2,3]}. \\

\texttt{D5} & Data Splitting & Split vectors/structs into scalars, e.g., \texttt{v=[1,2,3]} $\to$ \texttt{a=v[0], b=v[1], c=v[2]}. \\

\texttt{D6} & Change Member Order & Change the order of struct members, e.g., \texttt{struct S\{int a; bool b;\}} $\to$ \texttt{struct S\{bool b; int a;\}}. \\

\texttt{D7} & Change Variable Scope & Change scope of variables, e.g., block $\to$ local, local $\to$ global. \\

\texttt{D8} & Static-to-Dynamic Access & Replace static variable access with dynamic retrieval, e.g., \texttt{a=1} $\to$ \texttt{a=getValueA()}. \\

\texttt{C1} & Add Opaque Predicates & Insert opaque predicates and junk code into control flow, using meaningful names without indicating junk. \\

\texttt{C2} & Control Flow Flattening & Apply control flow flattening, encapsulating blocks in a dispatcher-controlled loop with switch-like/if-else jumps. \\

\texttt{C3} & Virtualization & Rewrite code using a stack-based virtual machine, replacing original logic with custom bytecode executed by a VM interpreter. \\

\bottomrule
\end{tabularx}

\label{tab:combo-instruction}
\end{table*}

\section{Detection Granularity at Existence Level}

For RQ3 in \S\ref{sec:experiment}, we adopt the binary existence setting from prior work, treating scores 2--4 as positive and only 1 as negative. As shown in \autoref{tab:detection-result-all-datasets-existence-evaluation}, this leads to a large increase in detection rate, but the gain is superficial and does not reflect the true LLM vulnerability-detection and anti-obfuscation capability.

\begin{table*}[b]
\centering
\caption{Detection successful rate across datasets. LLM Score 2,3,4 = Positive, Score 1 = Negative.}
\vspace{-1mm}
{
\tiny
\setlength\tabcolsep{3pt}
\renewcommand{\arraystretch}{0.6}
\begin{tabular}{l lccccccccccccccc}
\toprule
dataset & series & \multicolumn{6}{c}{qwen} & \multicolumn{4}{c}{llama} & \multicolumn{2}{c}{deepseek} & \multicolumn{3}{c}{openai} \\
\cmidrule(lr){3-8} \cmidrule(lr){9-12} \cmidrule(lr){13-14} \cmidrule(lr){15-17}
 & model & qn-7b & qn-14b & qn-32b & ds-qn-7b & ds-qn-14b & ds-qn-32b & lm-8b & lm-70b & ds-lm-8b & ds-lm-70b & ds-v3 & ds-r1 & gpt-3.5 & gpt-4o & o3-mini \\
\midrule
\multirow{13}{*}{\rotatebox{90}{\textbf{\smartbugs}}}
    & original & \ccell{0.94} & \ccell{0.98} & \ccell{1.00} & \ccell{0.98} & \ccell{0.95} & \ccell{0.95} & \ccell{1.00} & \ccell{1.00} & \ccell{0.96} & \ccell{1.00} & \ccell{1.00} & \ccell{1.00} & \ccell{0.99} & \ccell{1.00} & \ccell{0.98} \\
\cmidrule(lr){2-17}
    & L1 & \ccell{0.91} & \ccell{0.94} & \ccell{0.99} & \ccell{0.89} & \ccell{0.95} & \ccell{0.91} & \ccell{1.00} & \ccell{0.99} & \ccell{0.91} & \ccell{0.98} & \ccell{0.96} & \ccell{0.95} & \ccell{0.99} & \ccell{1.00} & \ccell{0.98} \\
    & L2 & \ccell{0.92} & \ccell{0.97} & \ccell{1.00} & \ccell{0.97} & \ccell{0.94} & \ccell{0.98} & \ccell{1.00} & \ccell{1.00} & \ccell{0.95} & \ccell{0.98} & \ccell{1.00} & \ccell{0.98} & \ccell{0.98} & \ccell{1.00} & \ccell{0.98} \\
    & L3 & \ccell{0.93} & \ccell{0.99} & \ccell{1.00} & \ccell{0.92} & \ccell{0.98} & \ccell{0.97} & \ccell{0.99} & \ccell{1.00} & \ccell{0.95} & \ccell{0.96} & \ccell{1.00} & \ccell{0.95} & \ccell{1.00} & \ccell{1.00} & \ccell{0.99} \\
    & L4+L5+L6 & \ccell{0.91} & \ccell{0.97} & \ccell{0.99} & \ccell{0.94} & \ccell{0.94} & \ccell{0.95} & \ccell{0.99} & \ccell{1.00} & \ccell{0.96} & \ccell{0.98} & \ccell{0.99} & \ccell{0.97} & \ccell{0.99} & \ccell{1.00} & \ccell{0.98} \\
    & L7 & \ccell{0.93} & \ccell{0.98} & \ccell{1.00} & \ccell{0.98} & \ccell{0.96} & \ccell{0.97} & \ccell{1.00} & \ccell{1.00} & \ccell{0.95} & \ccell{0.99} & \ccell{1.00} & \ccell{0.98} & \ccell{1.00} & \ccell{1.00} & \ccell{0.98} \\
    & L8 & \ccell{0.94} & \ccell{0.97} & \ccell{1.00} & \ccell{0.95} & \ccell{0.91} & \ccell{0.94} & \ccell{1.00} & \ccell{1.00} & \ccell{0.96} & \ccell{0.99} & \ccell{1.00} & \ccell{0.98} & \ccell{0.98} & \ccell{1.00} & \ccell{0.98} \\
    & D1+D2+D3 & \ccell{0.95} & \ccell{0.95} & \ccell{0.99} & \ccell{0.94} & \ccell{0.95} & \ccell{0.93} & \ccell{0.99} & \ccell{1.00} & \ccell{0.98} & \ccell{0.98} & \ccell{0.98} & \ccell{0.96} & \ccell{0.99} & \ccell{1.00} & \ccell{1.00} \\
    & D4+D5+D6+D7 & \ccell{0.96} & \ccell{0.97} & \ccell{1.00} & \ccell{0.93} & \ccell{0.97} & \ccell{0.98} & \ccell{1.00} & \ccell{1.00} & \ccell{0.95} & \ccell{0.98} & \ccell{1.00} & \ccell{0.96} & \ccell{0.99} & \ccell{1.00} & \ccell{0.98} \\
    & D8 & \ccell{0.89} & \ccell{0.95} & \ccell{0.99} & \ccell{0.95} & \ccell{0.93} & \ccell{0.91} & \ccell{1.00} & \ccell{1.00} & \ccell{0.97} & \ccell{0.98} & \ccell{0.99} & \ccell{0.97} & \ccell{0.99} & \ccell{1.00} & \ccell{0.98} \\
    & C1 & \ccell{0.91} & \ccell{0.95} & \ccell{0.99} & \ccell{0.88} & \ccell{0.89} & \ccell{0.94} & \ccell{1.00} & \ccell{1.00} & \ccell{0.93} & \ccell{0.98} & \ccell{0.99} & \ccell{0.95} & \ccell{0.98} & \ccell{1.00} & \ccell{0.98} \\
    & C2 & \ccell{0.95} & \ccell{0.96} & \ccell{1.00} & \ccell{0.95} & \ccell{0.92} & \ccell{0.95} & \ccell{0.98} & \ccell{1.00} & \ccell{0.98} & \ccell{0.98} & \ccell{0.99} & \ccell{0.98} & \ccell{1.00} & \ccell{1.00} & \ccell{0.95} \\
    & C3 & \ccell{0.95} & \ccell{1.00} & \ccell{1.00} & \ccell{0.96} & \ccell{0.95} & \ccell{0.94} & \ccell{0.99} & \ccell{1.00} & \ccell{0.98} & \ccell{0.98} & \ccell{1.00} & \ccell{0.99} & \ccell{1.00} & \ccell{1.00} & \ccell{0.97} \\
 
\midrule
\multirow{13}{*}{\rotatebox{90}{\textbf{\reposvulcpp}}}
    & original & \ccell{0.54} & \ccell{0.43} & \ccell{0.32} & \ccell{0.90} & \ccell{0.62} & \ccell{0.51} & \ccell{1.00} & \ccell{0.98} & \ccell{0.79} & \ccell{0.67} & \ccell{0.70} & \ccell{0.87} & \ccell{0.69} & \ccell{0.98} & \ccell{0.55} \\
\cmidrule(lr){2-17}
    & L1 & \ccell{0.74} & \ccell{0.64} & \ccell{0.65} & \ccell{0.83} & \ccell{0.55} & \ccell{0.55} & \ccell{1.00} & \ccell{0.99} & \ccell{0.62} & \ccell{0.73} & \ccell{0.62} & \ccell{0.88} & \ccell{0.89} & \ccell{0.99} & \ccell{0.74} \\
    & L2 & \ccell{0.46} & \ccell{0.54} & \ccell{0.68} & \ccell{0.90} & \ccell{0.64} & \ccell{0.64} & \ccell{0.99} & \ccell{0.98} & \ccell{0.76} & \ccell{0.65} & \ccell{0.80} & \ccell{0.89} & \ccell{0.89} & \ccell{0.99} & \ccell{0.62} \\
    & L3 & \ccell{0.46} & \ccell{0.71} & \ccell{0.60} & \ccell{0.87} & \ccell{0.64} & \ccell{0.65} & \ccell{0.99} & \ccell{0.98} & \ccell{0.69} & \ccell{0.74} & \ccell{0.86} & \ccell{0.92} & \ccell{0.92} & \ccell{0.98} & \ccell{0.65} \\
    & L4+L5+L6 & \ccell{0.52} & \ccell{0.54} & \ccell{0.62} & \ccell{0.92} & \ccell{0.62} & \ccell{0.69} & \ccell{1.00} & \ccell{0.98} & \ccell{0.77} & \ccell{0.82} & \ccell{0.75} & \ccell{0.93} & \ccell{0.83} & \ccell{0.98} & \ccell{0.76} \\
    & L7 & \ccell{0.60} & \ccell{0.55} & \ccell{0.71} & \ccell{0.89} & \ccell{0.70} & \ccell{0.56} & \ccell{1.00} & \ccell{0.98} & \ccell{0.82} & \ccell{0.69} & \ccell{0.86} & \ccell{0.86} & \ccell{0.89} & \ccell{0.98} & \ccell{0.60} \\
    & L8 & \ccell{0.67} & \ccell{0.86} & \ccell{0.94} & \ccell{0.96} & \ccell{0.83} & \ccell{0.85} & \ccell{0.99} & \ccell{1.00} & \ccell{0.88} & \ccell{0.82} & \ccell{0.99} & \ccell{0.99} & \ccell{0.90} & \ccell{1.00} & \ccell{0.80} \\
    & D1+D2+D3 & \ccell{0.65} & \ccell{0.68} & \ccell{0.81} & \ccell{0.85} & \ccell{0.77} & \ccell{0.67} & \ccell{1.00} & \ccell{0.96} & \ccell{0.82} & \ccell{0.83} & \ccell{0.81} & \ccell{0.87} & \ccell{0.96} & \ccell{0.98} & \ccell{0.73} \\
    & D4+D5+D6+D7 & \ccell{0.61} & \ccell{0.54} & \ccell{0.65} & \ccell{0.94} & \ccell{0.67} & \ccell{0.61} & \ccell{1.00} & \ccell{0.99} & \ccell{0.73} & \ccell{0.73} & \ccell{0.79} & \ccell{0.87} & \ccell{0.80} & \ccell{1.00} & \ccell{0.60} \\
    & D8 & \ccell{0.46} & \ccell{0.54} & \ccell{0.68} & \ccell{0.94} & \ccell{0.52} & \ccell{0.64} & \ccell{1.00} & \ccell{0.99} & \ccell{0.77} & \ccell{0.75} & \ccell{0.77} & \ccell{0.90} & \ccell{0.89} & \ccell{0.99} & \ccell{0.62} \\
    & C1 & \ccell{0.69} & \ccell{0.52} & \ccell{0.57} & \ccell{0.82} & \ccell{0.55} & \ccell{0.68} & \ccell{0.99} & \ccell{0.95} & \ccell{0.73} & \ccell{0.68} & \ccell{0.76} & \ccell{0.87} & \ccell{0.80} & \ccell{0.98} & \ccell{0.64} \\
    & C2 & \ccell{0.56} & \ccell{0.52} & \ccell{0.51} & \ccell{0.92} & \ccell{0.67} & \ccell{0.58} & \ccell{1.00} & \ccell{1.00} & \ccell{0.79} & \ccell{0.69} & \ccell{0.79} & \ccell{0.89} & \ccell{0.83} & \ccell{1.00} & \ccell{0.62} \\
    & C3 & \ccell{0.50} & \ccell{0.76} & \ccell{1.00} & \ccell{0.93} & \ccell{0.98} & \ccell{0.83} & \ccell{0.99} & \ccell{1.00} & \ccell{0.92} & \ccell{0.92} & \ccell{1.00} & \ccell{0.99} & \ccell{0.94} & \ccell{1.00} & \ccell{0.94} \\
 
\midrule
\multirow{13}{*}{\rotatebox{90}{\textbf{\reposvulpy}}}
    & original & \ccell{0.48} & \ccell{0.49} & \ccell{0.64} & \ccell{0.85} & \ccell{0.58} & \ccell{0.54} & \ccell{0.99} & \ccell{0.96} & \ccell{0.75} & \ccell{0.63} & \ccell{0.71} & \ccell{0.71} & \ccell{0.48} & \ccell{0.94} & \ccell{0.45} \\
\cmidrule(lr){2-17}
    & L1 & \ccell{0.77} & \ccell{0.77} & \ccell{0.83} & \ccell{0.82} & \ccell{0.54} & \ccell{0.54} & \ccell{1.00} & \ccell{0.96} & \ccell{0.66} & \ccell{0.64} & \ccell{0.72} & \ccell{0.79} & \ccell{0.68} & \ccell{0.95} & \ccell{0.61} \\
    & L2 & \ccell{0.44} & \ccell{0.62} & \ccell{0.75} & \ccell{0.83} & \ccell{0.60} & \ccell{0.55} & \ccell{1.00} & \ccell{0.95} & \ccell{0.72} & \ccell{0.68} & \ccell{0.79} & \ccell{0.83} & \ccell{0.55} & \ccell{0.95} & \ccell{0.52} \\
    & L3 & \ccell{0.62} & \ccell{0.72} & \ccell{0.80} & \ccell{0.85} & \ccell{0.62} & \ccell{0.54} & \ccell{1.00} & \ccell{0.98} & \ccell{0.78} & \ccell{0.65} & \ccell{0.82} & \ccell{0.81} & \ccell{0.85} & \ccell{0.98} & \ccell{0.56} \\
    & L4+L5+L6 & \ccell{0.47} & \ccell{0.57} & \ccell{0.74} & \ccell{0.86} & \ccell{0.62} & \ccell{0.52} & \ccell{1.00} & \ccell{0.97} & \ccell{0.71} & \ccell{0.71} & \ccell{0.77} & \ccell{0.81} & \ccell{0.53} & \ccell{0.97} & \ccell{0.57} \\
    & L7 & \ccell{0.55} & \ccell{0.71} & \ccell{0.86} & \ccell{0.89} & \ccell{0.70} & \ccell{0.66} & \ccell{0.99} & \ccell{0.98} & \ccell{0.80} & \ccell{0.78} & \ccell{0.86} & \ccell{0.86} & \ccell{0.76} & \ccell{0.99} & \ccell{0.67} \\
    & L8 & \ccell{0.59} & \ccell{0.79} & \ccell{0.94} & \ccell{0.93} & \ccell{0.74} & \ccell{0.75} & \ccell{1.00} & \ccell{1.00} & \ccell{0.90} & \ccell{0.76} & \ccell{0.96} & \ccell{0.93} & \ccell{0.69} & \ccell{1.00} & \ccell{0.77} \\
    & D1+D2+D3 & \ccell{0.62} & \ccell{0.76} & \ccell{0.85} & \ccell{0.81} & \ccell{0.61} & \ccell{0.58} & \ccell{1.00} & \ccell{0.96} & \ccell{0.76} & \ccell{0.71} & \ccell{0.80} & \ccell{0.82} & \ccell{0.82} & \ccell{0.97} & \ccell{0.58} \\
    & D4+D5+D6+D7 & \ccell{0.51} & \ccell{0.59} & \ccell{0.75} & \ccell{0.85} & \ccell{0.66} & \ccell{0.54} & \ccell{0.99} & \ccell{0.98} & \ccell{0.69} & \ccell{0.62} & \ccell{0.76} & \ccell{0.81} & \ccell{0.59} & \ccell{0.97} & \ccell{0.54} \\
    & D8 & \ccell{0.49} & \ccell{0.59} & \ccell{0.74} & \ccell{0.84} & \ccell{0.64} & \ccell{0.57} & \ccell{1.00} & \ccell{0.98} & \ccell{0.77} & \ccell{0.63} & \ccell{0.77} & \ccell{0.79} & \ccell{0.58} & \ccell{0.96} & \ccell{0.56} \\
    & C1 & \ccell{0.50} & \ccell{0.55} & \ccell{0.76} & \ccell{0.83} & \ccell{0.57} & \ccell{0.55} & \ccell{1.00} & \ccell{1.00} & \ccell{0.67} & \ccell{0.63} & \ccell{0.74} & \ccell{0.78} & \ccell{0.57} & \ccell{0.97} & \ccell{0.50} \\
    & C2 & \ccell{0.46} & \ccell{0.60} & \ccell{0.73} & \ccell{0.87} & \ccell{0.54} & \ccell{0.60} & \ccell{1.00} & \ccell{0.97} & \ccell{0.77} & \ccell{0.65} & \ccell{0.80} & \ccell{0.82} & \ccell{0.58} & \ccell{0.96} & \ccell{0.51} \\
    & C3 & \ccell{0.47} & \ccell{0.78} & \ccell{0.92} & \ccell{0.92} & \ccell{0.77} & \ccell{0.68} & \ccell{1.00} & \ccell{1.00} & \ccell{0.89} & \ccell{0.75} & \ccell{0.96} & \ccell{0.83} & \ccell{0.83} & \ccell{1.00} & \ccell{0.84} \\
 
\midrule
\multirow{13}{*}{\rotatebox{90}{\textbf{\primevulc}}}
    & original & \ccell{0.64} & \ccell{0.39} & \ccell{0.47} & \ccell{0.91} & \ccell{0.54} & \ccell{0.62} & \ccell{0.99} & \ccell{0.99} & \ccell{0.77} & \ccell{0.62} & \ccell{0.82} & \ccell{0.94} & \ccell{0.55} & \ccell{0.99} & \ccell{0.48} \\
\cmidrule(lr){2-17}
    & L1 & \ccell{0.78} & \ccell{0.80} & \ccell{0.82} & \ccell{0.89} & \ccell{0.55} & \ccell{0.64} & \ccell{1.00} & \ccell{1.00} & \ccell{0.75} & \ccell{0.66} & \ccell{0.75} & \ccell{0.98} & \ccell{0.90} & \ccell{1.00} & \ccell{0.69} \\
    & L2 & \ccell{0.59} & \ccell{0.65} & \ccell{0.73} & \ccell{0.92} & \ccell{0.62} & \ccell{0.58} & \ccell{0.99} & \ccell{1.00} & \ccell{0.70} & \ccell{0.61} & \ccell{0.88} & \ccell{0.97} & \ccell{0.74} & \ccell{1.00} & \ccell{0.57} \\
    & L3 & \ccell{0.50} & \ccell{0.78} & \ccell{0.71} & \ccell{0.93} & \ccell{0.63} & \ccell{0.61} & \ccell{0.99} & \ccell{0.99} & \ccell{0.85} & \ccell{0.60} & \ccell{0.88} & \ccell{0.97} & \ccell{0.94} & \ccell{1.00} & \ccell{0.69} \\
    & L4+L5+L6 & \ccell{0.60} & \ccell{0.71} & \ccell{0.79} & \ccell{0.91} & \ccell{0.64} & \ccell{0.67} & \ccell{1.00} & \ccell{1.00} & \ccell{0.85} & \ccell{0.71} & \ccell{0.92} & \ccell{0.94} & \ccell{0.71} & \ccell{1.00} & \ccell{0.78} \\
    & L7 & \ccell{0.69} & \ccell{0.69} & \ccell{0.77} & \ccell{0.89} & \ccell{0.58} & \ccell{0.68} & \ccell{1.00} & \ccell{1.00} & \ccell{0.79} & \ccell{0.62} & \ccell{0.87} & \ccell{0.95} & \ccell{0.82} & \ccell{0.99} & \ccell{0.57} \\
    & L8 & \ccell{0.62} & \ccell{0.90} & \ccell{0.96} & \ccell{0.94} & \ccell{0.81} & \ccell{0.80} & \ccell{0.99} & \ccell{1.00} & \ccell{0.90} & \ccell{0.83} & \ccell{0.99} & \ccell{0.92} & \ccell{0.86} & \ccell{1.00} & \ccell{0.86} \\
    & D1+D2+D3 & \ccell{0.76} & \ccell{0.81} & \ccell{0.93} & \ccell{0.91} & \ccell{0.73} & \ccell{0.71} & \ccell{1.00} & \ccell{1.00} & \ccell{0.83} & \ccell{0.73} & \ccell{0.89} & \ccell{0.98} & \ccell{0.92} & \ccell{1.00} & \ccell{0.78} \\
    & D4+D5+D6+D7 & \ccell{0.69} & \ccell{0.69} & \ccell{0.71} & \ccell{0.88} & \ccell{0.61} & \ccell{0.65} & \ccell{1.00} & \ccell{1.00} & \ccell{0.71} & \ccell{0.65} & \ccell{0.90} & \ccell{0.96} & \ccell{0.80} & \ccell{1.00} & \ccell{0.69} \\
    & D8 & \ccell{0.67} & \ccell{0.73} & \ccell{0.76} & \ccell{0.93} & \ccell{0.62} & \ccell{0.57} & \ccell{1.00} & \ccell{1.00} & \ccell{0.85} & \ccell{0.66} & \ccell{0.88} & \ccell{0.96} & \ccell{0.78} & \ccell{1.00} & \ccell{0.66} \\
    & C1 & \ccell{0.74} & \ccell{0.63} & \ccell{0.72} & \ccell{0.91} & \ccell{0.56} & \ccell{0.59} & \ccell{0.99} & \ccell{1.00} & \ccell{0.66} & \ccell{0.64} & \ccell{0.86} & \ccell{0.94} & \ccell{0.70} & \ccell{1.00} & \ccell{0.64} \\
    & C2 & \ccell{0.66} & \ccell{0.72} & \ccell{0.67} & \ccell{0.92} & \ccell{0.64} & \ccell{0.52} & \ccell{1.00} & \ccell{1.00} & \ccell{0.82} & \ccell{0.65} & \ccell{0.87} & \ccell{0.97} & \ccell{0.80} & \ccell{1.00} & \ccell{0.60} \\
    & C3 & \ccell{0.44} & \ccell{0.87} & \ccell{1.00} & \ccell{0.92} & \ccell{0.92} & \ccell{0.96} & \ccell{1.00} & \ccell{1.00} & \ccell{0.90} & \ccell{0.98} & \ccell{0.99} & \ccell{0.97} & \ccell{0.87} & \ccell{1.00} & \ccell{0.97} \\
 
\bottomrule
\end{tabular}
}
\label{tab:detection-result-all-datasets-existence-evaluation}
\vspace{-3.5mm}
\end{table*}

\end{document}